\documentclass[a4paper,12pt,english,oneside,final]{article}
\usepackage[utf8]{inputenc} 
\usepackage[T1]{fontenc} 
\usepackage[english]{babel}
\usepackage{setspace}
\usepackage{subfig}

\usepackage{listings}
\usepackage{xcolor}
\usepackage{lmodern}

\usepackage[natbibapa]{apacite} 

\usepackage{amsthm,amssymb,mathrsfs,amsmath,amsfonts}
\usepackage{mathtools}
\usepackage{nicefrac}

\usepackage{thmtools,thm-restate} 
\usepackage{euscript} 
\usepackage{verbatim}
\usepackage{caption}
\usepackage{enumitem}
\usepackage{microtype}
\usepackage{dsfont}
\usepackage{csquotes}
\usepackage{xcolor}
\usepackage{breqn}
\usepackage{tabulary}
\usepackage{booktabs}

\usepackage{graphicx}

\definecolor{coolblack}{rgb}{0.0, 0.18, 0.39}
\definecolor{codegreen}{rgb}{0,0.6,0}
\definecolor{codegray}{rgb}{0.5,0.5,0.5}
\definecolor{codepurple}{rgb}{0.58,0,0.82}
\definecolor{backcolour}{rgb}{0.95,0.95,0.92}
\usepackage[colorlinks=true,linkcolor=coolblack,
urlcolor=coolblack,
citecolor=coolblack,pdftex]{hyperref} 
\usepackage[all]{hypcap}  

\lstdefinestyle{mystyle}{
    backgroundcolor=\color{backcolour},   
    commentstyle=\color{codegreen},
    keywordstyle=\color{magenta},
    numberstyle=\tiny\color{codegray},
    stringstyle=\color{codepurple},
    basicstyle=\ttfamily\footnotesize,
    breakatwhitespace=false,         
    breaklines=true,                 
    captionpos=b,                    
    keepspaces=true,                 
    numbers=left,                    
    numbersep=5pt,                  
    showspaces=false,                
    showstringspaces=false,
    showtabs=false,                  
    tabsize=2
}

\usepackage{geometry}
\makeatletter

\newtheorem{proposition}{Proposition}

\newtheorem{lemma}{Lemma}
\newtheorem{definition}{Definition}
\newtheorem{observation}{Observation}

\theoremstyle{definition}
\newtheorem{step}{Step}

\usepackage[colorinlistoftodos]{todonotes}

\makeatother

\usepackage{babel}

\makeatletter
\def\@fnsymbol#1{\ensuremath{\ifcase#1\or \dagger\or \ddagger\or
		\mathsection\or \mathparagraph\or \|\or **\or \dagger\dagger
		\or \ddagger\ddagger \else\@ctrerr\fi}}
\makeatother

\begin{document}
	
	\pagestyle{plain}
	
	\author{Simon Lodato\thanks{Department of Economics, Middlesex University London, \textit{e-mail}: \href{mailto:s.lodato@mdx.ac.uk}{\sf s.lodato@mdx.ac.uk}} \and Christos Mavridis\thanks{Department of Economics, ``Gabriele d’Annunzio'' University of Chieti-Pescara, \textit{e-mail}: \href{mailto:christos.mavridis@unich.it}{\sf \mbox{christos.mavridis@unich.it}}.} \and Federico Vaccari\thanks{Department of Economics, University of Bergamo, \textit{e-mail}: \href{mailto:vaccari.econ@gmail.com}{\sf vaccari.econ@gmail.com}.}}
	\title{The Unelected Hand? Bureaucratic Influence and Electoral Accountability\thanks{We thank Emiel Awad, Benjamin Blumenthal, Stephen Calabrese, Gerard Domènech-Gironell, Paweł Dziewulski, Satoshi Fukuda, Elena Manzoni, Nobuhiro Mizuno, Federico Trombetta, Nikolas Tsakas, as well as seminar audiences at the 51st Meeting of the European Public Choice Society, the 2024 Workshop on Economic Theory at Bocconi University, the European Meeting on Game Theory 2024 (SING19), the GRASS XVIII at Collegio Carlo Alberto, the University of Sussex, the University of Pavia, the Catholic University of Milan, and the University of Bergamo for their useful comments. All errors are ours.}}

	\date{}
	\maketitle

        \begin{abstract}
		What role do non-elected bureaucrats play when elections provide imperfect accountability and create incentives for pandering? We develop a model where politicians and bureaucrats interact to implement policy. Both can either be good, sharing the voters' preferences over policies, or bad, intent on enacting policies that favor special interests. Our analysis identifies the conditions under which good bureaucrats choose to support, oppose, or force pandering. When bureaucrats wield significant influence over policy decisions, good politicians lose their incentives to pander, a shift that ultimately benefits voters. An intermediate level of bureaucratic influence over policymaking can be voter-optimal: large enough to prevent pandering but small enough to avoid granting excessive influence to potentially bad bureaucrats.
        \end{abstract}

	
	
	\thispagestyle{empty}


	\pagebreak
	
\begin{quote}
\textit{Something I've learnt not only during my time as Attorney, but also during my time as a Brexit minister, is that some of the biggest battles that you face as a minister are, in the nicest possible way, with Whitehall and internally with civil servants, as opposed to your political battles in the chamber. [...] Don’t take this as an opportunity to bash the civil service. But what I have seen time and time again, both in policymaking and in broader decision making, [is] that there is a Remain bias. I’ll say it. I have seen resistance to some of the measures that ministers have wanted to bring forward.}
\begin{flushright}
Suella Braverman, then Attorney General of the United Kingdom. 

Interview with The Sunday Telegraph, July 3, 2022.
\end{flushright}
\end{quote}

\section{Introduction}
National bureaucracies are an essential part of modern states, with bureaucrats---non-elected public employees or civil servants---playing a significant role in designing and implementing policies alongside elected politicians. The interaction, and sometimes conflict, between bureaucrats and politicians is interesting from both economic and social perspectives, as it may affect the accountability of elected politicians, potentially harming voters. Additionally, it can alter the intended policies of politicians, resulting in more or less beneficial outcomes for society. While politicians may not always possess extensive policymaking experience when they assume office, bureaucrats tend to be highly educated with specialist knowledge, especially those operating at the highest levels of a state's bureaucracy. They can tap into the bureaucracy's institutional memory and know-how, all of which are crucial assets for policymaking. However, these assets can also lead to bureaucracies becoming overly powerful to the extent that they can extract rents.

Political agency models, which explore how elections discipline politicians and enable voters to select their leaders, provide a framework for examining the role of bureaucrats in policymaking. These models highlight that elections may be imperfect instruments: sometimes bad politicians mimic good politicians to secure re-election. Furthermore, even good politicians may \textit{pander}, propose policies they know to be not socially optimal to enhance their re-election prospects. The presence of bureaucrats adds complexity to the policymaking process, potentially influencing the implemented policy and the effectiveness of elections to hold politicians accountable. Our paper focuses precisely on the extent to which bureaucrats should influence policymaking in an environment where voters' imperfect information can induce good politicians to pander. 

In particular, we identify conditions under which higher bureaucratic influence can benefit voters, shifting from an equilibrium where pandering takes place to one that does not. A good politician knows that if the bureaucrat who may influence the policy is good, their positive influence on policy will remain in the future, even if the politician themselves ends up being substituted with a bad politician. On the other hand, if the bureaucrat is bad, there's a high probability of a bad policy in the future, disincentivizing the good politician from choosing a bad policy in the present. Therefore, increased bureaucratic influence reduces good politicians' incentives to pander.

We analyze these ideas by exploring a two-period political agency model featuring ``good'' or ``bad'' politicians and ``good'' or ``bad'' bureaucrats. One of two possible states of the world is realized, which politicians and bureaucrats---but not voters---observe. Politicians then propose a policy, and bureaucrats can attempt to change it.  Voters prefer policies that match with the realized state of the world. Good politicians and good bureaucrats have preferences over policies that align with those of voters. In contrast, bad politicians and bad bureaucrats have private interests: regardless of the state, they receive additional, though uncertain, rents from implementing a particular policy. Moreover, politicians of both types benefit from office rents while in power. Voters, unable to observe the state of the world directly, can re-elect or replace incumbent politicians based only on the policies that are implemented. The good politicians' desire to remain in office incentivizes them to pander, and voters interpret the implementation of pandering policies as a signal that the incumbent politicians are good, leading to their re-election. 

The general idea behind equilibrium behavior is as follows. Voters understand that both politicians and bureaucrats may have private interests.\footnote{For example, they may be corrupt or captured by elites. \cite{boehmke2006} develop a model in which special interests, such as lobbies, directly target the bureaucracy to influence policy.} Voters associate policies like reduced taxes for high-income brackets, lower corporate taxes, or salary increases for unionized workers with these elites, as such measures often align with elite interests but are unpopular with the broader public. To avoid the appearance of corruption or elite capture, a forward-looking good politician may strategically avoid these policies, even if they are socially optimal, opting instead to pander to secure re-election. Once re-elected, good politicians can implement policies that are truly beneficial for society. However, this strategy can be disrupted by bureaucrats who influence policy implementation. 

We identify conditions under which good bureaucrats consistently attempt to correct current policymaking so that policy aligns with the state of the world. Such corrections may lead to the re-election of bad politicians, who would otherwise be replaced, and to the removal of good politicians, who would otherwise be re-elected. Good bureaucrats weigh the social costs of implementing a policy-state mismatch. The bureaucrat will seek to correct such behavior if the mismatch is relatively more socially costly when pandering occurs. Otherwise, the bureaucrat accommodates pandering instead, hoping this will increase the likelihood of a good politician being re-elected. Depending on the parameters, good bureaucrats may either seek to improve current policymaking or to enhance the re-election chances of good politicians. We find that, depending on the situation, they may achieve this by accommodating pandering or by forcing it on politicians unwilling to engage.

Central to our paper is the conflict of interest between elected and unelected policymakers, particularly between the good ones. The willingness of good politicians to pander is, along with other factors, influenced by bureaucratic power and office rents. A key finding is that higher bureaucratic influence can deter good politicians from pandering. Moreover, the level of bureaucratic influence that defines the switching point between equilibria in which good bureaucrats would pander and those in which they would not want to pander can also be voter-optimal, thus justifying an intermediate level of bureaucratic influence in policymaking. Such a level must be large enough to prevent pandering but small enough to avoid granting excessive influence to potentially corrupt bureaucrats. These findings contrast with the view in the literature that higher bureaucratic influence weakens political accountability \citep{ujhelyi2014,mccubbins1987administrative,Martin2021}.

In recent years, the topic of bureaucrats' ability to influence policy has garnered increasing attention.\footnote{Readers can turn to \cite{Gailmard2012} for a review of formal models of bureaucracy.} Bureaucrats can impact policy not only by voting in elections \citep{Forand2022} but also through their efforts \citep{Blumenthal2023, Yazaki2018, Li2023, Slough2022, Li2020}, their chosen level of capacity investments \citep{ting2011}, or by directly altering politicians' policies \citep{ujhelyi2014,Martin2021}. This line of work connects to the concepts of judicial review, where the judiciary has the power to overturn executive decisions \citep[see][]{Stephenson2011}, and hierarchical accountability, where elected intermediaries can remove policymakers on behalf of voters \citep{Vlaicu2016}. 

Much of the research on bureaucrats has focused on the effects of bureaucratic activity on the accountability of politicians. Elections allow voters to hold politicians accountable for their policies; however, this mechanism does not apply to bureaucrats. Moreover, the difficulty voters face in distinguishing between the influences of bureaucrats and politicians on policy makes disciplining politicians harder. Conversely, bureaucrats serve as vital components in the policymaking process.\footnote{Related to this is the concept of central bank independence \citep{dehaan2019}.} Their expertise is essential for policy implementation and can act as a bulwark against the whims of populist leaders \citep{sasso2021bureaucrats}. Additionally, bureaucratic efficiency can bolster voter confidence in effectively using tax revenues, despite concerns over bureaucratic size \citep{Forand2019}. However, it is also common for politicians to blame bureaucrats for policy failures or their own inadequacies \citep{Awad2023, Miller2022, fiorina1982}.

The possibility that elites or special interests may capture politicians has been well-analyzed in the literature (see \cite{schnakenberg2024} for an excellent survey on special interest influence). The idea in our paper is that these elites are engaged in a quid pro quo trade with the bad policymakers, who choose a policy preferred by the elites and get a reward from them in the form of rent. Since our model does not focus on the incentives and behavior of the elites,  they are not active players: they always reward bad policy makers in exchange for their preferred policy. \cite{schnakenberg2019}, in a model in which elites---interest groups in their nomenclature---are active players, show that this reward can be an equilibrium behavior on their part.

As mentioned above, in this paper, we associate \emph{pandering} with scenarios where politicians propose policies they know are not socially optimal, but which can enhance their re-election chances. This definition is consistent with the one given in Chapter 3 of \cite{besley2007}. In the literature, pandering has been studied as a signal of congruence with the preferences of the voters \citep{maskin2004politician, MaskinTirole2019, Trombetta2020, trombetta2022} or of competence \citep{Canes-Wrone2001}. In  \cite{maskin2004politician}, to pander, politicians choose a particular policy to signal they share policy preferences with the voters. In their setup, this policy is the ex-ante popular policy. In our setting, it is the policy that is not associated with captured policymaking.  

A similar mechanism to ours is at play in \cite{acemoglu2013political}, where a non-captured politician enacts a policy to the left of the one preferred by the median voter to signal that ``he is not beholden to the interests of the right.'' \cite{Trombetta2020} and \cite{trombetta2022} develop a setup similar to ours in which, all else equal, bad politicians prefer a specific policy regardless of the state of the world. In contrast, like voters, good politicians prefer policies that match the state. Good politicians may nevertheless choose to pander, that is, to avoid implementing the policy favored by bad politicians, even when appropriate, to signal that they are not bad.\footnote{The behavior of the good politician in our model may alternatively be described as \emph{posturing} in the sense used, for example, by \cite{Stephenson2011} or \cite{Bils2023}. We retain the term pandering, to emphasize the strategic selection of suboptimal policies aimed at re-election and to align with previous related works, such as the ones mentioned here.}

Similar to our work, \cite{Martin2021},  \cite{heo2024bureaucratic}, and \cite{Yazaki2018} study the relationship between accountability and bureaucratic influence. \cite{Martin2021} model policymaking as jointly determined by a politician and a bureaucrat, each of whom may be ``good'' or ``bad.'' Unlike their approach, where bureaucrats always select their preferred policies, our model treats these decisions as endogenous. This distinction is crucial as it broadens the scope for applications and outcomes of our model, including the study of pandering and collusion. For instance, \cite{Martin2021} argue and provide empirical evidence that increased bureaucratic influence decreases accountability, typically by weakening the ability of elections to select good politicians, a potentially negative outcome for voters. In contrast, our analysis allows for increased bureaucratic influence to improve accountability through shifting  the equilibrium from one where pandering occurs to one where it does not. Moreover, within the same equilibrium class, it does so by diminishing the re-election chances of bad politicians. Thus, bureaucratic influence can sometimes be beneficial for accountability. We find that this result can hold even when bureaucrats are more likely to be corrupt than politicians.

\cite{heo2024bureaucratic} examine the accountability problem where a politician must decide on reform implementation, subject to potential sabotage by bureaucrats. While bureaucrats introduce uncertainty into the voter's re-election decision, they are not policymakers and cannot propose reforms or assess their value independently.  \cite{Yazaki2018} considers a model in which bureaucrats can sabotage politicians by choosing the level of public good provision, which then serves as a signal of politician quality for the voters, potentially improving political selection. 

Close to the spirit of our paper, \cite{ujhelyi2014} explores a model with two periods, and good or bad bureaucrats and politicians. \cite{ujhelyi2014} examines the optimal punishment level, one class of the ``civil service laws,'' as a means to control an otherwise all-powerful bureaucracy: less control reduces the likelihood of achieving good policy in the first period and generally leads to worse second-period politicians. There are two important differences between our approach and Ujhelyi's. First, in our setting, voters do not observe the state of the world and rely only on the observed implemented policy for a signal of politician quality. Moreover, our bad policymakers have a particular favorite policy they want to implement regardless of the state of the world. Voters re-elect a politician if and only if they do not observe the policy favored by the bad politicians. As discussed earlier, these features may induce good politicians to pander, that is, to avoid implementing the bad policymakers’ preferred policy, even when appropriate. This creates a potential conflict between a good politician and a good bureaucrat, a tension absent in \cite{ujhelyi2014}.

We focus on the appropriate level of bureaucratic power and find that greater bureaucratic influence reduces good politicians’ incentives to pander, encouraging them to implement the appropriate policy. At the same time, it makes bad politicians less inclined to seek re-election. The former effect increases voter welfare in the first period, while the latter may improve selection, potentially raising the quality of second-period politicians. In this sense, our results stand in contrast to those of \cite{ujhelyi2014}.

The rest of this article is structured as follows: Section~\ref{sec:model} introduces the model. Section~\ref{sec:equilibrium} outlines and examines pandering and non-pandering equilibria and discusses their existence and implications. Section~\ref{sec:appl} is dedicated to applications. Specifically, Section~\ref{sec:benchmarks} explores two benchmarks: a dictatorial and a toothless bureaucracy. In Section~\ref{sec:cs}, we analyze the impact of bureaucratic influence on the voters' welfare. In Section~\ref{sec:electoralaccountability}, we study the effect of bureaucracy on electoral accountability and selection. Finally, Section~\ref{sec:conclusion} concludes. Proofs are in the Appendix.


\section{The model}\label{sec:model}

We analyze a two-period political agency model involving three players: a politician in charge ($P$), a bureaucrat ($B$), and a representative voter ($V$). Politicians and bureaucrats are \textit{policymakers}, and hereafter we sometimes refer to them as that. In each period $t\in\{1,2\}$, the incumbent politician and the bureaucrat put forward a policy proposal $q_t^j\in\{x,y\}$, $j\in\{P,B\}$.  Between periods, there is an election where the voter decides whether to replace the incumbent politician with a challenger.\footnote{To simplify notation, we do not consider the challenger as a player, since they take no strategic action.  If the incumbent is voted out, the challenger simply assumes office and becomes the new incumbent in the second period. Therefore, by using ``types,'' we treat the politician as a single player whose preferences may shift between periods. For simplicity, we denote the acting politician in both periods as $P$.}

\emph{State.---} A policy-relevant state of the world, $s_t\in \{x,y\}$, is drawn at the beginning of each period.\footnote{States and policies have the same labels in both periods for notational convenience only.} With probability $\rho\in(0,1)$, the state of the world in period $t$ is equal to $x$. The policymakers perfectly observe $s_t$ before choosing a policy proposal in period~$t$. The voter does not observe $s_1$ or any state-dependent payoffs before the election takes place. 

\emph{Policymaking.---} The period-$t$ implemented policy, $p_t$, depends on the policymakers proposals in the following way: when the politician and the bureaucrat propose the same policy, then $p_t$ will coincide with their proposals, i.e., $p_t=q_t^P=q_t^B$. Otherwise, the implemented policy will coincide with the bureaucrat's proposal with probability $\lambda\in\left(0,1\right)$,
and with the politician's proposal with probability $1-\lambda$. Score $\lambda$ represents the bureaucrat's ability to override the politician's proposals.

\emph{Election.---} Before the election, the voter observes the implemented policy, $p_1$, but does not observe the policymakers' proposals, $q_1^P$ and $q_1^B$. The voter's electoral decision after observing the period 1's implemented policy is $\nu(p_1)\in\{0,1\}$, where $\nu=1$ (resp.~$\nu=0$) indicates that the voter re-elects (resp.~replaces) the incumbent politician.\footnote{The voter's binary decision is without loss because we focus on pure-strategies only. Our model accommodates for an electorate with heterogeneous preferences as long as maximizing the electorate's welfare is equivalent to maximizing that of our representative (median) voter.}

\emph{Types.---} Policymakers can be either good ($g$) or bad ($b$). The bureaucrat's type, $\theta^B\in\{g,b\}$, is good with probability $\beta \in(0,1)$. The first-period politician's type, $\theta_1^P\in\{g,b\}$, is good with probability $\pi\in(0,1)$. The second-period politician's type, $\theta_2^P\in\{g,b\}$, depends on the voter's decision. If the voter re-elects the first-period incumbent, then the politician's type remains constant across periods, i.e., $\theta_2^P=\theta_1^P$. Otherwise, $\theta_2^P$ is drawn at the beginning of the second period according to the same distribution as $\theta_1^P$. Policymakers $P$ and $B$ privately know only their own type. The voter does not know the policymakers' types. The bureaucrat's type is constant across periods.

\emph{Payoffs.---} The payoffs players obtain from the policy implemented in period~$t$ depend on their own type and the period-$t$ state of the world. The voter's payoff equals to
\[
v(p_1,s_1) + \delta v(p_2,s_2),
\]
 where $\delta\in(0,1]$ is a common time-discount factor. With some abuse of notation, we define $p_t=s_t$ as a policy that matches the realized state, and $p_t \neq s_t$ as one that does not. The voter strictly prefers the implemented policy to match the same-period state.

Good policymakers ($\theta_t^P=g$ and $\theta^B=g$) share the same preferences over policies as the voter. This remains true for good politicians who are not in office, as they become voters themselves. By contrast, bad policymakers ($\theta_t^P=b$ and $\theta^B=b$) favour policy $y$ independently of the state: they extract rents $r_t^j\geq 0$ if the implemented policy is $p_t=y$, with $j\in\{P,B\}$ and $t\in\{1,2\}$. If $p_t=x$, then their period-$t$ rents are zero. Rents $r_t^j$ are distributed according to a cumulative distribution $F_t^j$, with mean $\mu_t^j$ and full support in $\left[0,\bar R_t^j\right]$, and are private information of policymaker~$j$ only. To simplify our analysis, we assume that bad policymakers receive a payoff of zero when out of office.\footnote{See 
\cite{besley2007} and  \cite{trombetta2022} for a similar treatment of out-of-office payoffs.} Policymakers, similarly to voters, discount future payoffs by $\delta$. Furthermore, both types of politicians get office rents $E \in \mathbb{R}$ in every period they are in charge.

\emph{Timeline.---} To sum up, the timing of the game is as follows. In period~1,
\begin{enumerate}[noitemsep]
    \item $s_1$, $\theta_1^P$, $\theta^B$, $r_1^P$, and $r_1^B$ are realized. State $s_1$ is observed only by the policymakers, $P$ and $B$, but not by the voter, $V$. $\theta_1^P$ and $r_1^P$ are private information of the politician, $P$. Likewise, $\theta^B$ and $r_1^B$ are private information of the bureaucrat, $B$;
    \item $P$ and $B$ sequentially choose policy proposals $q_1^j$, $j\in\{P,B\}$. $B$ observes the choice of $P$ before making their choice. Policy proposals are not observed by the voter, $V$. Given proposals $\left(q_1^P,q_1^B \right)$, the implemented policy $p_1$ is publicly realized;
    \item $V$ observes $p_1$ and decides whether to re-elect or replace $P$ with a challenger whose type is drawn from the same distribution as the incumbent's;
    \item Period-$1$ payoffs are obtained by the players.
\end{enumerate}
In period~$2$,
\begin{enumerate}[noitemsep]
    \item $s_2$, $r_2^P$, and $r_2^B$ are realized. State $s_2$ is observed by the policymakers, $P$ and $B$, but not by the voter, $V$. If the incumbent politician has been replaced with a challenger, then $\theta_2^P$ is realized and is private information of the politician in office, $P$. Rents $r_2^B$ and type $\theta^B$ are private information of the bureaucrat, $B$;
    \item $P$ and $B$ sequentially choose policy proposals $q_2^j$, $j\in\{P,B\}$. $B$ observes the choice of $P$ before making their choice. Given proposals $\left(q_2^P,q_2^B \right)$, the implemented policy $p_2$ is publicly realized;
    \item Period-$2$ payoffs are obtained by the players, and the game ends.
\end{enumerate}

\emph{Solution concept.---} The equilibrium concept is pure-strategy perfect Bayesian equilibrium (PBE). The analysis focuses on two classes of equilibria, \emph{pandering} and \emph{non-pandering}, which are defined in the following section.

\subsection{Discussion of the model}

While all the model's parameters are important in determining equilibrium behavior, the key parameter under scrutiny in this paper is $\lambda$. Formally, $\lambda$ is defined as the probability that disagreements between politicians and bureaucrats are resolved in favor of the latter’s choice. More descriptively, $\lambda$ represents the balance of power between politicians and bureaucrats, encapsulating their relative influence on shaping policy. We adopt a reduced-form approach to capture this tension in the policymaking process, treating $\lambda$ as an exogenous parameter. We then study its effect on equilibrium behavior and voter welfare as this parameter changes. This approach allows us to model both extreme cases and more nuanced scenarios. A \emph{spoils system}, where politicians appoint and retain full control over bureaucrats, corresponds to $\lambda \to 0$. In contrast, the independence of central bankers, who are fully shielded from political influence, corresponds to $\lambda \to 1$ \citep{rogoff1985optimal}.

An alternative approach would be to endogenize $\lambda$, for example, by determining it through a contest-like mechanism between the politician and the bureaucrat. While this would certainly be interesting, it is not the approach we pursue in this paper. By treating $\lambda$ as exogenous, we can analyze the resulting equilibrium behavior and welfare effects from institutional changes in bureaucratic influence, as might be performed by a social planner. Our approach avoids relying on a specific mechanism of determination and aligns with the view that bureaucratic influence can be externally imposed and (formally or informally) institutionalized \citep{raffler2022does}. This allows us to study the effects of external or exogenous variations in bureaucratic influence. In our context, $\lambda$ reflects the institutional framework and norms that determine how much influence bureaucrats have in shaping policy, and it varies based on the level of bureaucratic independence or insulation from political actors.

There are two additional modeling assumptions about the policymaking process worth discussing. First, we assume that the bureaucrat acts after the politician, allowing the bureaucrat to observe the politician's proposal before making a decision. This structure aligns with the interpretation that bureaucratic influence is associated with problems of imperfect monitoring or weak oversight. Politicians are responsible for the implementation of many policies but lack the time, resources, expertise, or capacity to implement all of them themselves. They must delegate part or all of this task to bureaucrats. Politicians instruct bureaucrats by indicating the policies they would like to see implemented (i.e., $q_t^P$), and bureaucrats have some freedom to deviate from these instructions due to bureaucratic influence. In an alternative specification where policymakers propose policies simultaneously, bureaucrats would not be able to update their priors about the politician's type and would instead act based on their prior. Our equilibrium characterization shows that this dynamic plays a crucial role in shaping bureaucratic behavior.  

Second, we assume that voters do not observe the policymakers' proposals before voting. This approach aligns with the idea that voters are unable to disentangle the actions of politicians from those of bureaucrats \citep{Martin2021}. Due to the complexity of policymaking, politicians cannot credibly signal their actions to voters, consistent with the notion that any such attempt amounts to cheap talk. Politicians and bureaucrats may blame each other for failures and claim credit for successes, further muddling communication. Crucially, if voters could observe the politicians' proposals, bureaucratic actions might influence the implemented policy but not electoral outcomes, and bureaucrats would have no bearing on electoral accountability. By contrast, our modeling approach enables an analysis of how bureaucratic influence also affects the selection of politicians, which is a key determinant of voter welfare.


\section{Equilibrium analysis}\label{sec:equilibrium}

This section develops our equilibrium characterization. In the second, and final, period of the game, policymakers naturally select their preferred policy, as there are no subsequent periods that influence their incentives (see Lemma~\ref{lemma:second} in Appendix~\ref{sec:app_cont_values}). Starting from this standard result, we step back to examine equilibrium play in the first period. We focus on informative equilibria, that is equilibria that improve the ability of voters to re-elect a good politician. The next definition provides a characterization of this class of equilibria.

\begin{definition}\label{def:informative}
    An informative equilibrium (IE) is a pure-strategy PBE in which the implemented policy conveys information about the politician's type, favoring outcome $p_1=x$ for re-election. That is, in an IE, after observing  implemented policy $x$, the voters' posterior belief that the politician is good is higher than their prior.
\end{definition}

Voters prefer a policy aligned with the state of the world. In period 2, only a good politician would select a policy that matches the state when the state is $x$. Furthermore, since there is no longer an impending election, the need for pandering disappears. Consequently, voters benefit from ensuring a good politician's tenure. To make this choice, voters seek an informative signal that the incumbent's type is good. Observing $p_1=x$ serves as such a signal, compelling the voter to favor re-electing the incumbent politician. 
 Next, in terms of the good politician behavior in the first period, we define two types of IE: pandering and non-pandering ones. 

We begin with pandering equilibria, in which a good politician may propose a policy they know to be suboptimal for the voter in order to signal their type and improve their re-election prospects. Specifically, we consider equilibria where, in the first period, good politicians always propose policy $x$ regardless of the state of the world. The following definition formalizes our notion of a pandering equilibrium.

\begin{definition}\label{def:pandering}
    A pandering equilibrium (PE) is an IE in which, in period~1, good politicians propose policy $x$ when the state is $y$. 
\end{definition}

Given the beliefs of the voters, a good politician would engage in pandering only if such behavior informatively signals their type, considering that voters cannot directly observe this characteristic. For the implemented policy to be informative, it must convey specific information about the politician's type. Because of bureaucratic interference, the implemented policy does not necessarily coincide with the politician's proposal. Characterizing the good bureaucrats' equilibrium behavior is an essential next step. This behavior is driven by the ratio of policy-state mismatch costs $\Delta$, which is given by
\[
\Delta \coloneqq \frac{v(y,y)-v(x,y)}{v(x,x)-v(y,x)}>0.
\]

We refer to $v(y,y)-v(x,y)$ as the mismatch costs in state $y$, and to $v(x,x)-v(y,x)$ as the mismatch costs in state $x$. When $\Delta>1$ (resp.~$\Delta<1$), the mismatch costs are larger in state $y$ (resp.~$x$) than in state $x$ (resp.~$y$).  If $\Delta$ is relatively high, the negative effect of a policy-state mismatch under state $y$, the \emph{pandering state}, is relatively higher than the negative effect of a policy-state mismatch under state $x$, giving  good bureaucrats strong incentives to try to avoid it by contesting pandering.

In practice, the costs associated with policy mismatches are often specific to particular states and can be highly asymmetric. \cite{trombetta2022}, pp. 132--133, provide a good example in this regard:
``...\emph{consider environmental Special Interest Groups (SIGs) arguing in favor of stricter (and costly) environmental policies irrespective of the true state of the world, i.e. whether they are needed or not. Probably, the cost of choosing the wrong environmental policy, when protection is needed, is higher than the cost of adopting restrictions, when they are not needed}.'' Many situations follow this pattern---that is, taking preventive measures when they are not needed might be far less costly than failing to take them when they are. 

An analogous extreme yet clarifying example would be enforcing public health measures in normal years versus during major health crises.
A similar principle applies in public finance: investing in tax administration and compliance systems during stable times might seem somewhat excessive, but it significantly reduces revenue losses and improves fiscal capacity when a crisis hits and public funds are urgently needed. In what follows, the relative costs of mismatches between policy and state will be crucial for the equilibrium behavior of good bureaucrats.

\subsection{Pandering equilibria with a correcting bureaucracy}

Definition~\ref{def:pandering} describes pandering equilibria through the behavior of good politicians. This definition is consistent with the predominant approach in the related literature focusing on the interaction between voters and politicians. We contribute to this literature by introducing bureaucrats, which requires us to characterize their equilibrium behavior. To this end, we begin by presenting and discussing in detail the PE where good bureaucrats seek to align policy with the state of the world. The next definition provides a characterization of this class of equilibria.
\begin{definition}\label{def:conj}
    A pandering equilibrium with a correcting bureaucracy (PECB) is a PE in which
    \begin{itemize}[nosep]
    \item Good bureaucrats always propose a policy that matches the state of the world;
    \item Bad bureaucrats always propose policy $y$ after politicians propose policy $x$;
    \item The proposal of bad bureaucrats after politicians propose $y$ is state-independent but depends on their realized rents;
    \item The proposal of bad politicians is state-independent but depends on their realized rents.
    \end{itemize}
\end{definition}

In PECB, good bureaucrats aim to align policy with the state of the world, employing a ``correcting'' strategy that may diverge from the incumbent politician's goals. A scenario of particular interest arises when a good bureaucrat and a good politician simultaneously hold office. In such instances, both policymakers and voters agree that the optimal policy corresponds to the state of the world. Interestingly, the concurrence of two good policymakers is insufficient to guarantee the adoption of such a socially optimal policy. The potential influence of bad politicians and the office motivation of good ones may result in the selection of an incorrect policy. In PECB, this situation occurs when the state is $y$. 

The behavior of bad policymakers in PECB is driven by their indifference to aligning policy with the state of the world. Their primary concern is accruing rents. Upon observing a politician's proposal of $q_1^P=x$, the bureaucrat infers that this politician is more likely to be good than the prior $\pi$ indicates.\footnote{This assessment follows from the observation that, in pandering equilibria, good politicians always propose $x$, whereas bad ones do not.} By attempting to shift the policy to $y$, the bad bureaucrat maximizes the chances of obtaining rents in period~1 while simultaneously pushing the voter to oust a good politician. Conversely, a politician's proposal of $q_1^P=y$ fully reveals that the politician is bad. In this case, the bad bureaucrat faces the following trade-off: either to confirm policy $y$, secure immediate rents and facilitate the replacement of the surely bad politician; or to challenge the politician's proposal, aiming to keep the bad politician in power for future rent opportunities, thereby sacrificing immediate gains for assured future benefits. This decision hinges on the value of period~1's rents, $r_1^B$.

Likewise, the bad politicians' decision in PECB also depends on their realized rents, $r_1^P$. If the period~$1$ rents from proposing $q_1^P=y$ are sufficiently high compared to the expected rents they can obtain in period~2, then bad politicians opt to propose $q_1^P=y$. By contrast, relatively low rents in period~$1$ incentivize bad politicians to seek re-election by proposing $q_1^P=x$ to have the opportunity to seize higher rents in the second term.

The next result outlines the necessary and sufficient conditions under which PECB exist.

\begin{proposition}\label{PECBexistence}\hyperref[PECPtechnicalconditions]{A PECB exists if and only if}
    
    \begin{itemize}[noitemsep]
        \item[i)] The politicians' office rents are sufficiently high;
        \item[ii)] The relative mismatch costs are relatively high;
        \item[iii)] The upper bound of the bad politician’s period-1 rent distribution is sufficiently large;
        \item[iv)] The bad bureaucrat’s expected rents in period 2 are sufficiently small.
    \end{itemize}
\end{proposition}

 The first condition indicates that pandering requires office rents to be high enough to incentivize good politicians to propose a suboptimal policy for re-election purposes. Importantly, this condition imposes a tighter restriction as the bureaucratic influence score $\lambda$ increases. The second condition influences the behavior of a good bureaucrat and is inherently met when mismatch costs in state $y$ exceed those in state $x$ or when relative mismatch costs are approximately symmetric. For example, the second condition is always satisfied when
\[
v(y,y)-v(x,y) \approxeq v(x,x)-v(y,x).
\]

\noindent That is, when $\Delta=1$ and $\delta \in (0,1)$, bureaucrats
try to correct pandering. The underlying reason is that the risk of electing bad politicians is discounted, so immediate policy costs consistently take precedence in bureaucrats' decision-making.

The final two conditions, $iii)$ and $iv)$, ensure that bad policymakers are motivated by short-term gains. By contrast, if the rents for the first period are insufficiently high, the equilibrium policy fails to convey useful information about the politician's type. In such scenarios, a bad politician would always aim for re-election by proposing $q_1^P=x$ to secure higher rents in the subsequent period. Likewise, bad bureaucrats would persistently strive to influence policies to ensure the re-election of bad politicians. Consequently, the implementation of policy $p_1=x$ would cease to serve as a reliable signal that the politician is good.  Conditions $iii)$ and $iv)$ ensure the existence of informative outcomes, allowing for pandering equilibria.

\subsubsection{Discussion of PECB}

The good politician's period-$1$ choice when the state is $x$ is straightforward: they propose $x$. Conditional on the bureaucrat's approval, doing so simultaneously ensures a policy-state match and re-election. However, in state $y$, the situation is more complicated because the good politician weighs the immediate policy impact, the desire for re-election and future influence over policy, and the benefits of office rents. Proposing $y$ aligns with the state, but it implies losing the election, missing out on future office rents, and risking a policy-state mismatch in the following period. Office rents are important here: if high enough, the politician might prioritize re-election over selecting the accurate policy, as outlined in condition $i)$ of Proposition~\ref{PECBexistence}. This condition determines whether an equilibrium involves pandering or not, significantly impacting voter welfare. Bureaucratic influence has a crucial role in such a condition: the larger $\lambda$ is, the higher office rents must be to support a pandering equilibrium.\footnote{For simplicity and ease of reading, the precise mathematical conditions in all propositions have been skipped, but they can be found in the proofs in the Appendix.}

Proposition~\ref{PECBexistence}'s condition $i)$ may be satisfied even with negative office rents, provided the mismatch cost in state $x$ significantly outweighs that in state $y$. The pandering condition is not met with sufficiently low probability of state $x$ ($\rho$), high chance of electing a good politician ($\pi$), or high bureaucratic influence ($\lambda$). Even if ousted, good politicians can be confident of avoiding a policy-state mismatch in period-$2$ when state $x$ is relatively unlikely, and the chance of having a good successor is relatively high. Moreover, the good politician knows that if the bureaucrat is a good one, increased $\lambda$ means it is more likely that in the next period there will not be a policy mismatch as the strong good bureaucrat will try to correct any proposed mismatches and likely succeed in it. Therefore, the value of being re-elected for the good politician decreases, because they understand that even if they are not in power next period the implemented policy will likely be the correct one.

This means that, freed from the burden of seeking re-election, the good politician can focus on implementing the appropriate policy in the first period. However, if the bureaucrat is bad, a higher $\lambda$ increases the likelihood of a policy-state mismatch in the next period. When the state of the world in the first period is $y$, re-election requires pandering. But pandering would result in implementing policy $x$, creating a mismatch in the first period in addition to a likely mismatch in the second. In this case, the good politician may prefer to secure at least one correct match by proposing $y$, even if this reduces their chances of re-election. Regardless of the bureaucrat’s type, which is unknown to the politician, increasing bureaucratic power reduces the good politician’s incentives to pander by lowering the value of holding office in the second period.

The bad politician's proposal is determined by a comparison between realized and expected rents, with no regard for the state of the world. If the period-$1$ realized rents, $r_1^P$, are relatively lower than period-$2$ expected rents, then the bad politician prefers to propose $q_1^P=x$ in the hope of getting re-elected and seizing possibly larger rents in period-$2$. Higher office and expected rents ($E$ and $\mu_2^P$) increase the chance that bad politicians propose policy $x$ in every state. Similarly, an increase in bureaucratic power ($\lambda$), a higher likelihood of the state being $x$ ($\rho$), and a greater inclination of bad bureaucrats to reject a proposal $q_1^P=y$ diminish the likelihood of bad politicians proposing policy $x$.

Good bureaucrats genuinely care about policy and are not motivated by re-election considerations. They adopt a ``correcting'' strategy when the state of the world is $x$: if the politician proposes $q_1^P=y$, they challenge the proposal. Policy $y$ under state $x$ is the only possible mismatch in the second period, and it is precisely the mismatch that a bad politician seeks to implement in the first. For the bureaucrat, the negative effect of a \textit{certain} mismatch in the first period outweighs the effect of its merely \textit{potential} occurrence in the second, leading them to attempt to correct the proposal, even at the cost of re-electing a surely bad politician. Nonetheless, the good bureaucrat knows they can again oppose mismatches if they arise in the future. Conversely, if policy $x$---the appropriate choice for the current state---is proposed, the good bureaucrat endorses it. The reasoning is equally straightforward when the state of the world is $y$ and the observed policy matches it: endorsing policy $y$ not only ensures the correct course of action but also facilitates the removal of a bad politician.

The decision for the good bureaucrat is more involved in scenarios where the state of the world is $y$, but the politician's proposal is $q_1^P=x$. Contesting proposal $x$ can ensure the implementation of an accurate policy for the current period. Yet, it risks displacing a potentially good politician, in which case the bureaucrat would be ``correcting'' pandering. The good bureaucrat opts to challenge the incorrect proposal provided that the negative impacts of a policy-state mismatch in state $x$ do not significantly exceed those in state $y$ (see the second condition in Proposition~\ref{PECBexistence}). This decision is underpinned by the rationale that the immediate cost of endorsing an inappropriate policy ($x$ in state $y$) is not justified by the potential to correct mismatches in the future. In doing this calculation, the bureaucrat accounts for the possibility of electing a good politician for the next term, coupled with the opportunity to amend any future policy mismatches.
 
An increase in the probability that an incoming politician is good ($\pi$), or a decrease in the likelihood that a bad politician will choose policy $x$, makes the observation of a proposal $q_1^P=x$ more informative. As a result, the good bureaucrat's expected value from contesting a proposal $q_1^P=x$ when the state is $s_1=y$ diminishes because of an increased belief that the incumbent politician is good. Furthermore, a lower probability of state $x$ ($\rho$) increases the bureaucrat's motivation to oppose an incorrect policy in the first period. This occurs for two reasons: first, the harmful second-period scenario (a bad politician facing state $x$) becomes less likely; second, a lower $\rho$ increases the probability that a bad politician would choose policy $x$, which reduces the bureaucrat’s posterior belief that the incumbent is good.

The impact of the bureaucratic influence on their willingness to challenge a wrong policy is ambiguous. On the one hand, a higher $\lambda$ enhances their capability to amend a policy mismatch in the future, encouraging a proactive stance in challenging inaccuracies immediately. On the other hand, a higher $\lambda$ implies a reduced probability that a bad politician would choose $x$ and, thus, an increased posterior belief that the politician is good. As a result, observing $q_1^P=x$ becomes stronger evidence that the politicians are good, reducing the appeal of contesting their proposal.

Bad bureaucrats are indifferent to the state of the world; their actions are solely motivated by the pursuit of immediate rents $\left(r_1^B\right)$ and the anticipation of future rents $\left(\mu_2^B\right)$. Upon observing a proposal $q_1^P=x$, they try to turn it into $y$, thereby securing today's rents and ensuring the removal of a politician who is likely to be good. The strategic interaction between policymakers is more involved when politicians propose $q_1^P=y$. The incumbent politician is surely bad, and their re-election would ensure future rents for the bad bureaucrat. Confirming policy $y$ might lead to the displacement of the bad politician, paving the way for a potentially good successor in the next term. Consequently, if period 1's rents are significantly higher compared to the anticipated future rents, the bad bureaucrat opts to endorse policy $y$, prioritizing the immediate financial gain while remaining open to the uncertainties of the future political landscape.

As the influence of bad bureaucrats increases (i.e., $\lambda$ increases), they become more inclined to endorse proposals $q_1^P=y$, leading to the removal of the incumbent politician. Despite the possibility of a good politician assuming office in the subsequent term, these bad bureaucrats are confident in their ability to influence policy changes as necessary. Additionally, the likelihood of a bad bureaucrat confirming policy $y$ decreases with an increase in the probability of electing a good politician in the next term or an increase in the probability that the forthcoming state will be $x$. The rationale behind this behavior is twofold: first, the prospect of securing future rents diminishes with the higher likelihood of a good politician's election, following the removal of the current bad one; second, the chance of extracting rents in the second period decreases with the likelihood of state $x$ occurring, which poses challenges from the perspective of the bad bureaucrat, further discouraging the confirmation of policy $y$.

\subsection{Pandering equilibria with a pandering bureaucracy}

The discussion of PECB highlights how the bureaucrats' behavior is sustained by sufficiently high relative mismatch costs, $\Delta$. This observation is relevant for the ``good'' type of bureaucrat, who has preferences over policies aligned with the voters'. If the mismatch costs in state $x$ are relatively high compared to those in state $y$, the equilibrium behavior of the good bureaucrat changes with respect to that prescribed by PECB. This section shows that, when $\Delta$ is sufficiently low, good bureaucrats' equilibrium behavior is supportive of political pandering.  

In the first period, and given a sufficiently low $\Delta$, good bureaucrats are willing to support policy proposals $q_1^P=x$ even when the state of the world is $y$. In this case, good bureaucrats are concerned about a bad politician holding office in the second period, especially if the state of the world will be $x$. A policy proposal $q_1^P=x$ signals to the bureaucrat that the incumbent politician is more likely to be good than the prior indicates. Thus, the bureaucrat would rather confirm such a proposal regardless of the state of the world, in the hope of minimizing mismatch costs in the second period.

Building on the previous observation, this section analyzes a scenario where, in the first period, the good bureaucrat supports the politician's policy proposal $q_1^P=x$ even if the state of the world is $s_1=y$.

\begin{definition}\label{def:PECB_xi_gamma}
    A pandering equilibrium with a pandering bureaucracy (PEPB) is a PE in which
    \begin{itemize}[noitemsep]
    \item Good bureaucrats propose a policy that matches the state of the world when politicians propose $y$;
    \item Good bureaucrats propose $x$ when politicians propose $x$;
    \item Bad bureaucrats always propose policy $y$ after politicians propose policy $x$;
    \item The proposal of bad bureaucrats after politicians propose $y$ is state-independent but depends on their realized rents;
    \item The proposal of bad politicians depends on the state and their realized rents.
    \end{itemize}
\end{definition}

The following proposition lists all the necessary and sufficient conditions for PEPB to exist.

\begin{proposition}\label{PEPBexistence} 
  \hyperref[PEPBtechnicalconditions]{A PEPB exists if and only if} 
    \begin{itemize}[noitemsep]
        \item[i)] The politicians' office rents are sufficiently high;
        \item[ii)] The relative mismatch costs are relatively low;
        \item[iii)] The upper bound of the bad politician’s period-1 rent distribution is sufficiently large;
        \item[iv)] The bad bureaucrat’s expected rents in period 2 are sufficiently small.
    \end{itemize}
\end{proposition}

 The first condition for existence ensures political pandering, and it is exactly the same as the one required in PECB. The key condition for the existence of PEPB is the second one in Proposition~\ref{PEPBexistence}, which requires the relative mismatch costs to be sufficiently low. As mentioned before, such a condition is not satisfied under approximately symmetric mismatch costs, that is, when $\Delta\approx 1$. More generally, it is not satisfied when the mismatch costs in state $y$ are sufficiently larger than those in state $x$. In equilibrium, the good bureaucrat can accommodate political pandering only if avoiding future mismatches when the state is $x$ is sufficiently profitable.
As in the PECB case, the final two conditions, $iii)$ and $iv)$, ensure that bad policymakers are motivated by short-term gains.

\subsection{Non-pandering equilibria}\label{sec:nonpequilibrium}

Having completed the analysis of pandering equilibria, we now turn our attention to the analysis of non-pandering equilibria---that is, IE in which the good politician does not have an incentive to pander. In these equilibria, the good politician chooses to match the policy with the state of the world, even when the state is $y$. The relevant definition follows.
\begin{definition}\label{def:NPE}
A non-pandering equilibrium (NPE) is an IE in which, in period~1, good politicians always propose a policy matching the state. 
\end{definition}

Our analysis shows that, in NPE, the strategy of good bureaucrats is qualitatively different to the one in PE. After observing $q_1^P = s_1 =  y$, they may either confirm $y$---a scenario we term with ``stand-firm'' bureaucracy---or propose $q_1^B = x$ to increase the likelihood of re-electing a good politician, which we refer to as a ``subversive'' bureaucracy. In what follows, we characterize NPE by analyzing these two cases separately.

\subsubsection{Non-pandering equilibria with a stand-firm bureaucracy}

We begin by defining non-pandering equilibria in the case of a stand-firm bureaucracy, which serves as the foundation for the analysis that follows.

\begin{definition}\label{def:non-pandering_SF}
    A non-pandering equilibrium with a stand-firm bureaucracy (NPE-SF) is an NPE where,
    \begin{itemize}[noitemsep]
    \item Good bureaucrats always propose a policy that matches the state of the world;
       \item Bad bureaucrats always propose policy $y$ after politicians propose a policy that matches the state of the world;
    
        \item The proposal of bad bureaucrats after politicians do not propose a policy matching the state is state-independent but depends on their realized rents;
        
        \item The proposal of bad politicians depends on the state, and their realized rents. 
    \end{itemize}
\end{definition}

The following proposition details all the necessary and sufficient conditions for a NPE-SF to exist.

\begin{proposition}\label{NPEexistence}
\hyperref[NPESFtechnicalconditions]{A NPE-SF exists if and only if}
    \begin{itemize}[noitemsep]
        \item[i)] The politicians' office rents are sufficiently low;
    \item[ii)] The relative mismatch costs are relatively high;
    \item[iii)] The upper bound of the bad politician’s period-1 rent distribution is sufficiently large;
        \item[iv)]  State $x$ is sufficiently likely to occur;
        \item[v)] The bad bureaucrat’s expected rents in period 2 are sufficiently small.
    \end{itemize}
    \end{proposition}

 Condition $i)$ in Proposition \ref{NPEexistence} is the flip-side of condition $i)$ in Propositions \ref{PECBexistence} and \ref{PEPBexistence}. In pandering equilibria, the willingness to pander and the desire to be re-elected coincide, and are both expressed by their condition $i)$. In particular, in such equilibria, it is precisely the pursuit of re-election that induces a good politician to pander. By contrast, in NPE-SF, the low office rents required by Proposition~\ref{NPEexistence} imply that the good politician has no real interest in re-election. Consequently, they do not seek re-election, either directly by pandering to voters or indirectly by signaling their type to the good bureaucrat. With no incentive to remain in office, the good politician abandons pandering altogether and simply aligns policy with the state of the world.
 
 Moreover, as in PECB, the good bureaucrat strongly prefers to match the policy to the state of the world when $\Delta$ is relatively high, as required by condition $ii)$. Condition $iii)$ mirrors its equivalent in Propositions~\ref{PECBexistence} and \ref{PEPBexistence}: the bad politician must be sufficiently motivated by short-term benefits so that the equilibrium remains informative. However, condition $iv)$ of Proposition~\ref{NPEexistence} is new. It requires state $x$ to be sufficiently likely, and if it fails, condition $v)$ cannot be satisfied. The intuition is as follows. The equilibrium cannot be informative if bad bureaucrats propose $x$ too frequently in period 1, reflecting excessively strong long-term incentives and a desire for the incumbent to be re-elected. When state $x$, the state under which a good politician in period 2 would be troublesome for the bad bureaucrat, is likely, the expected payoff of a bad bureaucrat in period 2 decreases, making going for policy $y$ in period 1 more attractive. On the other hand, if state $x$ is unlikely to occur, the bad bureaucrat's second-period expected payoff increases, incentivizing the bad bureaucrat to go for policy $x$ in the first period. 
 
 As a reminder,  condition $iv)$ is not required for the existence of PE. In NPE, good politicians do not always propose $x$, making policy $x$ less informative about the politician’s type. Thus, the informativeness requirement is stricter in NPE than in PE.

\subsubsection{Non-pandering equilibria with a subversive bureaucracy}

A subversive bureaucrat in NPE is the analogue of the pandering bureaucrat in a PE: they are good bureaucrats who care about the cost implications of implemented policy, but understand that minimizing overall mismatch costs does not necessarily go through trying to match policy with the state of the world in the first period; it may involve facilitating the re-election of a politician they expect to be good. The following definition illustrates the different behavior of good bureaucrats in this equilibrium compared to the NPE-SF.

\begin{definition}\label{def:non-pandering_SV}
    A non-pandering equilibrium with a subversive bureaucracy (NPE-SV) is an NPE where,
    \begin{itemize}[noitemsep]
    \item Good bureaucrats propose $x$ if the politician's proposal matches the state of the world when the state is $y$, otherwise they propose a policy matching the state;
       \item The bad politicians' and bad bureaucrats' strategies are the same as described by Definition~\ref{def:non-pandering_SF}.
    \end{itemize}
\end{definition}

The more interesting feature of these equilibria is that the good politician may or may not want to be re-elected. In fact there are two types of the non-pandering equilibria with a subversive bureaucracy. In the first type, a good politician does not seek re-election, and therefore does not pander, but the good bureaucrat intervenes to try to force their re-election. In the second type the good politician does seek re-election, but chooses to not pander. However, by choosing to not pander, they are signaling to the bureaucrats they are good, and the good bureaucrat intervenes to try to get them re-elected. We begin with the first type of equilibrium which we call non-pandering equilibrium with a \emph{forcing} subversive bureaucracy (NPE-FSV).

\begin{proposition}\label{NPE_FSV_existence}
\hyperref[obs:existence_NPE_SV]{An NPE-FSV exists if and only if}

    \begin{itemize}[noitemsep]
        \item[i)] The politicians' office rents are sufficiently low;
    \item[ii)] The relative mismatch costs are relatively low;
    \item[iii)] The upper bound of the bad politician’s period-1 rent distribution is sufficiently
large; 
\item[iv)]  State $x$ is sufficiently likely to occur;

\item[v)] Bureaucratic influence is sufficiently low;
     \item[vi)] The bad bureaucrat’s expected rents in period 2 are sufficiently small when $\lambda$ is low, and take intermediate values otherwise.
     \end{itemize}
    \end{proposition}

Similar to NPE-SF, the first condition indicates that also in NPE-FSV the good politician does not seek re-election. A novelty in this equilibrium class lies in condition $vi)$, which requires the bureaucrat’s expected period-2 rents, $\mu_2^B$, to fall within an intermediate range when bureaucratic influence is sufficiently high. Unlike earlier conditions, which required $\mu_2^B$ to be low in order to give bad bureaucrats short-term incentives and preserve informativeness, condition $vi)$ requires $\mu_2^B$ to be sufficiently high. The reason is that, under high $\lambda$, good politicians must be deterred from proposing $x$ in state $y$.

When bureaucratic influence is strong, the actions of a subversive bureaucracy become central to shaping politicians’ strategies. In this case, proposing $q_1^P=y$ can paradoxically increase the likelihood that the final implemented policy is $p_1=x$, since bureaucrats may overturn the proposal. Thus, politicians might strategically propose the policy they do not actually want to see implemented. This occurs when  $\mu_2^B$ is low enough to prevent bad bureaucrats from confirming $x$ in state $y$. To prevent such incentives, condition $vi)$ requires $\mu_2^B$ to be sufficiently high, ensuring that the best way to implement policy $x$ is to propose it.

In short, good politicians who prioritize correct policymaking over office rents prefer to propose the state-matching policy only if doing so reliably ensures its implementation. This requires bureaucrats to have sufficiently strong long-term incentives, captured by a high $\mu_2^B$, which condition $vi)$ guarantees. This requirement is relevant only when bureaucratic influence is high enough for subversive bureaucrats to overturn proposals of $y$ in state $y$.

We now turn to the second type of NPE-SV, which we term non-pandering equilibrium with an administrative subversive bureaucracy (NPE-ASV).

\begin{proposition}\label{prop:NPE_ASV_existence}
\hyperref[obs:existence_NPE_ASV]{An NPE-ASV exists if and only if}

    \begin{itemize}[noitemsep]
        \item[i)] The politicians' office rents are sufficiently high;
    \item[ii)] The relative mismatch costs are relatively low;
\item[iii)] Bureaucratic influence is sufficiently high;

        \item[iv)] The bad bureaucrat’s expected rents in period 2 are sufficiently small. 
    \end{itemize}
    \end{proposition}

Condition $i)$ is identical to condition $i)$ in Propositions~\ref{PECBexistence} and \ref{PEPBexistence} and the flip-side of that in Proposition~\ref{NPE_FSV_existence}. In this equilibrium, the good politician is indeed interested in being re-elected, as office rents are sufficiently high, however they do not try to pander. Instead, they match the policy to the state of the world. Under state $y$ and with low relative mismatch costs, with this strategy the good politician is signaling to the good bureaucrat that they are good which results to the good bureaucrat trying to change the policy to $x$ in order for the politician to be re-elected. With high bureaucratic influence, required by condition $iii)$, this strategy is not risky for the good politician, who ends up ``pandering'' through the bureaucracy.  Condition $vi)$ is also tighter here than in PE for intermediate values of $\lambda$, ensuring that condition $ii)$ remains feasible and that the good politician  achieves re-election by proposing $y$ rather than $x$.

This last equilibrium is of particular interest as it captures situations where non-elected officials effectively engage in pandering on behalf of the politicians. As we have discussed in the introduction, traditionally pandering has been understood as an inefficient strategy employed by politicians seeking re-election. That is, pandering is almost always tied to electoral incentives. Notably, here we show that a form of administrative pandering may also originate from actors who are not subject to electoral incentives.


\subsection{Characterization}

The five types of equilibria we identify encompass all possible pandering and non-pandering equilibria. Moreover, if an additional condition on $\lambda$ is satisfied, they also constitute all possible informative equilibria. We formalize this result in the next proposition.

\begin{proposition}
\label{characterization}
    Definitions \ref{def:conj}, \ref{def:PECB_xi_gamma}, \ref{def:non-pandering_SF}, and \ref{def:non-pandering_SV} provide a complete characterization of all PE and NPE. When $\lambda$ is sufficiently small, they provide a complete characterization of all IE.
\end{proposition}


\section{Applications}\label{sec:appl}

Previous work has examined the allocation of specific policy tasks between elected politicians and non-elected bureaucrats \citep{alesina2007bureaucrats,alesina2008bureaucrats,maskin2004politician,Stephenson2011}. The first part of this section connects to that line of research by analyzing, within the context of our model, the conditions under which voters prefer policymaking to be under the complete control of either politicians or bureaucrats. In the second part of this section, we show that restricting the analysis to these two extreme cases is limiting, as there are situations where voters' welfare is highest with an intermediate balance of power between policymakers.

\subsection{Benchmark: policy allocation}\label{sec:benchmarks}

This section compares two relevant and opposing benchmarks in which policymaking is completely controlled either by the politician or by the bureaucrat. This allows us to span the analysis from a canonical electoral accountability model, where bureaucratic influence is absent, to a judicial model, where electoral accountability is absent. In this way, we compare the two extremes of bureaucratic influence, as in, e.g., \cite{maskin2004politician}. 

In the equilibria we study, the bureaucracy can contest politicians’ proposals, though its authority is limited. When the bureaucracy is entirely powerless, or \emph{toothless}, we have $\lambda=0$, which replicates a standard pandering model without bureaucratic influence. In this case, bureaucrats’ decisions are irrelevant, and voters understand that policy outcomes depend directly on the politicians’ choices. A pandering equilibrium thus exists when the bureaucracy is toothless. By contrast, if the bureaucracy is all-powerful (i.e., $\lambda=1$), we obtain a \emph{dictatorship of the bureaucracy}: politicians’ decisions are irrelevant, and so are voters’ choices. In this case, a good bureaucrat implements the state-contingent policy, while a bad one always selects $y$.

We next calculate and compare the voter’s ex-ante equilibrium utility under the two benchmarks: a bureaucratic dictatorship and a toothless bureaucracy. In doing so, we assume parameters such that the good politician retains pandering incentives, while the bad politician pursues short-term interests. Further details and proofs are provided in Appendix~\ref{sec:app_bench}.

The next result shows that the voter may strictly prefer to relinquish electoral accountability and delegate power to a bureaucratic dictatorship when the pandering state $y$ is sufficiently likely, or when pandering is highly inefficient. Conversely, the voter prefers a toothless bureaucracy---removing any scope for bureaucratic correction---when bureaucrats are sufficiently likely to be corrupt or misaligned.

\begin{proposition}\label{prop:bench1}
    The voter always prefers a dictatorial bureaucracy to a toothless one when $\rho$ is sufficiently low. Otherwise, the voter prefers a dictatorial bureaucracy over a toothless one if and only if $\beta$ is sufficiently high.
\end{proposition}

Furthermore, the next result shows that a higher probability of bureaucratic alignment with the voter always increases the relative attractiveness of a dictatorial bureaucracy. By contrast, a higher probability of political alignment has the opposite effect if and only if pandering under a toothless bureaucracy is expected to be sufficiently rare. Stronger re-election incentives, as captured by higher $E$ or $\mu_2^P$, enhance the appeal of a toothless bureaucracy, and thus of electoral accountability, provided that $\rho$ is sufficiently high. Once again, the mechanism behind this result is the inefficiency generated by pandering. 

\begin{proposition}\label{prop:bench2}
    The voter’s relative preference for a dictatorial over a toothless bureaucracy is always increasing in $\beta$. It is decreasing in $\pi$, $E$ and in $\mu_2^P$ if and only if $\rho$ is sufficiently high.
\end{proposition}

These two propositions illustrate how the tension between electoral accountability and pandering inefficiency is resolved at the extremes by granting full control to bureaucrats, provided they are sufficiently aligned with the voter.

It is not uncommon, particularly in developed democracies, for monetary policy to be delegated primarily, sometimes exclusively, to bureaucrats such as independent central bankers ($\lambda \rightarrow 1$). Our model shows that voter ex-ante welfare is higher under a dictatorship of the bureaucracy than under a toothless bureaucracy, conditional on the probability of appointing a good bureaucrat ($\beta$) being sufficiently high. In the case of monetary policy, the ability to evaluate outcomes through a simple and observable measure such as inflation may typically increase the likelihood of appointing good bureaucrats.

However, within our framework, the relative rarity of states in which inflationary policies are optimal might lead us to think that pandering is unlikely to arise in monetary policy, which in turn would weaken the case for an all-powerful central banker. Still, pandering can occur in monetary policy, for example by keeping interest rates artificially low before elections to stimulate short-term growth and reduce unemployment. That is more so in fiscal policy, which is highly state-dependent and therefore especially vulnerable to pandering.

In general, when the pandering state is unlikely (high $\rho$), a toothless or fully dependent bureaucracy ($\lambda \rightarrow 0$) may enhance voter welfare, provided that good bureaucrats are sufficiently rare (i.e., $\beta$ is low enough).

Our model highlights that between these two extreme benchmark cases lies a continuum of possibilities in which bureaucrats exert a limited degree of influence over policymaking. Our equilibrium characterization already illustrates how such influence shapes the players’ behavior. In the remainder of the paper, we show that these intermediate cases carry important welfare implications. Our findings underscore the value of considering partial bureaucratic influence rather than focusing exclusively on the extreme benchmark cases analyzed in this section.

\subsection{Voter's welfare}
\label{sec:cs}

In the characterization presented in Section~\ref{sec:equilibrium}, we outline the policymakers' equilibrium behavior. In this part of the paper, we use our previous findings to analyze how the voters' ex-ante equilibrium welfare varies with bureaucratic influence, represented by  parameter $\lambda$. Given the complexity of the welfare function, we resort to numerical and graphical analyses, with the exception of Propositions~\ref{prop:jump} and \ref{prop:selection}. First, we introduce a few assumptions regarding the model's parameters. As demonstrated in the Appendix, our selection of parameters enables the analysis of voters' ex-ante welfare across the entire range of $\lambda$. 

\noindent{\bf Assumptions.} To perform comparative statics, we assume
\begin{itemize}[nosep]
    \item That the rents obtained by bad policymakers for implementing $y$ follow a uniform distribution. That is,
    \[
    F_t^j \sim \mathcal{U}[0,2] \implies \mu_t^j=1, \;\text{ for every }\; t=\{1,2\} \;\text{ and }\; j=\{P,B\};
    \]
    \item That $\delta=v(y,y)=1$, and $v(x,y)=v(y,x)=0$;
    \item To analyze PECB and PEPB, that $E=1$; to analyze NPE-SF, that $E=.85$;
    \item To analyze PECB and NPE-SF, that $v(x,x)=1$; to analyze PEPB, that  $v(x,x)=500$.
\end{itemize}

Section~\ref{sec:benchmarks} presents two critical benchmarks: the scenario where the bureaucracy is toothless ($\lambda=0$) and the scenario where it is dictatorial ($\lambda=1$). Under our parametric assumptions, we obtain that the voter's welfare is $1+\pi\rho$ with a toothless bureaucracy, whereas it is $2[1-\rho(1-\beta)]$ with a dictatorial bureaucracy. As a result, the voter prefers a dictatorial to a toothless bureaucracy provided that bureaucrats are sufficiently aligned, that is, when\footnote{Threshold $\tilde\beta$ can be smaller than $\pi$, implying that the voter may prefer a dictatorial bureaucracy even when bureaucrats are more likely to be corrupt than politicians. The intuition is that political alignment induces pandering, which is inefficient. Indeed, $\tilde\beta < \pi \iff \rho < \frac{1}{2 - \pi}$, since pandering becomes more inefficient the less likely state $x$ is.}   
\begin{equation}\label{eq:indifference}
\beta>\tilde\beta\coloneqq\frac{\pi}{2} - \left( \frac{1-2\rho}{2\rho} \right).
\end{equation}

However, reality typically lies in between those two extremes, and bureaucrats have some positive but limited influence over policymaking. It is relevant to understand what are the possible effects of increased bureaucratic interference. The following analysis concerns the intermediate cases where $\lambda\in(0,1)$. The Appendix additionally contains multiple figures showing how the voters' ex-ante welfare is affected by bureaucratic interference for (relatively) low, medium, and high levels of $\beta$, $\pi$, and $\rho$.

\begin{figure}[htbp]
    \centering
    \begin{minipage}[t]{.5\linewidth}%
    \includegraphics[width=1\linewidth]{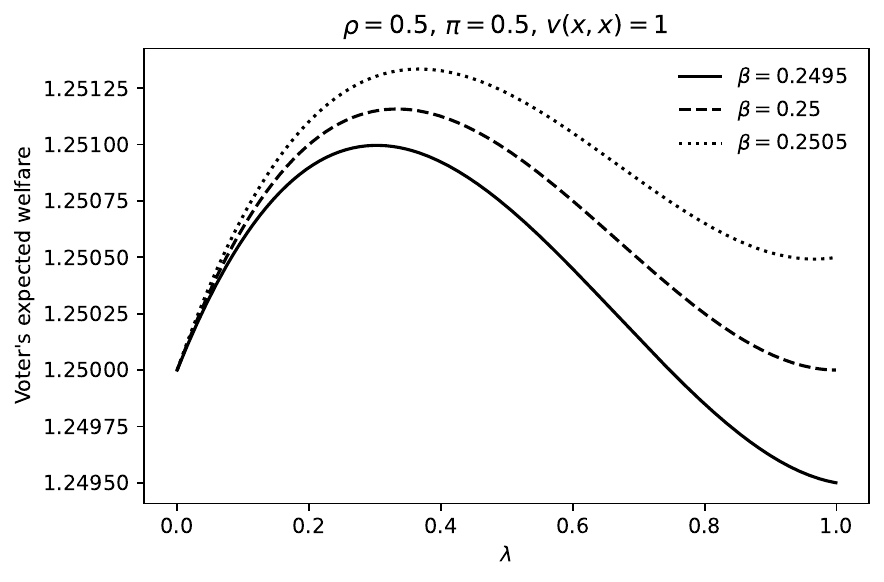}
    \end{minipage}\hfill%
    \begin{minipage}[t]{.5\linewidth}%
    \includegraphics[width=1\linewidth]{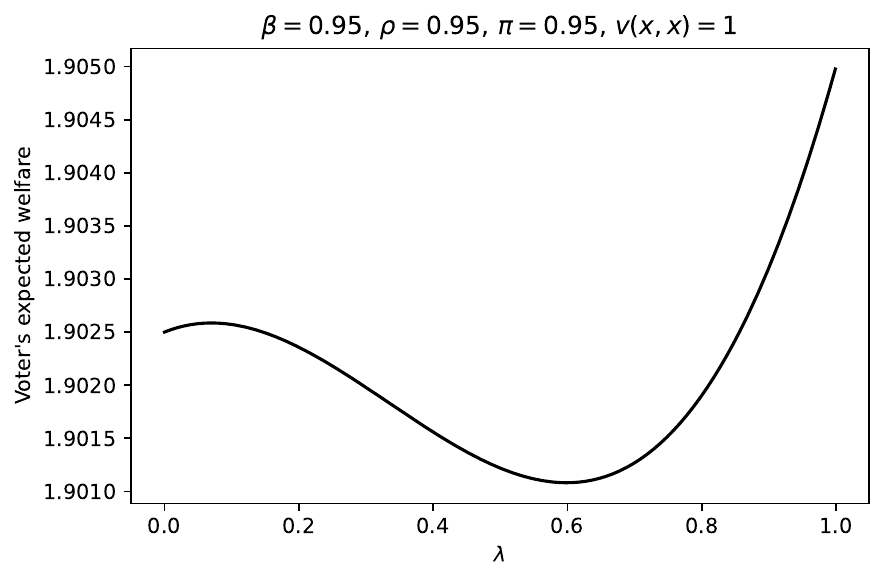}
    \end{minipage}%
    \caption{Effect of bureaucratic influence on voter welfare in a PECB. Bureaucratic influence, represented by $\lambda$, can have a non-monotonic effect on the voter's welfare. Depending on parameters, intermediate values of $\lambda$ can maximize or minimize welfare.}
    \label{fig:intermediate}
\end{figure}

We begin by analysing PECB. The left-hand side panel of Figure~\ref{fig:intermediate} depicts the case where $\beta$ takes values around $\tilde\beta$. When $\beta=\tilde\beta=0.25$, voters are indifferent between full bureaucratic and full political control over policymaking. The figure shows that, in these cases, the voters' welfare is maximized for an intermediate level of bureaucratic interference. This result highlights an inherent trade-off: higher bureaucratic influence mitigates the detrimental effects of pandering, but at the same time it weakens accountability. The right-hand side panel of Figure~\ref{fig:intermediate} shows that the effects of such a trade-off on the voters' welfare can be non-trivial. In some cases, intermediate levels of bureaucratic interference can yield local maxima and global minima in the voters' welfare.

\begin{figure}[htbp]
    \centering
    \begin{minipage}[t]{.5\linewidth}
    \includegraphics[width=1\linewidth]{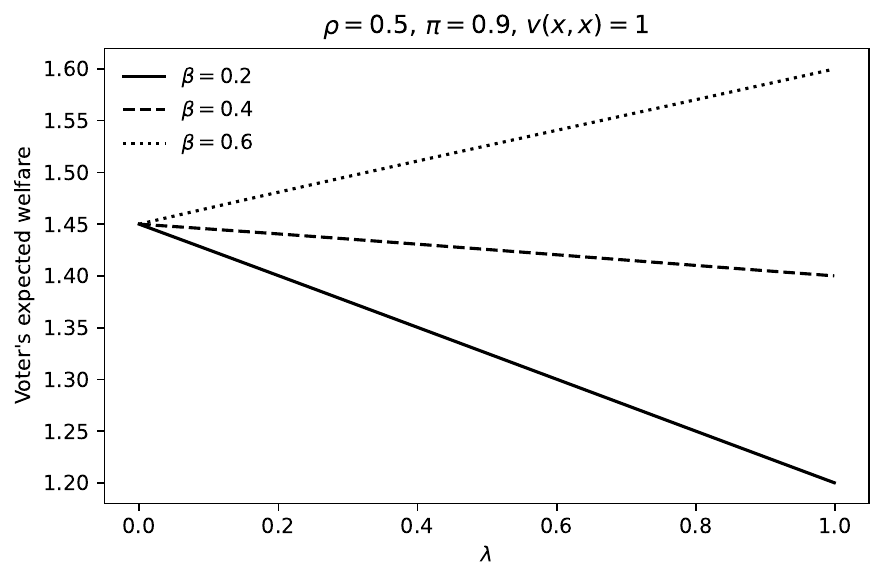}
    \end{minipage}\hfill
    \begin{minipage}[t]{.5\linewidth}
    \includegraphics[width=1\linewidth]{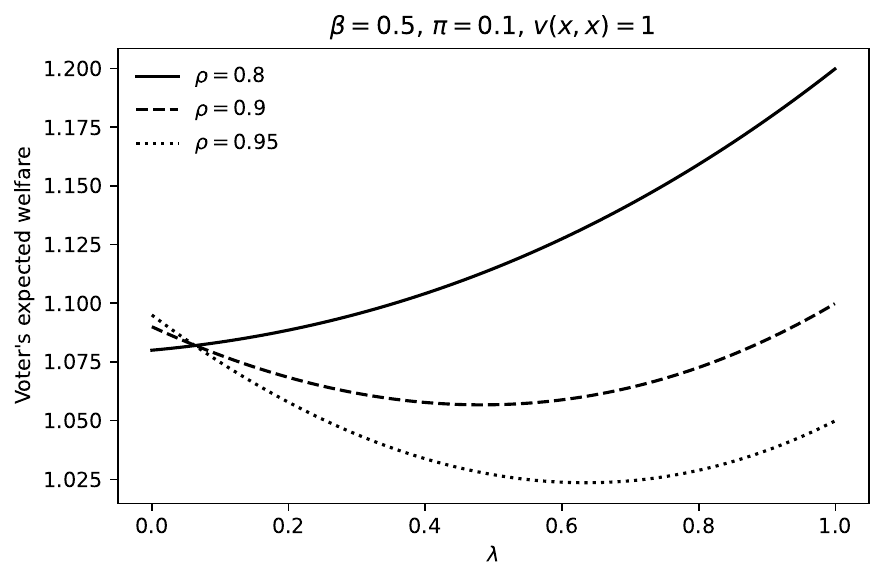}
    \end{minipage}
    \caption{Effect of bureaucratic influence on voter welfare in a PECB. Bureaucratic influence can have a monotonic effect on the voter's welfare. Intermediate values of $\lambda$ may never maximize or minimize welfare, which can be always convex in $\lambda$.}
    \label{fig:extreme}
\end{figure}

Figure~\ref{fig:extreme} tells different stories. The left-hand side panel displays situations where increased bureaucratic influence has a monotonic effect---positive or negative---on the voters' welfare. The right-hand side panel shows situations where the welfare function is always convex in $\lambda$. Moreover, the right-hand side panel of Figure~\ref{fig:intermediate} and the left-hand side panel of Figure~\ref{fig:extreme} (dotted line) depict situations whereby the voters' welfare is maximized under a dictatorial bureaucracy, even though bureaucrats are more or equally likely to be corrupted than politicians. On the other hand, the right-hand side panel of Figure~\ref{fig:extreme} (dotted line) shows a case where the voters' welfare is maximized under a toothless bureaucracy, even though bureaucrats are less likely to be corrupt than politicians.

\begin{figure}[htbp]
    \centering
    \begin{minipage}[t]{.5\linewidth}
    \includegraphics[width=1\linewidth]{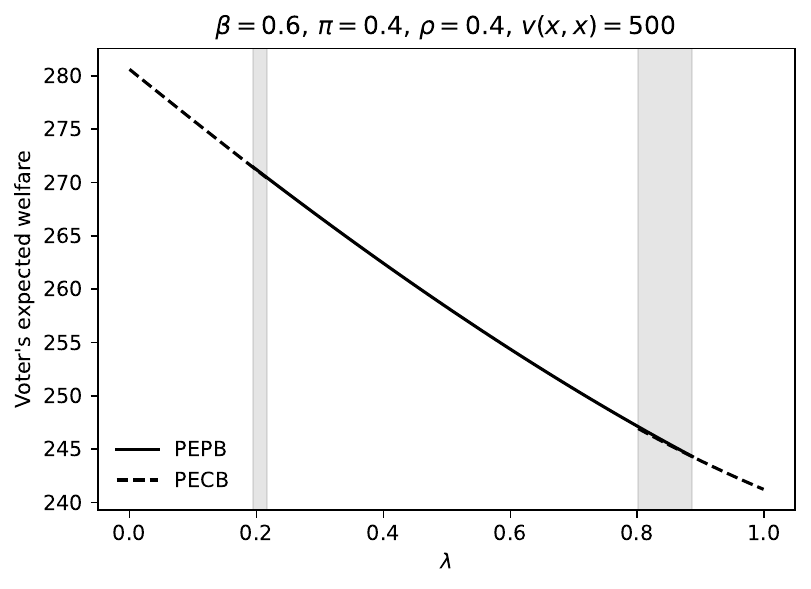}
    \end{minipage}\hfill
    \begin{minipage}[t]{.5\linewidth}
    \includegraphics[width=1\linewidth]{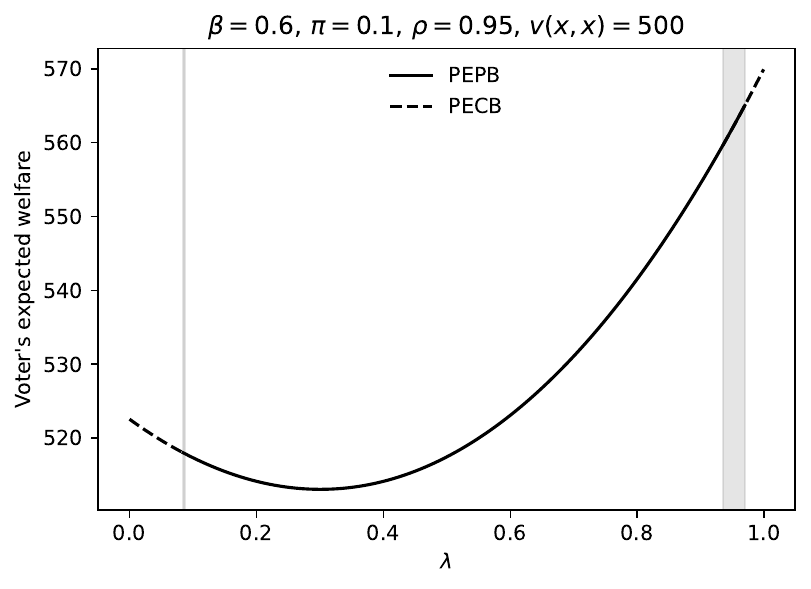}
    \end{minipage}
    \caption{Effect of bureaucratic influence on voter welfare in PECB and PEPB. Voter welfare is shown with a solid line for values of $\lambda$ where the equilibrium is a PEPB, and with a dashed line where the equilibrium is a PECB. The grey-shaded regions indicate values of $\lambda$ for which PECB and PEPB co-exist.}
    \label{fig:PEPB}
\end{figure}

To analyze PEPB, we need to change the relative mismatch costs, $\Delta$. We do so by setting $v(x,x)=500$. The condition on $\Delta$, ensuring we are analyzing a PEPB, requires $\lambda$ taking intermediate values. Figure~\ref{fig:PEPB} shows two qualitatively different cases. In the left-hand side panel, increased bureaucratic influence always damages voters. The voter's welfare is maximized under full political control, where the equilibrium is a PECB. In the right-hand side panel, the voters' welfare is convex in $\lambda$ and reaches a global minimum for an intermediate value of bureaucratic influence, under which the equilibrium is a PEPB.

Lastly, the analysis of NPE-SF requires sufficiently low office rents. To this end, we set $E=.85$. The equilibrium is without political pandering when bureaucrats have sufficiently high bureaucratic influence. Specifically, under our parameter configuration, the existence of NPE-SF requires
\[
\lambda>\ell= 1-\frac{1-E}{\rho(1-\pi)}.
\]
The voter’s equilibrium welfare exhibits a discontinuity at $\ell$, the threshold above which politicians cease to pander. 

\begin{figure}[htbp]
    \centering
    \begin{minipage}[t]{.5\linewidth}
    \includegraphics[width=1\linewidth]{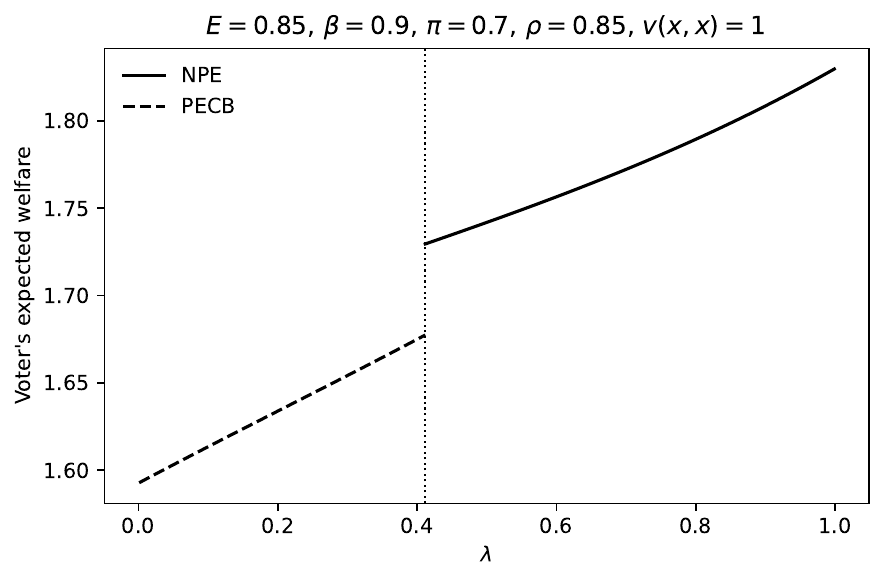}
    \end{minipage}\hfill
    \begin{minipage}[t]{.5\linewidth}
    \includegraphics[width=1\linewidth]{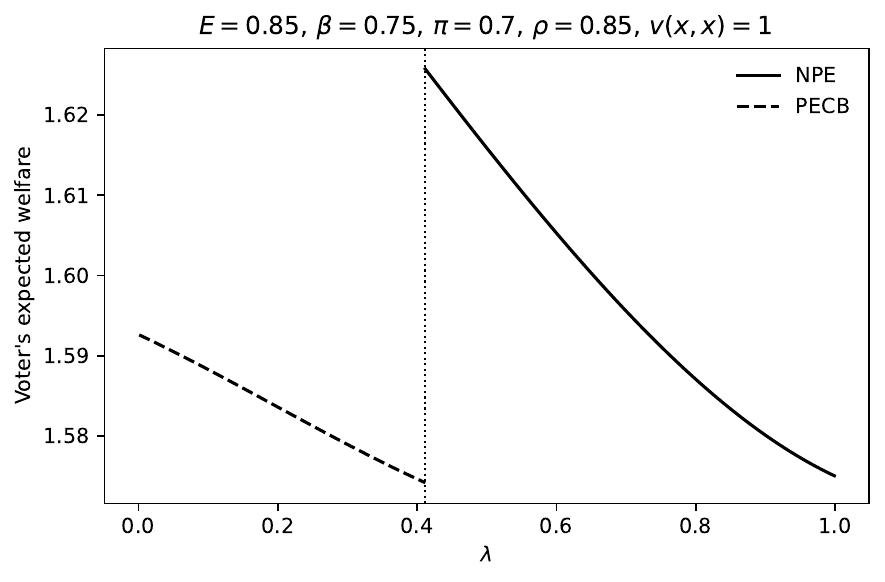}
    \end{minipage}
    \caption{Effect of bureaucratic influence on voter welfare in PECB and NPE. Voter welfare exhibits a discontinuous jump when greater bureaucratic influence shifts the equilibrium from PECB to NPE. Equilibrium welfare is shown with a dashed line for values of $\lambda$ corresponding to PECB and with a solid line for values of $\lambda$ corresponding to NPE.}
    \label{fig:NPE}
\end{figure}

When bureaucrats are sufficiently influential, good politicians stop pandering and instead seek to align policy with the state. Figure~\ref{fig:NPE} illustrates this discontinuity in two cases. The left-hand panel shows a scenario in which greater bureaucratic influence always benefits voters. By contrast, the right-hand panel depicts a situation where bureaucrats are more susceptible to capture than politicians. In this case, voter welfare improves only when $\lambda$ crosses $\ell$, peaking at the intermediate value $\lambda=\ell$.

The graphical analysis performed in this section reveals two relevant channels through which some positive but limited bureaucratic influence is optimal from the voters' perspective. The left-hand side panel of Figure~\ref{fig:intermediate} considers cases where the voter is approximately indifferent between full political and full bureaucratic control over policymaking (i.e., $\lambda=0$ and $\lambda=1$, respectively). The right-hand side panel of Figure~\ref{fig:NPE} considers cases where bureaucratic influence effectively deters political pandering. In both scenarios, the voter's welfare is maximized for some intermediate $\lambda$. Our analysis identifies two potential channels behind the widespread acceptance and informal institutionalization of bureaucratic influence over policymaking. Furthermore, our model suggests that such an influence is an effective way to combat political pandering. Through this lens, bureaucrats may play a crucial role in steering policymaking towards more substantive outcomes.

We conclude this section with the following analytical result, which shows that the welfare discontinuity illustrated in Figure~\ref{fig:NPE} holds more generally across a broader range of parameters.
\begin{proposition}\label{prop:jump}
    Under our parameter configuration, the voter’s welfare exhibits a strictly positive discontinuous jump when a marginal increase in $\lambda$ shifts the equilibrium from a PECB to an NPE-SF.
\end{proposition}

\subsection{Electoral accountability}
\label{sec:electoralaccountability}
Elections play a crucial role in holding politicians accountable: they discipline political behavior and weed out bad politicians, thereby mitigating both moral hazard and adverse selection problems. However, bureaucratic influence over policymaking means that voters may be unsure about the extent to which policy performance reflects the actual choices of politicians. This uncertainty is generally believed to weaken electoral accountability \citep{Martin2021}. Our results challenge this perspective.

In the previous section, we analyzed cases where the voters' welfare increases in bureaucratic influence. Specifically, condition $i)$ in Propositions~\ref{PECBexistence} to \ref{NPEexistence} indicates that the inefficient phenomenon of political pandering is not supported in equilibrium for sufficiently high $\lambda$ (see condition~\eqref{eq:pandering} in Appendix). Figure~\ref{fig:NPE} illustrates an instance where bureaucratic influence effectively deters political pandering. When the equilibrium shifts from a pandering to a non-pandering one, the voter's welfare increases discontinuously, recording an improvement in good politicians' behavior. Thus, bureaucratic influence can enhance political accountability by disciplining good politicians and eliminating inefficiencies.

Accountability is also about selection. Can bureaucratic influence improve this aspect of elections as well? Consider a PECB. In equilibrium, a good politician is sometimes removed when pandering under a good bureaucracy. Likewise, bad politicians can end up being re-elected when good bureaucrats correct policymaking toward $x$. This argument suggests that an aligned bureaucracy may be detrimental to political selection.

\begin{figure}[htbp]
    \centering
    \begin{minipage}[t]{.5\linewidth}
    \includegraphics[width=1\linewidth]{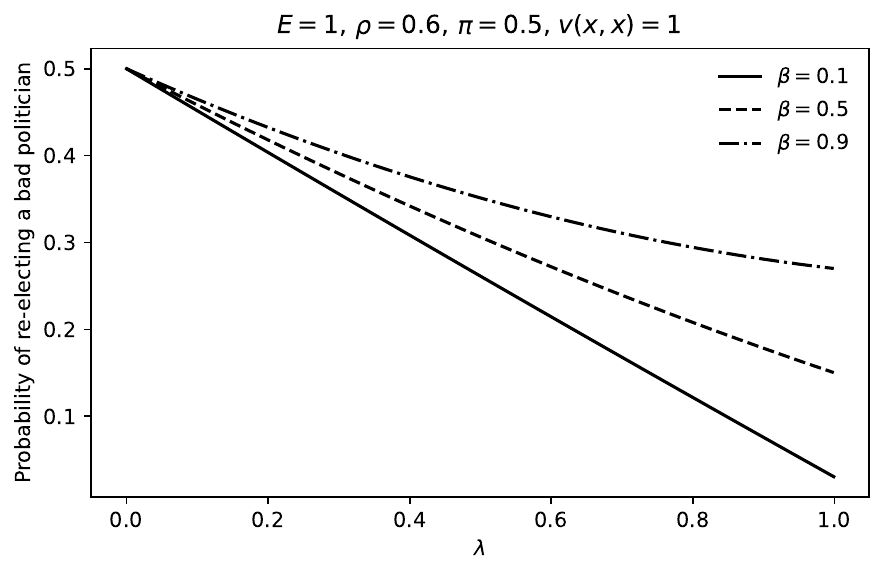}
    \end{minipage}\hfill
    \begin{minipage}[t]{.5\linewidth}
    \includegraphics[width=1\linewidth]{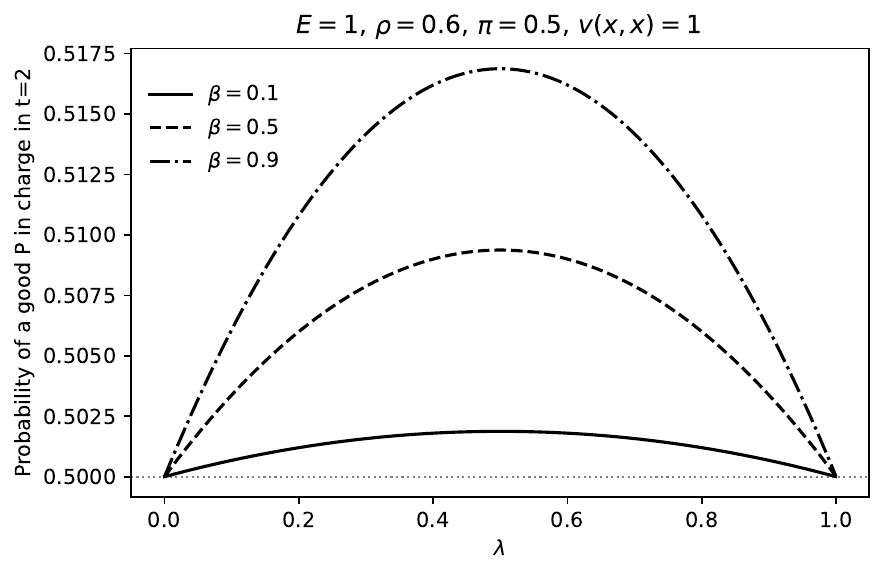}
    \end{minipage}
    \caption{Effect of bureaucratic influence over political selection in a PECB. The left panel depicts the probability of re-electing a bad politician. The right panel depicts the probability of having a good politician in charge in the second period.}
    \label{fig:selection}
\end{figure}

The panels in Figure~\ref{fig:selection} depict the relationship between measures of political selection and bureaucratic influence over policymaking within a PECB. The left-hand side panel focuses on the probability of re-electing a bad politician. While this is only a partial measure of selection, it helps to disentangle the overall effect of bureaucratic influence on electoral outcomes. The right panel focuses on the probability that a good politician holds office in the second period, which we take as our key measure of selection.

The left panel shows that, under our parametric configuration, greater bureaucratic influence helps weed out bad politicians from office. However, this effect is mitigated by bureaucratic quality. The reason is that, in PECB, good bureaucrats engage in corrective behavior, which can force bad politicians in office while driving out good politicians who pander.

The right panel illustrates two key effects of bureaucratic influence. First, any intermediate level of bureaucratic influence delivers better political selection than either a toothless or a dictatorial bureaucracy. Second, an increase in bureaucratic influence improves political selection, provided that such influence remains sufficiently low. These two observations hold regardless of the bureaucracy’s quality, as measured by $\beta$. The main takeaway is that bureaucratic interference in the policymaking process can unambiguously enhance electoral accountability in addition to improving the voter’s welfare.

The next result shows that the insights from the right panel of Figure~\ref{fig:selection} hold over a broader range of parameters.
\begin{proposition}\label{prop:selection}
    In PECB, and under our parametric configuration, the probability that a good politician holds office in the second period increases in $\lambda$ for all $\lambda<\frac{1}{2}$, and decreases in $\lambda$ otherwise.
\end{proposition}


\section{Concluding remarks}\label{sec:conclusion}

The influence of modern bureaucracies on policymaking is indisputable. Yet the nature of this influence, whether consistently positive or not, remains uncertain. Like politicians, bureaucrats can range from corrupt or ineffectual to upright and decisive. The complex interplay of incentives between politicians and bureaucrats can weaken elections as a tool for holding politicians accountable. At the same time, this interplay has the potential to generate policies that benefit society. The fact that bureaucrats cannot be disciplined through elections is a double-edged sword, bringing both challenges and advantages.

In our model, good politicians share the same policy preferences as voters. However, electoral pressures may compel them to intentionally choose a sub-optimal policy to secure re-election. By contrast, bad policymakers care only about obtaining rents from enacting specific policies, regardless of their suitability. In this context, good bureaucrats face a dilemma: whether to accept an undesirable policy today to increase the likelihood of a correct policy tomorrow, or to insist on a correct policy today at the risk of enabling an undesirable one tomorrow. Because of electoral incentives, the presence of two good policymakers who favor the socially optimal policy does not guarantee its consistent implementation across all periods and states.

In our model we call policymakers \emph{bad} when their interests are misaligned with those of voters. For example, some external elites may have promised them rewards for implementing a specific policy. This is a modeling assumption that we consider reasonable. However, our findings would still hold if we instead assumed that implementing a particular policy generated uncertain income for bad policymakers, for instance through insider trading or by influencing the award of state contracts. The results would also remain valid if we dispensed with rents altogether and assumed that bad policymakers are purely ideological, applying a one size fits all approach to policy. As long as they derive uncertain utility from implementing their preferred policy, voters would still favor a flexible good politician over the inflexible bad one.

Our paper contributes to the literature on the impact of bureaucracy on policymaking and on the role bureaucrats may play in shaping policies and, by extension, electoral outcomes. Bureaucrats can be seen either as unelected meddlers or as a safeguard against populism. While good bureaucrats are typically preferable to bad ones, we show that even good bureaucrats may face the dilemma of whether to oppose a good politician.

One of our main findings provides a rationale for granting unelected bureaucrats some influence over policymaking, though such influence may need to be limited. Political pandering, a major source of inefficiency, can be mitigated through bureaucratic intervention. Whether pandering arises depends not only on the rents and perks of holding office but also on the degree of bureaucratic control. We identify a threshold level of bureaucratic influence beyond which politicians are fully discouraged from pandering, with a positive effect on voter welfare. The existence of an optimal, intermediate level of bureaucratic power extends the earlier literature, which has primarily focused on identifying which policies are more efficiently implemented by politicians alone and which are better left to bureaucrats.

\clearpage
\appendix

\section{Appendix}\label{appendix}

\subsection{Preliminaries}

\subsubsection{Strategies}

Since the game involves three players, each with different information, it is useful to clearly outline their strategies, recalling that we focus exclusively on pure strategies.

A politician's proposal in period-$t$ is a function
\[
q^P_t:\{g,b\}\times\{x,y\}\times \left[0,\bar R^P_t\right]\to \{x,y\},
\]
such that $q^P_t\left(\theta_t^P,s_t,r_t^P\right)$ is $P$'s policy proposal in period-$t$ when her type is $\theta_t^P$, the state is $s_t$, and realized rents are $r_t^P$. A bureaucrat's proposal in period-$t$ is a function
\[
q^B_t:\{g,b\}\times\{x,y\}\times \left[0,\bar R^B_t\right] \times\{x,y\} \to \{x,y\},
\]
such that $q^B_t\left(\theta^B,s_t,r_t^B,q_t^P\right)$ is $B$'s policy proposal in period-$t$ when her type is $\theta^B$, the state is $s_t$, the realized rents are $r_t^B$, and the politician's proposal is $q_t^P$. We will sometimes use $q_t^j$ as a shortcut to denote the above strategies, with $j\in\{P,B\}$ and $t\in\{1,2\}$. 

The voter's electoral decision is a function\footnote{Specifically, $\nu$ denotes the probability of re-election. Since we restrict attention to equilibria where the realized policy is informative about the politician's type (see Definition~\ref{def:pandering}), restricting $\nu$ to be either $0$ or $1$ is without loss of generality.}
\[
\nu:\{x,y\}\to\{0,1\},
\]
such that $\nu(p_1)$ denotes $V$'s electoral choice after observing policy $p_1$. 

\emph{Beliefs.---} The voter's and bureaucrat's posterior beliefs that the politician type is good are denoted by $\Pi_V(p_1)$ and $\Pi_B(p_1,s_1)$, respectively. A posterior belief function for the voter is a mapping $\Pi_V:\{x,y\}\to[0,1]$ such that $\Pi_V(p_1)$ denotes $V$'s posterior belief that the politician type is good. Similarly, $\Pi_B:\{x,y\}\times\{x,y\}\to[0,1]$ denotes $B$'s posterior belief that the politician's type is good.

\subsubsection{Equilibrium continuation values}\label{sec:app_cont_values}

This section presents the players’ continuation values, which are common across all equilibria and serve as a reference for the subsequent analysis. To derive these values, we first characterize equilibrium play in the second period. The starting point is the canonical result that, in every equilibrium, policymakers implement their preferred policy in the second period. 
\begin{lemma}\label{lemma:second}
    In every perfect Bayesian equilibrium (PBE), the policymakers' strategies in $t=2$ are, for $j\in\{P,B\}$, 
    \begin{equation*}
 q_2^j = 
  \begin{cases} 
   s_2 & \text{if } \theta_2^j=g \\
   y       & \text{if } \theta_2^j=b
  \end{cases}.
\end{equation*}
\end{lemma}
The proof of Lemma~\ref{lemma:second} is straightforward and thus omitted.


\bigskip
\noindent{\bf Bureaucrats' continuation values.}

\noindent First, we calculate the good bureaucrat's expected payoff in different scenarios. If a good politician is re-elected, a good bureaucrat expects to get
\[
\delta V_{gB}^{gP} \coloneqq \delta\left[ \rho v(x,x) + (1-\rho) v(y,y) \right].
\]
If a bad politician is re-elected, a good bureaucrat expects to get
\[
\delta V_{gB}^{bP} \coloneqq  \delta \left\{ \rho\left[ \lambda v(x,x) + (1-\lambda) v(y,x) \right] + (1-\rho)v(y,y) \right\}.
\]
However, recall that the bureaucrat does not know the politician's type. If the politician is removed in favor of the challenger (which, in the equilibria we consider, happens when $p_1=y$), a good bureaucrat expects to get
\[
\delta V_{gB}^\pi \coloneqq \delta \left[ \pi V_{gB}^{gP} + (1-\pi)V_{gB}^{bP} \right].
\]

Next, we calculate the bad bureaucrat's expected payoffs in different scenarios. If a good politician is re-elected, the bad bureaucrat expects to obtain
\[
\delta V_{bB}^{gP} \coloneqq \delta\left[ \rho\lambda\mu_2^B + (1-\rho)\mu_2^B\right]=\delta\mu_2^B\left[1-\rho(1-\lambda)\right].
\]
 If a bad politician is re-elected, the bad bureaucrat expects to obtain
\[
\delta V_{bB}^{bP}\coloneqq\delta\mu_2^B.
\]
If the politician is removed and replaced with an unknown challenger, the bad bureaucrat expects to get
\[
\delta V_{bB}^\pi\coloneqq\delta\left[ \pi V_{bB}^{gP} + (1-\pi) V_{bB}^{bP} \right]=\delta\mu_2^B [1-\pi\rho(1-\lambda)].
\]


\bigskip
\noindent{\bf Politicians' continuation values.}

\noindent The good politician's expected payoff from re-election with a good bureaucrat is
\[
\delta V_{gP}^{gB} \coloneqq \delta\left[ \rho v(x,x) + (1-\rho)v(y,y) + E \right].
\]
The good politician's expected payoff from re-election with a bad bureaucrat is
\[
\delta V_{gP}^{bB} \coloneqq \delta\left\{ \rho\left[ \lambda v(y,x) + (1-\lambda)v(x,x) \right] + (1-\rho)v(y,y) + E  \right\}.
\]
Notice that $V_{gP}^{bB} =  V_{gP}^{gB}-\rho\lambda\left( v(x,x)-v(y,x) \right)$.

The good politician's expected payoff from being replaced with an unknown challenger and with a good bureaucrat is
\begin{align*}
\delta U_{gP}^{gB} \coloneqq &\delta \Big\{ \pi\left[ \rho v(x,x) + (1-\rho)v(y,y) \right] \\
&+ (1-\pi)\left[ \rho\left( \lambda v(x,x) + (1-\lambda)v(y,x) \right) + (1-\rho)v(y,y) \right] \Big\}.
\end{align*}
The above expression can be re-written as
\[
U_{gP}^{gB}= \rho[\pi+(1-\pi)\lambda] v(x,x) + (1-\rho) v(y,y) + (1-\pi)(1-\lambda)\rho v(y,x).
\]
The good politician's expected payoff from being replaced with an unknown challenger and with a bad bureaucrat is
\begin{align*}
\delta U_{gP}^{bB}\coloneqq &\delta\Big\{ \pi\left[ \rho \left( \lambda v(y,x) + (1-\lambda)v(x,x) \right) + (1-\rho)v(y,y) \right] \\
& + (1-\pi)\left[ \rho v(y,x) + (1-\rho)v(y,y) \right] \Big\}.
\end{align*}
The above equation can be re-written as
\[
U_{gP}^{bB}= \pi\rho (1-\lambda) v(x,x) + (1-\rho)v(y,y) + \rho [(1-\pi) +\pi\lambda] v(y,x).
\]

We can further notice that $U_{gP}^{bB}=U_{gP}^{gB}-\rho\lambda[v(x,x)-v(y,x)]$. We obtain the following relationship, which will be useful in later calculations,
\[
V_{gP}^{gB}-V_{gP}^{bB}=U_{gP}^{gB}-U_{gP}^{bB}=\rho\lambda[v(x,x)-v(y,x)].
\]
Lastly, we observe that $V_{gP}^{gB}>V_{gP}^{bB}$, $U_{gP}^{gB}>U_{gP}^{bB}$, and $V_{gP}^{gB}>U_{gP}^{gB}$. 

We turn our attention to bad politicians. The re-election utility a bad politician gets when there is a good bureaucrat is
\[
\delta V_{bP}^{gB}\coloneqq\delta\left\{ \rho  (1-\lambda)\mu_2^P + (1-\rho)\mu_2^P  + E\right\} = \delta \left[ (1-\rho\lambda)\mu_2^P + E \right].
\]
The re-election utility a bad politician gets when there is a bad bureaucrat is
\[
\delta V_{bP}^{bB}\coloneqq\delta\left( \mu_2^P + E \right).
\]

Finally, we assume that the expected utility of a bad politician upon being replaced is normalized to zero, since they are indifferent to policy outcomes. That is,
\[
U_{bP}^{gB}=U_{bP}^{bB}=0.
\]


\subsection{Pandering equilibria}\label{sec:app_PE}

\subsubsection{Belief updating in PECB}\label{sec:app_beliefsPECB}

This section outlines the players’ beliefs in PECB and their implications. We begin with the following definition, which specifies the expected strategies of bad policymakers in PECB.

\begin{definition}
    Consider a PECB. From the perspective of voters and politicians, the probability that a bad bureaucrat proposes $x$ after the politician has proposed $y$ is state-independent and denoted by $\xi$, where
        \[
        \xi \coloneqq Pr\left( q_1^B=x \mid \theta^B=b,q_1^P=y,s_1\right);
        \]
    From the perspective of voters and bureaucrats, the probability that a bad politician proposes $x$ is state-independent and denoted by $\gamma$, where, for every $s_1$,
        \[
        \gamma \coloneqq Pr\left( q_1^P=x \mid \theta_1^P=b, \cdot \right).
        \]
\end{definition}

 Next, we examine the voter’s beliefs about the politician’s type given the implemented policy. The voter expects a proposal $q_1^P=x$ ($q_1^P=y$) to materialize in the same policy $p_1=x$  ($p_1=y$) with some probability $X_V$ ($Y_V$), and to convert to $p_1=y$ ($p_1=x$) with the remaining probability. From the voter's viewpoint,
\[
X_V=\beta \rho \lambda+(1-\lambda),
\]
\[
  Y_V= \beta\left(1-\rho\lambda\right) + (1-\beta)\left( 1-\xi\lambda \right).
\]


Given these probabilities, and upon observing an implemented policy $p_1$, voters use Bayes' rule to update their posterior belief that the politician's type is good, which is
\[
\Pi_V(x)=\frac{\pi X_V}{\pi X_V + \left(1-\pi\right)\left[\gamma X_V + \left(1-\gamma\right)\left(1-Y_V\right)\right]}.
\]

The voter's posterior is consistent with the PECB outlined by Definition~\ref{def:conj} provided that $\Pi_V(x) > \pi $. Indeed, the voter re-elects the incumbent after observing $p_1=x$, and ousts the incumbent otherwise.\footnote{The inequalities $\Pi_V(x) > \pi$ and $\Pi_V(y) < \pi$ both lead to the same condition. This holds not only here, but also in the belief updating process of the other equilibria.} Moreover, the implemented policy must be informative about the politician's type, which is a requirement for our definition of PE (Definition~\ref{def:pandering}). The inequalities $\Pi_V(x)>\pi$ hold true when $\gamma<1$ and $X_V > 1-Y_V$. We obtain the condition expressed by Observation~\ref{obs:cond}.
\begin{observation}\label{obs:cond}
    The existence of PECB requires $\gamma<1$ and
    \[
     \xi<\frac{1-\lambda}{\lambda(1-\beta)}.
    \]
    The latter inequality binds if and only if $\lambda>\frac{1}{2-\beta}$.
\end{observation}
The above observation indicates that, in expectation, the bad bureaucrat should not try to convert proposals $q_1^P=y$ into a policy $p_1=x$ too often. If the probability $\xi$ were to be excessively high, then the implemented policy $p_1=x$ would no longer constitute an informative signal that the politician is good. In that case, sequential rationality would imply that voters replace politicians when the implemented policy turns out to be $x$. If the condition in Observation~\ref{obs:cond} is satisfied with equality, then $p_1=x$ would not change the voter's prior, making the equilibrium non-informative.

We now turn our attention to the bureaucrat's posterior beliefs about the politician's type, $\Pi_B\left( q_1^P , s_1 \right)$. After observing $q_1^P=x$, $B$'s posterior belief that $P$'s type is good is,
\begin{equation}\label{eq:beliefs_PECB1}
\Pi_B(x,x)=\Pi_B(x,y)=\frac{\pi}{\pi+(1-\pi)\gamma}.
\end{equation}
We obtain that $\Pi_B(x,\cdot)>\pi$ as long as the probability that the bad politician proposes $x$ in the first period is $\gamma<1$ which, as we observed, must hold true in PECB. Since in this conjectured PECB only the bad politician proposes $q_1^P=y$ with positive probability, we also have that
\begin{equation}\label{eq:beliefs_PECB2}
\Pi_B (y,x) = \Pi_B (y,y) =  0.
\end{equation}

\subsubsection{Belief updating in PEPB}\label{sec:app_beliefsPEPB}

This section studies beliefs in PEPB. As before, we begin with the following definition.

\begin{definition}
    Consider a PEPB. From the perspective of voters and politicians, the probability that a bad bureaucrat proposes $x$ after the politician has proposed $y$ is denoted by $\xi$, as outlined in Definition~\ref{def:PECB_xi_gamma}. From the perspective of voters and bureaucrats, the probability that a bad politician proposes $x$ is state-dependent and denoted by $\gamma(s_1)$, where
        \[
        \gamma(s_1) \coloneqq Pr\left( q_1^P=x \mid \theta_1^P=b, s_1 \right).
        \]
\end{definition}

We can recalculate the probabilities from the voters' viewpoint as follows
\[
X_V=\beta \lambda+(1-\lambda),
\]
\[
Y_V=  \beta\left(1-\rho\lambda\right) + (1-\beta)\left( 1-\xi\lambda \right).
\]
Given these probabilities, and upon observing an implemented policy $p_1$, voters use Bayes' rule to update their posterior belief that the politician's type is good, which is
\[
\Pi_V(x)=\frac{\pi X_V}{\pi X_V + \left(1-\pi\right)\left[\hat \gamma X_V + \left(1-\hat\gamma\right)\left(1-Y_V\right)\right]},
\]
where $\hat\gamma$ is defined as
\[
\hat\gamma\coloneqq\rho\gamma(x) + (1-\rho)\gamma(y).
\]
As we shall see, in a PEPB it is possible that $\gamma(x)\neq \gamma(y)$.

The voters' beliefs are consistent with the strategies outlined in PEPB if and only if \(\Pi_V(x) > \pi \), which holds true  $\hat\gamma <1$ and \(X_V>1-Y_V\). Compared to PECB, PEPB require a less stringent condition on the ex-ante likelihood that bad bureaucrats attempt converting a proposal \(q_1^P=y\) into policy \(p_1=x\). This softening occurs because, in PEPB, a good bureaucrat supports policy proposals \(q_1^P=x\) under any circumstance, making policy \(x\) a stronger  indicator of the politicians' type, as such proposals are more often set forth by good politicians. The inequality $X_V>1-Y_V$ yields the following restriction,
\[
\xi<\frac{1-\lambda}{\lambda(1-\beta)}+(1-\rho)\frac{\beta}{1-\beta}
\]
However, incorporating the good politician’s equilibrium behavior refines this condition, making it as stringent as the one in the PECB case (i.e., Observation~\ref{obs:cond})---as shown in Step~\ref{step:PECB_goodP_x_sx}, which applies equally to the PEPB. 

The following observation provides requirements ensuring that the conjectured PEPB is informative, and it already includes the additional equilibrium restriction from Step~\ref{step:PECB_goodP_x_sx}.
\begin{observation}\label{obs:2}
    The existence of PEPB requires $\hat\gamma<1$ and
    \[
      \xi<\frac{1-\lambda}{\lambda(1-\beta)}.
    \]
    The latter inequality binds if and only if $\lambda>\frac{1}{2-\beta}$.
\end{observation}
We now turn our attention to the bureaucrat's posterior beliefs about the politician's type, i.e., $\Pi_B\left( q_1^P,s_1 \right)$. After observing $q_1^P$ in state $s_1$, $B$'s posterior belief that $P$'s type is good is, for $s_1\in\{x,y\}$,
\begin{equation}\label{eq:beliefs_PEPB1}
\Pi_B(x,s_1)=\frac{\pi}{\pi+(1-\pi)\gamma(s_1)}\geq\pi,
\end{equation}
\begin{equation}\label{eq:beliefs_PEPB2}
\Pi_B(y,s_1)=0.
\end{equation}
We obtain that $\Pi_B(x,s_1)\geq\pi$ always, and $\Pi_B(x,s_1)>\pi$ as long as the probability that the bad politician proposes $x$ in state $s_1$ is $\gamma(s_1)< 1$ which, as we observed, must hold true in at least one of the two states. By Step~\ref{step:PEPB_goodB_x_sy}, $\gamma(y)<1$ must hold in PEPB, implying $\Pi_B(x,y)>\pi$.


\subsubsection{Proof of Proposition~\ref{PECBexistence}}

We prove the proposition using a series of steps and observations. We examine, in order, the behaviors of the good bureaucrat, the bad bureaucrat, the good politician, and the bad politician. Steps~\ref{step:PECB_goodB_x}--\ref{step:PECB_badP_y} and Observations~\ref{obs:cond}--\ref{obs:PECB_gamma} complete the proof of Proposition~\ref{PECBexistence}.


\bigskip

\noindent{\bf Good bureaucrats in PECB.}


\begin{step}\label{step:PECB_goodB_x}
    Consider the case where $\theta^B=g$, $s_1=q_1^P=x$. Upon observing $q_1^P=x$ in $s_1=x$, proposing $q_1^B=x$ (thereby confirming the politician proposal and inducing $p_1=x)$ grants the bureaucrat an expected payoff of
    \[
v(x,x)+\delta\left[ \Pi_B(x,\cdot) V_{gB}^{gP} + (1-\Pi_B(x,\cdot))V_{gB}^{bP} \right].
\]
Alternatively, the bureaucrat can contest the politician's proposal by setting forth $q_1^B=y$, obtaining
\begin{multline*}
    \lambda \left[ v(y,x) + \delta\left( \pi V_{gB}^{gP} + (1-\pi)V_{gB}^{bP} \right)  \right] \\
    + (1-\lambda) \left[ v(x,x)+\delta\left(  \Pi_B(x,\cdot) V_{gB}^{gP} + (1-\Pi_B(x,\cdot))V_{gB}^{bP} \right)  \right]
\end{multline*}
Since $v(x,x)>v(y,x)$, $\Pi_B(x,\cdot)>\pi$, and $V_{gB}^{gP}>V_{gB}^{bP}$, we obtain that in a PECB, a good bureaucrat's best reply to $q_1^P=s_1=x$ is $q^B_1\left(g,x,r_1^B,x\right)=x$.
\end{step}
    

\begin{step}\label{step:PECB_goodB_y_sx}
    Consider the case where $\theta^B=g$, $q^P_1=y$, and $s_1=x$. Proposing $q_1^B=y$ yields the bureaucrat a bad outcome today, but guarantees that the surely bad politician is removed from office and replaced with a random challenger. The bureaucrat's expected payoff from proposing $q_1^B=y$  in this case is
\[
v(y,x) +\delta\left[ \pi V_{gB}^{gP} + (1-\pi)V_{gB}^{bP} \right] = v(y,x) + \delta V_{gB}^\pi.
\]
Differently, proposing $q_1^B=x$ may yield to the bureaucrat a better outcome in period 1, at the cost of potentially inducing the re-election of a surely bad politician. The bureaucrat's expected payoff from proposing $q_1^B=x$ in this case is
\[
\lambda \left[v(x,x) + \delta V_{gB}^{bP} \right] + (1-\lambda)\left[v(y,x) + \delta \left( \pi V_{gB}^{gP} + (1-\pi)V_{gB}^{bP} \right) \right]. 
\]
The bureaucrat's expected payoff from proposing $q_1^B=x$ is higher than that from proposing $q_1^B=y$ if
\[
v(x,x)-v(y,x)>\delta\pi\left( V_{gB}^{gP} - V_{gB}^{bP}  \right) = \delta\pi \left[ \rho (1-\lambda) \left( v(x,x) - v(y,x) \right) \right],
\]
which is always true as $\delta,\pi,\rho,\lambda\in(0,1)$. Therefore, in a PECB, a good bureaucrat's best reply to $q_1^P=y\neq s_1=x$ is $q^B_1\left(g,x,r_1^B,y\right)=x$.
\end{step}


\begin{step}\label{step:PECB_goodB_x_sy}
    Consider the case where $\theta^B=g$, $q_1^P=x$, and $s_1=y$. The good bureaucrat prefers a realized policy $p_1=y$ in state $y$. Confirming $q_1^B=x$ yields the bureaucrat a bad policy outcome in the first period. However, it also yields the re-election of a politician that is more likely to be good than the challenger, as $\Pi_B(x,\cdot)>\pi$. The bureaucrat's expected payoff from proposing $q_1^B=x$ is
\[
v(x,y) + \delta \left[ \Pi_B(x,\cdot) V_{gB}^{gP} + (1-\Pi_B(x,\cdot))V_{gB}^{bP} \right].
\]
Differently, proposing $q_1^B=y$ may give the bureaucrat a better policy outcome in the first period at the cost of potentially replacing a good politician. The bureaucrat's expected payoff from proposing $q_1^B=y$ is
\begin{multline*}
\lambda\left[ v(y,y) + \delta\left( \pi V_{gB}^{gP} + (1-\pi)V_{gB}^{bP} \right) \right] \\
+ (1-\lambda)\left[ v(x,y) + \delta \left(\Pi_B(x,\cdot) V_{gB}^{gP} + \left(1-\Pi_B(x,\cdot)\right)V_{gB}^{bP} \right) \right].
\end{multline*}
As a result, proposing $q_1^B=y$ is optimal for the bureaucrat if and only if
\[
v(y,y) + \delta\left[ \pi V_{gB}^{gP} + (1-\pi)V_{gB}^{bP} \right] \geq v(x,y) + \delta \left[\Pi_B(x,\cdot) V_{gB}^{gP} + (1-\Pi_B(x,\cdot))V_{gB}^{bP} \right].
\]
Rearranging, we obtain
\[
v(y,y)-v(x,y) \geq \delta (\Pi_B(x,\cdot)-\pi)\left[ V_{gB}^{gP} - V_{gB}^{bP} \right] = \delta (\Pi_B(x,\cdot)-\pi)\rho(1-\lambda)[v(x,x)-v(y,x)].
\]
Since $\Pi_B(x,\cdot)>\pi$, and $\delta,\rho,\lambda,\Pi_B(x,\cdot),\pi\in(0,1)$, we have that $\delta (\Pi_B(x,\cdot)-\pi)\rho(1-\lambda)\in (0,1)$. If the condition is not satisfied, then it is optimal for the bureaucrat to propose $q^B_1\left(g,y,r_1^B,x\right)=x$. Therefore, in a PECB, a good bureaucrat's best reply to $q_1^P=x\neq s_1=y$ is $q^B_1\left(g,y,r_1^B,x\right)=y$ if and only if
\[
v(y,y)-v(x,y) \geq \delta (\Pi_B(x,\cdot)-\pi)\rho(1-\lambda)[v(x,x)-v(y,x)],
\]
This condition can be re-written as
\[
    \Delta \geq \Delta_{CB}\coloneqq \delta\rho(1-\lambda)(\Pi_B(x,\cdot)-\pi).
    \]
When $\Delta = \delta\rho(1-\lambda)(\Pi_B(x,\cdot)-\pi)$, the good bureaucrat is indifferent between proposing $x$ and $y$. We assume that, in this knife-edge case, the good bureaucrat challenges $x$ by proposing $y$.
\end{step}


\begin{step}\label{step:PECB_goodB_y}
    Consider the case where $\theta^B=g$ and $q_1^P=s_1=y$. Recall that, in a PECB, $\Pi_B(y,\cdot)=0$. In this case, that the good bureaucrat chooses to confirm policy $y$  follows from the observation that, by proposing $q_1^B=x$, the bureaucrat generates with positive probability both a policy-state mismatch and the re-election of a surely bad politician. By contrast, proposing $q_1^B=y$ surely leads to a good policy outcome and the replacement of a bad politician. Thus, in a PECB, a good bureaucrat's best reply to $q_1^P=s_1=y$ is $q^B_1\left(g,y,r_1^B,y\right)=y$.
\end{step}

    
\bigskip

\noindent{\bf Bad bureaucrats in PECB.}


\begin{step}\label{step:PECB_badB_x}
    Consider the case where $\theta^B=b$ and $q_1^P=x$. The bureaucrat's posterior about the politician's type being good is $\Pi_B(x,\cdot)>\pi$. The bad bureaucrat's expected utility when proposing $q_1^B=x$ is
    \[
    \delta\left\{ \Pi_B(x,\cdot)\left[ \rho\lambda\mu_2^B + (1-\rho)\mu_2^B \right] + \left( 1-\Pi_B(x,\cdot) \right)\mu_2^B \right\} = \delta\mu_2^B\left[ 1-\Pi_B(x,\cdot)\rho(1-\lambda) \right].  
    \]
    On the other hand, their expected utility when proposing $q_1^B=y$ is
    \[
    \lambda\left(r_1^B+\delta V_{bB}^\pi\right)+(1-\lambda)\delta\mu_2^B\left[ 1-\Pi_B(x,\cdot)\rho(1-\lambda) \right].
    \]
    After some simplification, we see that when $q_1^P=x$, the bad bureaucrat is better off when proposing $y$ rather than $x$ when
    \[
    r_1^B\geq \delta\mu_2^B\rho(1-\lambda)\left(\pi-\Pi_B(x,\cdot)\right).
    \]
    The right-hand side of this inequality is negative (since $\Pi_B(x,\cdot)>\pi$). As a result, the above condition is always satisfied, implying that in a PECB the bad bureaucrat's best response to $q_1^P=x$ is always $q_1^B=y$ (for every $s_1\in\{x,y\}$).
\end{step}


\begin{step}\label{step:PECB_badB_y}
     Consider the case where $\theta^B=b$ and $q_1^P=y$. The bureaucrat's posterior about the politician's type being good is $\Pi_B(y,\cdot)=0$. The bad bureaucrat's expected utility when proposing $q_1^B=x$ is
    \[
    \lambda\delta V_{bB}^{bP} +(1-\lambda)\left(r_1^B+\delta V_{bB}^\pi\right).
    \]
    The bad bureaucrat's expected utility when proposing $q_1^B=y$ is
    \[
    r_1^B +\delta V_{bB}^\pi.
    \]
    Therefore, in a PECB, and for every $s_1\in\{x,y\}$, the bad bureaucrat's best response to  $q_1^P=y$ is $q_1^B=x$ if and only if
    \[
    r_1^B<\delta\pi\rho(1-\lambda)\mu_2^B.
    \]
    It follows that the probability that a bad bureaucrat proposes policy $x$ when $q_1^P=y$ is $\xi$, where
    \begin{equation}\label{eq:xi_PE}
    \xi\coloneqq Pr\left( r_1^B<\delta\pi\rho(1-\lambda)\mu_2^B \right)=F_1^B\left( \delta\pi\rho(1-\lambda)\mu_2^B  \right).
    \end{equation}
\end{step}

\bigskip

Because the realization of $r_1^B$ is stochastic, the following observation is in order. In the conjectured equilibrium, $\xi=1$  unless the upper bound on bad bureaucrats’ rents satisfies
    \[
    \bar R_1^B>\delta\pi\rho(1-\lambda)\mu_2^B.
    \]
If this condition holds, then $\xi<1$.

However, from Section~\ref{sec:app_beliefsPECB}, we have an equilibrium restriction requiring that $\xi$, which must satisfy $\xi<\frac{1-\lambda}{\lambda(1-\beta)}\in[0,+\infty)$. This upper bound is binding only when
\[
\frac{1-\lambda}{\lambda(1-\beta)}\leq 1, 
\]
which occurs if and only if
\[
\lambda\geq \frac{1}{2-\beta}\in\left[\nicefrac{1}{2},1\right].
\]

If $\lambda < \frac{1}{2-\beta}$, no additional condition is needed, as $\xi$ can take any value in $[0,1]$. However, when $\lambda\geq \frac{1}{2-\beta}$, we must ensure that $\xi<\frac{1-\lambda}{\lambda(1-\beta)}$. From Step~\ref{step:PECB_badB_y}, we know that
    \[
    \xi \coloneqq F_1^B\left( \delta\pi\rho(1- \lambda)\mu_2^B \right),
    \]
    so the condition becomes 
    \[
    F_1^B\left( \delta\pi\rho(1- \lambda)\mu_2^B \right)<\frac{1-\lambda}{\lambda(1-\beta)}.
    \]
By inverting the CDF $F_1^B$, we obtain the following observation.
\begin{observation}\label{obs:PECB_xi}
    In PECB, when $\lambda \geq \frac{1}{2 - \beta}$, it must hold that
    \[
    \xi<\frac{1-\lambda}{\lambda(1-\beta)},
    \]
    or, written alternatively,
    \begin{equation}\label{eq:mu_PE}
        \mu_2^B < \bar \mu_{PE} \coloneqq \frac{\left(F_1^{B}\right)^{-1}\left( \frac{1-\lambda}{\lambda(1-\beta)}\right)}{\delta\pi\rho(1-\lambda)}.
    \end{equation}
\end{observation}

This observation is key to the existence of PECB. It implies that bad bureaucrats must retain short-term incentives when they hold sufficiently high influence over policymaking: they must be tempted often enough to secure period-1 rents by replacing a surely bad politician, even at the cost of forgoing expected rents in the second period.


\bigskip

\noindent{\bf Good politicians in PECB.}

\begin{step}\label{step:PECB_goodP_x_sx}
Suppose that $\theta^P=g$ and $s_1=x$. By selecting $q_1^P=x$, the good politician obtains in expectation
\begin{displaymath}
\beta\left[ v(x,x) + \delta V_{gP}^{gB} \right] + (1-\beta)\left[ \lambda\left( v(y,x)+ \delta U_{gP}^{bB} \right) + (1-\lambda)\left( v(x,x) + \delta V_{gP}^{bB} \right) \right].
\end{displaymath}
By selecting $q_1^P=y$, the good politician obtains
\begin{multline*}
\beta\left\{  \lambda \left[ v(x,x) +\delta V_{gP}^{gB} \right] + (1-\lambda) \left[ v(y,x)+ \delta U_{gP}^{gB} \right]\right\} \\
+ (1-\beta) \xi \left[ \lambda \left( v(x,x) + \delta V_{gP}^{bB} \right) + (1-\lambda) \left( v(y,x)+ \delta U_{gP}^{bB} \right)\right] \\
+ (1-\beta)(1-\xi)\left[ v(y,x) + \delta U_{gP}^{bB} \right].
\end{multline*}

Recall that $v(x,x)>v(y,x)$, $V_{gP}^{gB}>U_{gP}^{gB}$, and $V_{gP}^{bB}>U_{gP}^{gB}$. After re-arranging and simplifying,  the first expected utility (from $q_1^P=x$) is always greater than the second (from $q_1^P=y$) when
\begin{multline*}
\left[ (1-\lambda) - (1-\beta)\xi\lambda \right](v(x,x)-v(y,x))\\
\geq (1-\beta)[\xi\lambda - (1-\lambda)]\delta\left( V_{gP}^{bB} - U_{gP}^{bB} \right) - \beta(1-\lambda)\delta\left( V_{gP}^{gB} - U_{gP}^{gB} \right).
\end{multline*}
By isolating $\xi$, we get the following inequality
\begin{displaymath}
    \xi \leq \left(\frac{1-\lambda}{\lambda(1-\beta)} \right) \left\{\frac{\left[v(x,x)-v(y,x)\right]+\delta\left[ \beta\left( V_{gP}^{gB} - U_{gP}^{gB} \right) +(1-\beta)\left( V_{gP}^{bB} - U_{gP}^{bB} \right) \right]}{\left[v(x,x)-v(y,x)\right] + \delta\left( V_{gP}^{bB} - U_{gP}^{bB} \right)} \right\}.
\end{displaymath}
The second fraction in the inequality's right-hand side is equal to 1 when
\[
V_{gP}^{gB}-U_{gP}^{gB}= V_{gP}^{bB}-U_{gP}^{bB},
\]
which, as we have already shown, holds true. The inequality boils down to 
\[
\xi\leq \frac{1-\lambda}{\lambda(1-\beta)},
\]
which we know is satisfied in our equilibrium by the conditions we are imposing to have informative equilibria, as per Section~\ref{sec:app_beliefsPECB}. Therefore, in PECB, when the state is $x$ the good politician prefers to propose $x$.
\end{step}


\begin{step}\label{step:PECB_goodP_x_sy}
     Suppose that $\theta^P=g$ and  $s_1=y$. By selecting $q_1^P=x$, the good politician obtains
    \begin{multline*}\label{eq:goodP}
        \beta \left\{ \lambda \left[ v(y,y) + \delta U_{gP}^{gB} \right] + (1-\lambda)\left[ v(x,y) + \delta V_{gP}^{gB} \right] \right\} \\
        + (1-\beta) \left\{ \lambda \left[ v(y,y) + \delta U_{gP}^{bB} \right] + (1-\lambda)\left[ v(x,y) + \delta V_{gP}^{bB} \right] \right\}. 
    \end{multline*}
     By selecting $q_1^P=y$, the good politician obtains
\begin{multline*}
\beta(v(y,y)+\delta U^{gB}_{gP})\\
+(1-\beta)\left[\xi(\lambda(v(x,y)+\delta V^{bB}_{gP} )+(1-\lambda)(v(y,y)+\delta U^{bB}_{gP} )   ) +(1-\xi)(v(y,y)+\delta U^{bB}_{gP} ) \right].
\end{multline*}
Using the condition $\xi\leq \frac{1-\lambda}{\lambda(1-\beta)}$, we obtain that the good politician prefers proposing $x$ to $y$ when
\begin{equation}\label{eq:pandering}
    \delta E \geq v(y,y)-v(x,y) - \delta\rho( 1-\pi)(1-\lambda)\left[ v(x,x)-v(y,x) \right].
  \end{equation}
Therefore, in a PECB, the good politician's optimal proposal in state $s_1=y$ is $q_1^P=x$ if and only if office rents satisfy the above inequality. The right-hand side of the inequality is non-negative if
  \begin{equation*}
v(y,y)-v(x,y) - \delta\rho( 1-\pi)(1-\lambda)\left[ v(x,x)-v(y,x)\right]\geq 0.
\end{equation*}
Rearranging, and using the definition of $\Delta$, the condition becomes
\begin{equation*}
\Delta \geq \delta\rho( 1-\pi)(1-\lambda).
\end{equation*}

Recall that in our conjectured equilibrium, the following condition holds: $\Delta > \delta\rho\left(\Pi_B(x,\cdot)-\pi\right)(1-\lambda)$. Notice that the former implies the latter because $\Pi_B(x,\cdot)<1$. Under payoff symmetry ($\Delta=1$), both are trivially satisfied.
 
Lastly, we can rearrange the condition on office rents and express it in terms of $\Delta$. By doing so, we can see that the conjectured PECB equilibrium requires
\[
\Delta\leq \frac{\delta E}{v(x,x)-v(y,x)}+\delta\rho( 1-\pi)(1-\lambda).
\]
Therefore, for the conjectured PECB to exist, we require that 
\[
\frac{\delta E}{v(x,x)-v(y,x)}+\delta\rho( 1-\pi)(1-\lambda)\geq \Delta>\delta\left(\Pi_B(x,\cdot)-\pi\right)\rho(1-\lambda).
\]
which trivially holds for $E\geq 0$, and may even hold for some negative values of $E$.
\end{step}


\bigskip

\noindent{\bf Bad politicians in PECB.}

\begin{step}\label{step:PECB_badP_x}
    Suppose $\theta^P=b$ and $s_1=x$. By proposing $q_1^P=x$, the bad politicians' expected payoff is
    \[
    \beta \delta V_{bP}^{gB} + (1-\beta)\left[ \lambda r_1^P + (1-\lambda) \delta V_{bP}^{bB} \right].
    \]
    By proposing $q_1^P=y$, their expected payoff is
    \[
    \beta \left[ \lambda \delta V_{bP}^{gB} + (1-\lambda)r_1^P \right] + (1-\beta) \left\{ \xi \left[ \lambda \delta V_{bP}^{bB} + (1-\lambda)r_1^P \right] + (1-\xi)r_1^P \right\}.
    \]
    Therefore, the bad politician prefers to propose $x$ in state $x$ when the former payoff is at least as high as the latter. By isolating rents $r_1^P$, and invoking the restriction $\xi\leq\frac{1-\lambda}{\lambda(1-\beta)}$, this is the case when  
    \[
    r_1^P \leq \delta \left(\mu_2^P + E \right) - \frac{\delta\beta\rho\lambda(1-\lambda)}{1-\lambda\left[ 1+\xi(1-\beta) \right]}\mu_2^P.
    \]
The inequality's right-hand side is strictly positive  provided that
    \[
    \xi < \frac{1-\lambda}{\lambda(1-\beta)} + \underbrace{\left(\frac{\beta}{1-\beta}\right)\frac{(1-\lambda)\rho\mu_2^P}{\mu_2^P + E}}_{>0}.
    \]
    Since $\xi < \frac{1-\lambda}{\lambda(1-\beta)}$, the above condition is automatically satisfied. Therefore, in a PECB, the bad politician's optimal proposal in state $s_1=x$ is $q_1^P=x$ if and only if $r_1^P$ is sufficiently low.

    It follows that the probability that a bad politician proposes policy $x$ when $s_1=x$ is $\gamma$, where
    \begin{multline*}
    \gamma\coloneqq Pr\left( r_1^P<\delta \left(\mu_2^P + E\right) - \frac{\delta\beta\rho\lambda(1-\lambda)}{1-\lambda\left[ 1+\xi(1-\beta) \right]}\mu_2^P \right) \\
    =F_1^P\left( \delta \left(\mu_2^P + E\right) - \frac{\delta\beta\rho\lambda(1-\lambda)}{1-\lambda\left[ 1+\xi(1-\beta) \right]}\mu_2^P   \right).
    \end{multline*}
\end{step}

\begin{step}\label{step:PECB_badP_y}
    Suppose $\theta^P=b$ and $s_1=y$. Proposing $x$ gives the bad politician
    \[
    \beta\left[ \lambda r_1^P + (1-\lambda)\delta V_{bP}^{gB}\right] + (1-\beta)\left[ 
\lambda r_1^P + (1-\lambda)\delta V_{bP}^{bB} \right].
    \]
    By contrast, the bad politicians' expected payoff when proposing $y$ in state $y$ is
    \[
    \beta r_1^P + (1-\beta)\left\{ \xi\left[ \lambda \delta V_{bP}^{bB} + (1-\lambda)r_1^P \right] + (1-\xi)r_1^P \right\}.
    \]
    We obtain that in state $y$ the bad politician prefers to propose $x$ when
     \[
    \beta (1-\lambda) \delta V_{bP}^{gB} + (1-\beta)[1-\lambda(1+\xi)]\delta V_{bP}^{bB} \geq \left\{ 1-\lambda\left[ 1 + \xi(1-\beta) \right] \right\}r_1^P.
    \]
    The above is exactly the same condition we have found in state $x$. Therefore, in a PECB, the bad politician's optimal proposal in state $s_1=y$ is $q_1^P=x$ if and only if
    \[
    r_1^P \leq \delta \left(\mu_2^P + E\right) - \frac{\delta\beta\rho\lambda(1-\lambda)}{1-\lambda\left[ 1+\xi(1-\beta) \right]}\mu_2^P.
    \]
    It follows that the probability that a bad politician proposes policy $x$ when $s_1=y$ is exactly the same as when the state is $s_1=x$, that is, $\gamma$.
\end{step}

\bigskip

The following observation is in order.

\begin{observation}\label{obs:PECB_gamma}
    For the conjectured PECB to exist, we need that
    \[
        \bar R_1^P > \delta \left(\mu_2^P + E\right) - \frac{\delta\beta\rho\lambda(1-\lambda)}{1-\lambda\left[ 1+\xi(1-\beta) \right]}\mu_2^P.
    \]
\end{observation}
If the above condition does not hold, then the bad politician would always choose $x$, resulting in $\gamma=1$. In such a case, the equilibrium would be uninformative because we would have $\Pi_V(x)=\Pi_B(x,\cdot)=\pi$ (see Observation~\ref{obs:cond} and the following discussion). From Step~\ref{step:PECB_badP_x} and \ref{step:PECB_badP_y}, we obtain that the probability $\gamma$ is
    \begin{equation}\label{eq:PECB_gamma}
        \gamma\coloneqq F_1^P\left( \delta \left(\mu_2^P + E\right) - \frac{\delta\beta\rho\lambda(1-\lambda)}{1-\lambda\left[ 1+\xi(1-\beta) \right]}\mu_2^P \right).
    \end{equation}

The condition in Observation~\ref{obs:PECB_gamma} ensures that $\gamma<1$, and follows straight-forwardly from the fact that, by definition,  $F_1^P\left(\bar R_1^P\right)\triangleq 1$.

Collecting the results of this proof we get the following technical conditions that correspond to the verbal ones in the statement of the proposition in the text:

\begin{observation}
\label{PECPtechnicalconditions}
A PECB exists if and only if
\begin{itemize}[noitemsep]
        \item[i)] $\delta E\geq  v(y,y)-v(x,y) - \delta\rho( 1-\pi)(1-\lambda)\left[ v(x,x)-v(y,x) \right]$;
        \item[ii)]    
       $ \Delta \geq  \delta \rho(1-\lambda)\left[ \Pi_B(x,\cdot) - \pi \right];$
        
        \item[iii)]  $\bar R_1^P >  \delta \left(\mu_2^P + E \right) - \frac{\beta\rho\lambda(1-\lambda)}{1-\lambda\left[ 1+\xi(1-\beta) \right]}\delta\mu_2^P$;     
        \item[iv)]   
       $ \mu_2^B  <  \frac{\left(F_1^{B}\right)^{-1}\left( \frac{1-\lambda}{\lambda(1-\beta)}\right)}{\delta\pi\rho(1-\lambda)}.$
       \qed
\end{itemize}
\end{observation}



\subsubsection{Proof of Proposition~\ref{PEPBexistence}}
The proof of the existence of a PEPB follows steps analogous to those of the PECB. Steps~\ref{step:PECB_goodB_x}, \ref{step:PECB_goodB_y}, and \ref{step:PECB_badB_x} differ only in that the posteriors $\Pi_B(x,\cdot)$ and $\Pi_B(y,\cdot)$, defined by~\eqref{eq:beliefs_PECB1} and \eqref{eq:beliefs_PECB2}, are replaced by $\Pi_B(x,x)$ and $\Pi_B(y,y)$, defined by~\eqref{eq:beliefs_PEPB1} and \eqref{eq:beliefs_PEPB2}, respectively. Since the results remain unchanged, we do not repeat them here. Only Steps~\ref{step:PECB_goodB_x_sy}, \ref{step:PECB_goodP_x_sy}, and \ref{step:PECB_badP_y} differ substantially, and shall be revised accordingly; all other steps are identical and therefore omitted for brevity. These results and Observation~\ref{obs:PEPB_bounds} complete the proof.


\begin{step}\label{step:PEPB_goodB_x_sy}
    Consider the case where $\theta^B=g$, $q_1^P=x$, and $s_1=y$. This step follows from Step~\ref{step:PECB_goodB_x_sy} for PECB. A key difference is that the bureaucrat's posterior is now $\Pi_B(x,y)$.  In PEPB, the good bureaucrat confirms $q_1^P=x$ when $s_1=y$ if and only if
    \[
    \Delta \leq  \Delta_{PB}\coloneqq\delta\rho(1-\lambda)(\Pi_B(x,y)-\pi).
    \]
   The inequality's right-hand side is strictly positive provided that $\Pi_B(x,y)>\pi$, which requires $\gamma(y)<1$. Otherwise, a PEPB could not exist, because $\Delta>0$ by definition. Therefore, $\gamma(y)<1$ is necessary for the existence of PEPB. 
\end{step}


\begin{step}
    This step replaces Step~\ref{step:PECB_goodP_x_sy} for PECB. Conjecture a PEPB, and suppose that $\theta^P=g$ and  $s_1=y$. By selecting $q_1^P=x$, the good politician obtains
   \begin{align*}\label{eq:goodPPEPB}
        \beta \left[ v(x,y) + \delta V_{gP}^{gB} \right] 
        + (1-\beta) \left\{ \lambda \left[ v(y,y) + \delta U_{gP}^{bB} \right] + (1-\lambda)\left[ v(x,y) + \delta V_{gP}^{bB} \right] \right\}. 
    \end{align*}
    The difference with respect to the payoff obtained under the same circumstance by the politician in PECB is that now the good bureaucrat supports pandering, i.e., she confirms $x$ in state $y$. By selecting $q_1^P=y$, the good politician obtains in expectation
\begin{multline*}
\beta \left[v(y,y)+\delta U^{gB}_{gP} \right]\\
+(1-\beta)\Big\{\xi\left[\lambda\left(v(x,y)+\delta V^{bB}_{gP} \right)+(1-\lambda)\left(v(y,y)+\delta U^{bB}_{gP} \right) \right] \\
+(1-\xi)\left(v(y,y)+\delta U^{bB}_{gP} \right) \Big\}.
\end{multline*}
We have that the good politician prefers proposing $x$ to $y$ when the former utility is greater than the latter. By imposing the restriction $\xi\leq \frac{1-\lambda}{\lambda(1-\beta)}$ (from Step~\ref{step:PECB_goodP_x_sx}), this condition simplifies to
\begin{equation*}
    \delta E \geq v(y,y)-v(x,y) - \delta\rho( 1-\pi)(1-\lambda)\left[ v(x,x)-v(y,x) \right],
  \end{equation*} 
which is the same as in PECB.
\end{step}


\begin{step}\label{step:PEPB_badP_y}
    This step replaces Step~\ref{step:PECB_badP_y} for PECB. Conjecture a PEPB, and suppose that $\theta^P=b$ and $s_1=y$. Proposing $x$ gives the bad politician
    \[
    \beta\delta V_{bP}^{gB} + (1-\beta)\left[ 
\lambda r_1^P + (1-\lambda)\delta V_{bP}^{bB} \right].
    \]
    Proposing $y$ in state $y$ gives the bad politician
    \[
    \beta r_1^P + (1-\beta)\left\{ \xi\left[ \lambda \delta V_{bP}^{bB} + (1-\lambda)r_1^P \right] + (1-\xi)r_1^P \right\}.
    \]
   By using $\xi\leq \frac{1-\lambda}{\lambda(1-\beta)}$, we find that the bad politician's optimal proposal in state $s_1=y$ is $q_1^P=x$ if and only if
    \[
    r_1^P \leq \delta (\mu_2^P + E) - \frac{\delta\beta\rho\lambda}{1-\lambda(1+\xi)(1-\beta) }\mu_2^P.
    \]
     It follows that the probability that a bad politician proposes policy $x$ when $s_1=y$ is $\gamma(y)$, where
    \begin{multline}\label{eq:PEPB_gammay}
    \gamma(y)\coloneqq Pr\left( r_1^P<\delta (\mu_2^P + E) - \frac{\delta\beta\rho\lambda}{1-\lambda(1+\xi)(1-\beta) }\mu_2^P \right) \\
    =F_1^P\left( \delta (\mu_2^P + E) - \frac{\delta\beta\rho\lambda}{1-\lambda(1+\xi)(1-\beta) }\mu_2^P   \right).
    \end{multline}
\end{step}


\bigskip

An observation is in order. The thresholds governing good bureaucrats’ behavior in PE differ between PECB and PEPB (see Step~\ref{step:PEPB_goodB_x_sy}). Specifically,
\[
\Delta_{PB} > \Delta_{CB} \iff \frac{\pi}{\pi + (1-\pi)\gamma(y)}>\frac{\pi}{\pi + (1-\pi)\gamma} \iff \gamma >\gamma(y) \iff \xi<\frac{1-\lambda}{\lambda}.
\]
These inequalities lead directly to the next observation.

\begin{observation}\label{obs:PE_Deltas}
    The existence of PEPB requires $\gamma(y)<1$. Moreover, we have that $\Delta_{PB}>\Delta_{CB}$ if and only if 
    \[
    \xi<\frac{1-\lambda}{\lambda}.
    \]
\end{observation}

A further observation follows. Step~\ref{step:PECB_badP_x} applies to PEPB as well, defining $\gamma(x)$ in PEPB as
\[
\gamma(x)\coloneqq F_1^P\left( \delta \left(\mu_2^P + E\right) - \frac{\delta\beta\rho\lambda(1-\lambda)}{1-\lambda\left[ 1+\xi(1-\beta) \right]}\mu_2^P   \right).
\]
However, Step~\ref{step:PEPB_badP_y} shows that such a probability differs when the state is $s_1=y$, leading potentially to $\gamma(x)\neq\gamma(y)$. The ranking of $\gamma(s_1)$ depends on $\lambda$. Given that $\gamma(x)=\gamma$, this ranking is outlined by Observation~\ref{obs:PE_Deltas}, as $\Delta_{PB}>\Delta_{CB} \iff \gamma>\gamma(y)$. Section~\ref{sec:app_beliefsPEPB} shows that a necessary condition to have an informative PEPB is $\hat\gamma<1$ (from $\Pi_V(x)>\pi$). Step~\ref{step:PEPB_goodB_x_sy} shows that $\gamma(y)<1$ is also necessary, which implies $\hat\gamma<1$. Observation~\ref{obs:PEPB_bounds} outlines the sufficient and necessary condition ensuring $\gamma(x)<1$ by drawing from Step~\ref{step:PEPB_badP_y}.

\begin{observation}\label{obs:PEPB_bounds}
 Existence of PEPB requires
    \begin{displaymath}
        \bar R_1^P >\delta \left(\mu_2^P + E \right) - \frac{\delta\beta\rho\lambda}{1-\lambda(1+\xi)(1-\beta)}\mu_2^P.
    \end{displaymath}
    This condition implies $\gamma(y)<1$ and $\hat\gamma<1$.
\end{observation}

Collecting the results of this proof we get the following technical conditions that correspond to the verbal ones in the statement of the proposition in the text: 

\begin{observation}
\label{PEPBtechnicalconditions}
A PEPB exists if and only if
    \begin{itemize}[noitemsep]
        \item[i)] $\delta E\geq  v(y,y)-v(x,y) - \delta\rho( 1-\pi)(1-\lambda)\left[ v(x,x)-v(y,x) \right]$;
        \item[ii)]
        $\Delta \leq  \delta \rho(1-\lambda)\left[ \Pi_B(x,y) - \pi \right]$;
        \item[iii)]  $\bar R_1^P >\ \delta \left(\mu_2^P + E \right) - \frac{\beta\rho\lambda}{1-\lambda(1+\xi)(1-\beta)}\delta\mu_2^P$;
        \item[iv)]   $ \mu_2^B  <  \frac{\left(F_1^{B}\right)^{-1}\left( \frac{1-\lambda}{\lambda(1-\beta)}\right)}{\delta\pi\rho(1-\lambda)}.$
    \end{itemize}

\qed
\end{observation}

\subsection{Non-pandering equilibria}

This section outlines the players’ beliefs in NPE and their implications. We begin with the following definition, which specifies the expected strategies of bad policymakers in NPE.

\begin{definition}
    Consider an NPE. From the perspective of voters and politicians, the probability that a bad bureaucrat proposes $x$ after the politician's proposal does not match the state is state-independent, and it is denoted by $\psi$, where
        \[
        \psi \coloneqq Pr\left( q_1^B=x \mid \theta^B=b,q_1^P\neq s_1,s_1\right);
        \]
    From the perspective of voters and bureaucrats, the probability that a bad politician proposes $x$ is state-dependent and denoted by $\gamma(s_1)$, where
        \[
        \gamma(s_1) \coloneqq Pr\left( q_1^P=x \mid \theta_1^P=b, s_1 \right).
        \]
\end{definition}

\subsubsection{Belief updating in NPE-SF}\label{sec:app_NPE_SF_beliefs}

This section analyzes belief updating in NPE-SF. Since the good politician’s strategy in NPE is state-dependent, we abandon the previous notation (i.e., $X_V$ and $Y_V$) in favor of a more convenient one. Let $\mathcal{X}_{V,\omega}^{\theta_1^P}$ denote the probability that, from the voter’s perspective, the \emph{equilibrium strategy} of a politician of type $\theta_1^P$ ultimately results in the implementation of policy $p_1 = x$, while considering also the bureaucrat's strategy. The subscript $\omega$ specifies the bureaucrat’s equilibrium behavior, which can be either stand-firm ($SF$) or subversive ($SV$).

Consider an NPE-SF. From the voter’s perspective, the resulting probabilities are as follows
\[
\mathcal{X}_{V,SF}^{g} = \rho\left[ 1-\lambda(1-\beta) \right],
\]
\begin{align*}
\mathcal{X}_{V,SF}^{b}  & = \gamma(x)\mathcal{X}_{V,SF}^{g}  +  \left( 1-\gamma(x) \right) \rho  \lambda \left[ \beta + (1-\beta)\psi  \right] \\
       & +  \gamma(y)(1-\rho)\left\{   \beta(1-\lambda) + (1-\beta)\left[ \psi + (1-\psi)(1-\lambda)  \right] \right\}.
\end{align*}
Given these probabilities, and upon observing an implemented policy $p_1$, voters use Bayes' rule to update their posterior belief that the politician's type is good, which yields
\[
\Pi_V(x)=\frac{\pi \mathcal{X}_{V,SF}^{g}}{\pi \mathcal{X}_{V,SF}^{g} + \left(1-\pi\right)\mathcal{X}_{V,SF}^{b}}. 
\]

Before proceeding with the analysis we define the following functions,
\begin{equation}\label{eq:gamma_ratio}
    \Gamma\coloneqq  \frac{\rho(1-\gamma(x)) - (1-\rho)\gamma(y)}{\rho(1-\gamma(x)) + (1-\rho)\gamma(y)}<1.
\end{equation}
\begin{equation}\label{eq:lambda_ratio}
    \Lambda\coloneqq  \frac{\rho(1-\gamma(x)) - (1-\rho)\gamma(y)}{\rho\left[ 2(1-\gamma(x)) -\beta(1-\gamma(x)-\gamma(y)) \right]-\beta\gamma(y)}.
\end{equation}

The voter's posterior beliefs are consistent with our definition of NPE provided that $\Pi_V(x) > \pi $. These inequalities hold if and only if the conditions outlined by the following observation are met.
\begin{observation}\label{obs:NPE_beliefs}
    Existence of NPE-SF requires
    \begin{equation}\label{eq:gammax}
        \gamma(x)<1,
    \end{equation}
    \begin{equation}\label{eq:rho_gamma}
        \rho > \hat\rho \coloneqq\frac{\gamma(y)}{1-\gamma(x)+\gamma(y)},
    \end{equation}
    \begin{equation}\label{eq:psi_NPE_SF}
        \text{and }\; \psi<\left(\frac{1-\lambda}{\lambda(1-\beta)}\right)\Gamma.
    \end{equation}
    When conditions~\eqref{eq:gammax} and \eqref{eq:rho_gamma} hold, we have $\Gamma\in(0,1)$ and $\Lambda\in(0,1)$. Since $\psi\leq 1$, condition~\eqref{eq:psi_NPE_SF} is binding if and only if $\lambda > \Lambda$.
\end{observation}

When $\lambda\leq \Lambda$ and $\rho>\hat\rho$, then $(1-\lambda)\Gamma\geq \lambda(1-\beta)$, and thus condition~\eqref{eq:psi_NPE_SF} does not bind as $\psi\leq 1$ always. Condition~\eqref{eq:rho_gamma} ensures that $\Gamma> 0$, and condition~\eqref{eq:gammax} makes condition~\eqref{eq:rho_gamma} feasible, in the sense that $\hat\rho<1$.

We now turn our attention to the bureaucrat's posterior beliefs, as indicated by $\Pi_B\left(q_1^P,s_1\right)$. Because the good politician's strategy is now state-dependent, the bureaucrat's beliefs naturally depend both on the state and on the politician's proposal. Given that $\gamma(x)<1$ is necessary for the existence of NPE with stand-firm bureaucracy, we obtain that
\begin{align*}
&\Pi_B(x,y)=\Pi_B(y,x)=0, \\    
&\Pi_B(x,x)=\frac{\pi}{\pi+(1-\pi)\gamma(x)}>\pi, \\    
&\Pi_B(y,y)=\frac{\pi}{\pi+(1-\pi)[1-\gamma(y)]}\geq\pi.
\end{align*}

\subsubsection{Belief updating in NPE-SV}\label{sec:app_NPE_SV_beliefs}

\noindent Consider an NPE with a \emph{subversive} bureaucracy. We obtain
\[
\mathcal{X}_{V,SV}^{g} = \mathcal{X}_{V,SF}^{g} + (1-\rho)\beta\lambda,
\]
\[
\mathcal{X}_{V,SV}^{b} = \mathcal{X}_{V,SF}^{b} + (1-\rho)(1-\gamma(y))\beta\lambda.
\]
The above probabilities are higher then in the stand-firm case because a subversive bureaucracy \emph{forces} pandering by trying to convert proposals $q_1^P=y$ into policies $p_1=x$.

The posterior probability $\Pi_V(x)$ is thus similar as in the previous case, that is,
\[
\Pi_V(x)=\frac{\pi \mathcal{X}_{V,SV}^{g}}{\pi \mathcal{X}_{V,SV}^{g} + \left(1-\pi\right)\mathcal{X}_{V,SV}^{b}}. 
\]

Before proceeding, we define the following functions,
\begin{equation}\label{eq:NPE_FP_hatgamma}
\hat\Gamma \coloneqq  \frac{(1-\rho)\gamma(y)}{\rho(1-\gamma(x)) + (1-\rho)\gamma(y)}\in(0,1),
\end{equation}
\begin{equation}\label{eq:tildelambda}
\tilde\Lambda \coloneqq  \frac{\rho(1-\gamma(x))-(1-\rho)\gamma(y)}{\rho(1-\gamma(x)) - (1+\beta)(1-\rho)\gamma(y)},
\end{equation}
where 
\[
\tilde\Lambda<1 \iff \rho< \frac{(1+\beta)\gamma(y)}{1-\gamma(x)+(1+\beta)\gamma(y)},\; \text{and }\; \frac{(1+\beta)\gamma(y)}{1-\gamma(x)+(1+\beta)\gamma(y)}>\hat\rho.
\]
The usual condition ensuring the conjectured equilibria are informative, i.e., $\Pi_V(x)>\pi$, yields the following observation.

\begin{observation}\label{obs:NPE_SV_beliefs}
    The existence of NPE with a subversive bureaucracy requires
    \begin{equation}\label{eq:psi_NPE_SV_beliefs_2}
        \text{either }\; \rho\geq \hat\rho, \text{ or} \; \rho<\hat\rho \; \text{ and } \; \lambda > \tilde\Lambda,
    \end{equation}
    \begin{equation}\label{eq:psi_NPE_SV_beliefs}
        \text{and }\; \psi< \hat \Psi_{Subv} \coloneqq \left(\frac{1-\lambda}{\lambda(1-\beta)}\right)\Gamma + \left(\frac{\beta}{1-\beta}\right)\hat\Gamma,
    \end{equation}
    where condition~\eqref{eq:psi_NPE_SV_beliefs_2} ensures that $\hat \Psi_{Subv}>0$.
\end{observation}

We now turn our attention to the bureaucrat's posterior beliefs, as indicated by $\Pi_B\left(q_1^P,s_1\right)$. Given that, as of now, $\gamma(s_1)\in[0,1]$, we obtain that, in NPE with a subversive bureaucracy,
\begin{align*}
&\Pi_B(x,y)=\Pi_B(y,x)=0, \\    
&\Pi_B(x,x)=\frac{\pi}{\pi+(1-\pi)\gamma(x)}\geq\pi, \\    
&\Pi_B(y,y)=\frac{\pi}{\pi+(1-\pi)[1-\gamma(y)]}\geq\pi.
\end{align*}
Later on, Observation~\ref{obs:NPESV_Delta} shows that in NPE with a subversive bureaucracy it must be that $\gamma(y)<1$. This implies that, in those equilibria, $\Pi_B(y,y)>\pi$.

\subsubsection{Proof of Proposition~\ref{NPEexistence}}

As we did before for the PE we prove Proposition~\ref{NPEexistence} by using a series of steps and observations.

\bigskip

\noindent{\bf Good bureaucrats in NPE-SF.}

\begin{step}\label{step:NPE_gB_sqx}
    Consider the case where $\theta^B=g$, $s_1=q_1^P=x$. Proposing $q_1^B=x$ grants the bureaucrat an expected payoff of
    \[
v(x,x)+\delta\left[ \Pi_B(x,x) V_{gB}^{gP} + (1-\Pi_B(x,x))V_{gB}^{bP} \right].
\]
By proposing $q_1^B=y$, the bureaucrat obtains
\begin{multline*}
    \lambda \left[ v(y,x) + \delta\left( \pi V_{gB}^{gP} + (1-\pi)V_{gB}^{bP} \right)  \right] \\
    + (1-\lambda) \left[ v(x,x)+\delta\left(  \Pi_B(x,x) V_{gB}^{gP} + (1-\Pi_B(x,x))V_{gB}^{bP} \right)  \right].
\end{multline*}
Since $v(x,x)>v(y,x)$, $\Pi_B(x,x)>\pi$, and $V_{gB}^{gP}>V_{gB}^{bP}$, we find that, in NPE, a good bureaucrat's best reply to $q_1^P=s_1=x$ is $q^B_1\left(g,x,r_1^B,x\right)=x$.
\end{step}


\begin{step}\label{step:NPE_gB_qy_sx}
    Consider the case where $\theta^B=g$, $q^P_1=y$, and $s_1=x$. The bureaucrat's expected payoff from proposing $q_1^B=y$ is
\[
v(y,x) +\delta\left[ \pi V_{gB}^{gP} + (1-\pi)V_{gB}^{bP} \right] = v(y,x) + \delta V_{gB}^\pi.
\]
By contrast, proposing $q_1^B=x$ in this case gives
\[
\lambda \left[v(x,x) + \delta V_{gB}^{bP} \right] + (1-\lambda)\left[v(y,x) + \delta V^{\pi}_{gB}\right]. 
\]
The bureaucrat's expected payoff from proposing $q_1^B=x$ is higher than that from proposing $q_1^B=y$ if and only if
\[
v(x,x)-v(y,x)>\delta\pi\left( V_{gB}^{gP} - V_{gB}^{bP}  \right) = \delta\pi \rho (1-\lambda) \left[ v(x,x) - v(y,x) \right],
\]
which is always true as $\delta,\pi,\rho,\lambda\in(0,1)$. Therefore, in NPE, a good bureaucrat's best reply to $q_1^P=y\neq s_1=x$ is $q^B_1\left(g,x,r_1^B,y\right)=x$.
\end{step}


\begin{step}\label{step:NPE_gB_qx_sy}
     Consider the case where $\theta^B=g$, $q_1^P=x$, and $s_1=y$. The bureaucrat's expected payoff from proposing $q_1^B=x$ is
\[
v(x,y) + \delta V_{gB}^{bP} .
\]
Proposing $q_1^B=y$ gives
\[
\lambda\left[ v(y,y) + \delta\left( \pi V_{gB}^{gP} + (1-\pi)V_{gB}^{bP} \right) \right] + (1-\lambda)\left[ v(x,y) + \delta V_{gB}^{bP} \right].
\]
By comparing the two payoffs, it follows that it is always optimal for the bureaucrat to propose $q^B_1\left(g,y,r_1^B,x\right)=y$.
\end{step}


\begin{step}
\label{step:gb-condition-NPE-SF}
    Consider the case where $\theta^B=g$ and $q_1^P=s_1=y$. The bureaucrat's expected payoff from proposing $q_1^B=x$ is
\begin{multline*}
    \lambda\left[ v(x,y) + \delta\left( \Pi_B(y,y) V_{gB}^{gP} + (1-\Pi_B(y,y))V_{gB}^{bP} \right) \right] \\
    + (1-\lambda)\left[v(y,y)+\delta\left( \pi V_{gB}^{gP} + (1-\pi)V_{gB}^{bP} \right) \right].
\end{multline*}
Proposing $q_1^B=y$ gives
\[
v(y,y) + \delta\left( \pi V_{gB}^{gP} + (1-\pi)V_{gB}^{bP} \right).
\]
The bureaucrat faces a trade-off between matching the state in period-$1$ and increasing the re-election chances of a good politician, given that $\Pi_B(y,y)\geq\pi$. They will propose $y$ when
\[
v(y,y) - v(x,y) \geq \left( \Pi_B(y,y)-\pi \right)\delta\left[V_{gB}^{gP}-V_{gB}^{bP} \right].
\]
By replacing $V_{gB}^{gP}$ and $V_{gB}^{bP}$, the inequality becomes
\[
v(y,y) - v(x,y) \geq \left( \Pi_B(y,y)-\pi \right)\delta\rho (1-\lambda)(v(x,x)-v(y,x)).
\]
It is optimal for the bureaucrat to propose $q^B_1\left(g,y,r_1^B,y\right)=y$ if and only if
\[
\Delta \equiv \frac{v(y,y) - v(x,y)}{v(x,x)-v(y,x)} \geq \left( \Pi_B(y,y)-\pi \right)\delta \rho (1-\lambda) .
\]
This condition is satisfied when, e.g., $ \Delta \approx 1 $. This step highlights how the good bureaucrat's equilibrium behavior differs depending on the relative mismatch costs. We have ``standing-firm'' or ``subversive'' bureaucracy when $\Delta$ is sufficiently high or low, respectively.
\end{step}

\begin{observation}\label{obs:NPE_Delta}
    The existence of NPE with a stand-firm bureaucracy requires    
$\gamma(x)<1$  and  $\Delta \geq \delta \rho (1-\lambda)\left[ \Pi_B(y,y)-\pi \right].$

\end{observation}


\bigskip

\noindent{\bf Bad bureaucrats in NPE-SF.}


\begin{step}\label{step:NPE_bB_qx_sx}
    Consider the case where $\theta^B=b$, $s_1=x$, and $q_1^P=x$. The bureaucrat's posterior about the politician's type being good satisfies $\Pi_B(x,x)\geq\pi$ (and $\Pi_B(x,x)>\pi$ in NPE with a stand-firm bureaucracy). The bad bureaucrat's expected utility when proposing $q_1^B=x$ is
    \[
    \delta\left\{ \Pi_B(x,x)\left[ 1-\rho(1-\lambda)\right] \mu_2^B  + \left( 1-\Pi_B(x,x) \right)\mu_2^B \right\} = \delta\mu_2^B\left[ 1-\Pi_B(x,x)\rho(1-\lambda) \right].
    \]
    The expected utility from proposing $q_1^B=y$ is
    \[
    \lambda\left(r_1^B+\delta V_{bB}^\pi\right)+(1-\lambda)\delta\mu_2^B\left[ 1-\Pi_B(x,x)\rho(1-\lambda) \right].
    \]
    The bad bureaucrat is better off when proposing $y$ rather than $x$ when
    \[
    r_1^B\geq \delta\mu_2^B\rho(1-\lambda)\left(\pi-\Pi_B(x,x)\right).
    \]
    The right-hand side of this inequality is non-positive because $\Pi_B(x,x)\geq\pi$ (and it is strictly negative in NPE with a stand-firm bureaucracy). As a result, the above condition is always satisfied, implying that the bad bureaucrat's best response to $q_1^P=x=s_1$ is always $q_1^B=y$.
\end{step}


\begin{step}\label{step:NPE_bB_qx_sy}
    Consider the case where $\theta^B=b$, $s_1=y$, and $q_1^P=x$. The bureaucrat's posterior about the politician's type being good is $\Pi_B(x,y)=0$. The bad bureaucrats' expected utility when proposing $q_1^B=x$ is $ \delta \mu_2^B$. Their expected utility when proposing $q_1^B=y$ is
    \[
    \lambda\left(r_1^B+\delta V_{bB}^\pi\right)+(1-\lambda)\delta\mu_2^B.
    \]
    In this case, bad bureaucrats are better off when proposing $x$ rather than $y$ when
    \[
    r_1^B < \delta\mu_2^B \pi \rho(1-\lambda).
    \]
    It follows that the probability that a bad bureaucrat proposes policy $x$ when $q_1^P=x$ and  $s_1=y$ is $\psi$, where\footnote{The probabilities $\xi$ and $\psi$ take the same value. However, they refer to the likelihood of different events, since bad bureaucrats behave differently in NPE than in PECB. For this reason, distinct notation is used to represent them.}
    \begin{equation}\label{eq:NPE_psi}
    \psi \coloneqq Pr\left( r_1^B < \delta\mu_2^B \pi \rho(1-\lambda) \right) = F_1^B\left( \delta\mu_2^B \pi \rho(1-\lambda) \right).
    \end{equation}
\end{step}


\begin{step}\label{step:NPE_bB_qy_sx}
    Consider the case where $\theta^B=b$, $s_1=x$, and $q_1^P=y$. The bureaucrat's posterior about the politician's type being good is $\Pi_B(y,x)=0$. The bad bureaucrat's expected utility when proposing $q_1^B=x$ is
    \[
     \lambda \delta \mu_2^B+(1-\lambda)\left(r_1^B+ \delta V^{\pi}_{bB}\right).
    \]
    Its expected utility when proposing $q_1^B=y$ is
    \[
   r_1^B+ \delta V^{\pi}_{bB}.
    \]
    The bad bureaucrat is better off when proposing $x$ rather than $y$ when
    \[
    r_1^B < \delta\mu_2^B \pi \rho(1-\lambda),
    \]
    which results in the same condition as in Step~\ref{step:NPE_bB_qx_sy}.
\end{step}


\begin{step}\label{step:NPE_bB_qsy}
    Consider the case where $\theta^B=b$ and $q_1^P=s_1=y$. The bureaucrat's posterior about the politician's type being good is $\Pi_B(y,y)\geq\pi$. The bad bureaucrat's expected utility when proposing $q_1^B=x$ is
    \[
    \lambda \delta \left[\Pi_B(y,y) V_{bB}^{gP}+(1-\Pi_B(y,y))V_{bB}^{bP}\right] +(1-\lambda)\left(r_1^B+\delta V_{bB}^\pi\right).
    \]
    The bad bureaucrat's expected utility when proposing $q_1^B=y$ is
    \[
    r_1^B +\delta V_{bB}^\pi.
    \]
    Therefore, in NPE, the bad bureaucrat's best response to  $q_1^P=y=s_1$ is $q_1^B=y$ if and only if
    \[
    r_1^B\geq \delta\mu_2^B\rho(1-\lambda)\left(\pi-\Pi_B(y,y)\right),
    \]
which is always satisfied for non-negative rents because $\Pi_B(y,y)\geq\pi$.
\end{step}

\bigskip

\noindent{\bf Good politicians in NPE-SF.}


\begin{step}\label{step:NPE_gP_sx}
    Suppose that $\theta^P=g$ and $s_1=x$. By selecting $q_1^P=x$, the good politician obtains in expectation
\begin{displaymath}
\beta\left[ v(x,x) + \delta V_{gP}^{gB} \right] + (1-\beta)\left[ \lambda\left( v(y,x)+ \delta U_{gP}^{bB} \right) + (1-\lambda)\left( v(x,x) + \delta V_{gP}^{bB} \right) \right].
\end{displaymath}
By selecting $q_1^P=y$, the good politician obtains
\begin{multline*}
\beta\left\{  \lambda \left[ v(x,x) +\delta V_{gP}^{gB} \right] + (1-\lambda) \left[ v(y,x)+ \delta U_{gP}^{gB} \right]\right\} \\
+ (1-\beta) \psi \left[ \lambda \left( v(x,x) + \delta V_{gP}^{bB} \right) + (1-\lambda) \left( v(y,x)+ \delta U_{gP}^{bB} \right)\right] \\
+ (1-\beta)(1-\psi)\left[ v(y,x) + \delta U_{gP}^{bB} \right].
\end{multline*}

Recall that $v(x,x)>v(y,x)$, $V_{gP}^{gB}>U_{gP}^{gB}$, and $V_{gP}^{bB}>U_{gP}^{gB}$. After re-arranging and simplifying,  the first utility (from $q_1^P=x$) is always greater than the second (from $q_1^P=y$) when
\begin{multline*}
\left[ (1-\lambda) - (1-\beta)\psi\lambda \right](v(x,x)-v(y,x))\\
\geq (1-\beta)[\psi\lambda - (1-\lambda)]\delta\left( V_{gP}^{bB} - U_{gP}^{bB} \right) - \beta(1-\lambda)\delta\left( V_{gP}^{gB} - U_{gP}^{gB} \right).
\end{multline*}
By isolating $\psi$, we get the following inequality,
\begin{displaymath}
    \psi \leq \left(\frac{1-\lambda}{\lambda(1-\beta)}\right) \left\{ \frac{\left[v(x,x)-v(y,x)\right]+\delta\left[ \beta\left( V_{gP}^{gB} - U_{gP}^{gB} \right) +(1-\beta)\left( V_{gP}^{bB} - U_{gP}^{bB} \right) \right]}{\left[v(x,x)-v(y,x)\right] + \delta\left( V_{gP}^{bB} - U_{gP}^{bB} \right)} \right\}.
\end{displaymath}
The second fraction in the right-hand side of the above inequality is equal to 1 if
\[
V_{gP}^{gB}-U_{gP}^{gB}= V_{gP}^{bB}-U_{gP}^{bB},
\]
which, as we have shown in Section~\ref{sec:app_cont_values}, holds true. Therefore, in NPE, the good politician prefers to propose $x$ when the state is $x$ if and only if
\[
    \psi \leq \frac{1-\lambda}{\lambda(1-\beta)}.
\]
\end{step}

\bigskip

The following observation is in order. Steps~\ref{step:NPE_bB_qx_sy} and~\ref{step:NPE_bB_qy_sx} define $\psi$, the probability that a bad bureaucrat proposes policy $x$ when $q_1^P\neq s_1$. Step~\ref{step:NPE_gP_sx} establishes an equilibrium restriction requiring that $\psi \leq \frac{1-\lambda}{\lambda(1-\beta)}$. This mirrors the situation encountered in the analysis of the PECB, which led to Observation~\ref{obs:PECB_xi}. The same procedure applies here and is omitted for brevity. While the probabilities $\psi$ and $\xi$ refer to different events, they take the same value. Recall that $\frac{1-\lambda}{\lambda(1-\beta)}<1$ if and only if $\lambda \geq \frac{1}{2 - \beta}$, thus this restriction is in force only for sufficiently high $\lambda$. We obtain the following result.

\begin{observation}\label{obs:NPE_psi}
In NPE, it must hold that
\[
    \psi \leq \frac{1-\lambda}{\lambda(1-\beta)}.
\]
This restriction binds if and only if $\lambda \geq \frac{1}{2 - \beta}$. It applies in addition to those stated in Observation~\ref{obs:NPE_beliefs} for a stand-firm bureaucracy and in Observation~\ref{obs:NPE_SV_beliefs} for a subversive bureaucracy. Since in NPE we have $\Gamma\in(0,1)$ and $\Lambda < \frac{1}{2-\beta}$, the restrictions in Observation~\ref{obs:NPE_beliefs} imply the ones stated here for the stand-firm bureaucracy case.  
\end{observation}


\bigskip


\begin{step}\label{step:NPE_gP_SF}
    Suppose that $\theta^P=g$ and $s_1=y$. By selecting $q_1^P=x$, the politician obtains
    \begin{multline*}\label{eq:goodP}
        \beta \left\{ \lambda \left[ v(y,y) + \delta U_{gP}^{gB} \right] + (1-\lambda)\left[ v(x,y) + \delta V_{gP}^{gB} \right] \right\}  + (1-\beta)\psi\left[v(x,y)+\delta V^{bB}_{gP} \right] \\
        +(1-\beta)(1-\psi)\left\{\lambda\left[v(y,y)+\delta U^{bB}_{gP}\right] + (1-\lambda)\left[ v(x,y) + \delta V^{bB}_{gP}\right]\right\}. 
    \end{multline*}
By selecting $q_1^P=y$, the politician obtains
\[
\beta\left[v(y,y)+\delta U^{gB}_{gP}\right]
+(1-\beta)\left[v(y,y)+\delta U^{bB}_{gP}\right]
\]
The latter payoff is larger than the former if and only if
\[
\underbrace{\left\{ 1-\lambda[1-\psi(1-\beta)] \right\}}_{>0}\left\{ v(y,y)-v(x,y) - \delta\rho( 1-\pi)(1-\lambda)\left[ v(x,x)-v(y,x) \right] -\delta E  \right\}\geq0.
\]
It follows that the politician prefers to propose $q_1^P=y=s_1$ if and only if
\begin{equation}\label{eq:non-pandering}
    \delta E \leq v(y,y)-v(x,y) - \delta\rho( 1-\pi)(1-\lambda)\left[ v(x,x)-v(y,x) \right].
  \end{equation}
Condition~\eqref{eq:non-pandering} is necessary for the good politician to avoid pandering, and it complements the pandering condition found in Step~\ref{step:PECB_goodP_x_sy}, inequality~\eqref{eq:pandering}, characterizing PE.

In NPE with a stand-firm bureaucracy, the good politician's optimal proposal in state $s_1=y$ is $q_1^P=y$ if and only if office rents satisfy condition~\eqref{eq:non-pandering}. 
\end{step}



\bigskip

\noindent{\bf Bad politicians in NPE-SF.}


\begin{step}\label{step:NPE_badP_sx}
    Consider a NPE, and suppose that $\theta^P=b$ and $s_1=x$. By proposing $q_1^P=x$, the bad politician's expected payoff is
    \[
    \beta \delta V_{bP}^{gB} + (1-\beta)\left[ \lambda r_1^P + (1-\lambda) \delta V_{bP}^{bB} \right].
    \]
    By proposing $q_1^P=y$, their expected payoff is
    \[
    \beta \left[ \lambda \delta V_{bP}^{gB} + (1-\lambda)r_1^P \right] + (1-\beta) \left\{ \psi \left[ \lambda \delta V_{bP}^{bB} + (1-\lambda)r_1^P \right] + (1-\psi)r_1^P \right\}.
    \]
    
    The bad politician prefers to propose $x$ in state $x$ when the former payoff is at least as high as the latter. By isolating rents $r_1^P$ and invoking $\psi\leq\frac{1-\lambda}{\lambda(1-\beta)}$ (see Observation~\ref{obs:NPE_psi}), this is the case when
    \[
    r_1^P \leq \delta \left(\mu_2^P + E \right) - \frac{\delta\beta\rho\lambda(1-\lambda)}{1-\lambda\left[ 1+\psi(1-\beta) \right]}\mu_2^P \triangleq \left(F_1^P\right)^{-1}\left( \gamma(x) \right).
    \]
The inequality's right-hand side is strictly positive  provided that
    \[
    \psi < \frac{1-\lambda}{\lambda(1-\beta)} + \underbrace{\left(\frac{\beta}{1-\beta}\right)\left[\frac{(1-\lambda)\rho\mu_2^P}{\mu_2^P + E}\right]}_{>0}.
    \]
The above condition is always satisfied as per Observation~\ref{obs:NPE_psi}, implying $\gamma(x)>0$.

In NPE, the probability $\gamma(x)$ is
\begin{equation}\label{eq:NPE_gammax}
\gamma(x) := F_1^P\left( \delta \left(\mu_2^P + E \right) - \frac{\delta\beta\rho\lambda(1-\lambda)}{1-\lambda\left[ 1+\psi(1-\beta) \right]}\mu_2^P \right).
\end{equation}
\end{step}


\bigskip


\begin{step}\label{step:NPE_SF_badP_sy} Suppose that $\theta^P=b$ and $s_1=y$. Proposing $x$ gives the bad politician
    \[
    \beta\left[ \lambda r_1^P + (1-\lambda)\delta V_{bP}^{gB}\right] + (1-\beta) \left\{ (1-\psi) \left[ (1-\lambda) \delta V_{bP}^{bB} + \lambda r_1^P \right] + \psi \delta V_{bP}^{bB} \right\}.
    \]
    By contrast, proposing $y$ gives the bad politician
    \[
    \beta r_1^P + (1-\beta) r_1^P=r_1^P
    \]
    It follows that the bad politician prefers to propose $x$ in state $y$ if and only if
    \[
    r_1^P \leq \delta \left(\mu_2^P + E\right) - \frac{\delta\beta\rho\lambda(1-\lambda)}{1-\lambda\left[ 1-\psi(1-\beta) \right]}\mu_2^P \triangleq \left(F_1^P\right)^{-1}\left( \gamma(y)  \right).
    \]
Notice that the condition on rents found in Step~\ref{step:NPE_badP_sx} implies the condition above, implying that, in NPE with a stand-firm bureaucracy, we have $0<\gamma(x)<\gamma(y)\leq 1$.

In NPE with a stand-firm bureaucracy, the probability $\gamma(y)$ is strictly positive, and it is given by
    \begin{equation}\label{eq:NPE_SF_gammay}
        \gamma(y) = F_1^P\left( \delta \left(\mu_2^P + E\right) - \frac{\delta\beta\rho\lambda(1-\lambda)}{1-\lambda\left[ 1-\psi(1-\beta) \right]}\mu_2^P  \right)>0.
    \end{equation}
    Moreover, $0<\gamma(x)<\gamma(y)$.
\end{step}


\begin{observation}\label{obs:NPE_SF_gammax}
    For NPE with a stand-firm bureaucracy to exist, we need that $\gamma(x)<1$ (see Observation~\ref{obs:NPE_beliefs}). From Step~\ref{step:NPE_badP_sx}, we obtain that such a condition is met if and only if
    \[
        \bar R_1^P > \delta \left(\mu_2^P + E \right) - \frac{\delta\beta\rho\lambda(1-\lambda)}{1-\lambda\left[ 1+\psi(1-\beta) \right]}\mu_2^P.
    \]
\end{observation}


\bigskip

\noindent The collection of Observations~\ref{obs:NPE_beliefs}, \ref{obs:NPE_Delta}, \ref{obs:NPE_psi}, and \ref{obs:NPE_SF_gammax} along with Step \ref{step:NPE_gP_SF} yield the necessary and sufficient conditions for the existence of NPE with a stand-firm bureaucracy, which we summarize in the following observation whose conditions
correspond to the verbal ones in the statement of the proposition in the main text:

\begin{observation}
\label{NPESFtechnicalconditions}
    NPE with a stand-firm bureaucracy exist if and only if

\begin{itemize}[noitemsep]
        \item[i)] $\delta E<  v(y,y)-v(x,y) - \delta\rho( 1-\pi)(1-\lambda)\left[ v(x,x)-v(y,x) \right]$;
\item[ii)] $\Delta \geq \delta \rho (1-\lambda)\left[ \Pi_B(y,y)-\pi \right];$
    
    \item[iii)]
    $\bar R_1^P > \delta \left(\mu_2^P + E \right) - \frac{\delta\beta\rho\lambda(1-\lambda)}{1-\lambda\left[ 1+\psi(1-\beta) \right]}\mu_2^P;$
        \item[iv)] $\rho>\hat\rho;$
\item[v)] $    \psi<\left(\frac{1-\lambda}{\lambda(1-\beta)}\right)\Gamma,
$ which is  binding if and only if $\lambda>\Lambda$. 
\end{itemize}

\end{observation}

\qed

\subsubsection{Proof of Proposition \ref{NPE_FSV_existence} }
The proof of the existence of an NPE-FSV follows steps analogous to those of the NPE-SF.  Step \ref{step:gb-condition-NPE-SF} naturally differs only with respect to the direction of its final inequality, which results in Observation \ref{obs:NPESV_Delta}.  Steps \ref{step:NPE_gP_SF} and \ref{step:NPE_SF_badP_sy} differ substantially and have to be revised. All other steps are identical and therefore omitted for brevity. These results complete the proof.
\bigskip
\begin{observation}\label{obs:NPESV_Delta}
    
The existence of NPE with a subversive bureaucracy requires
\[
\gamma(y)<1 \; \text{ and } \; \Delta \leq \delta \rho (1-\lambda)\left[ \Pi_B(y,y)-\pi \right].
\]
\end{observation}


\begin{step}\label{step:NPE_gP_SB}
   Suppose that $\theta^P=g$ and $s_1=y$. In this case, the politician's payoff when proposing $q^P_1=x$ is the same as in Step~\ref{step:NPE_gP_SF}. However, when proposing $q^P_1=y$, the politician's expected payoff is
\[
\beta \left\{\lambda \left[v(x,y)+\delta V^{gB}_{gP}\right] +(1-\lambda)\left[v(y,y)+\delta U^{gB}_{gP}\right]\right\}
+(1-\beta)\left[v(y,y)+\delta U^{bB}_{gP}\right].
\]
Define the \emph{pandering payoff} function as
\[
P \coloneqq v(y,y)-v(x,y) - \delta\rho( 1-\pi)(1-\lambda)\left[ v(x,x)-v(y,x) \right] -\delta E.
\]
When $s_1=y$, the good politician's payoff from proposing $q^P_1=y$ is higher than that from proposing $q^P_1=x$ if and only if
\begin{equation}\label{eq:NPE_SV_condition}
    \underbrace{\left\{ \beta(2\lambda -1) - (1-\beta)\left[ 1-\lambda(1-\psi) \right] \right\}}_{\coloneqq G} P \leq 0.
\end{equation}
The above condition implies that either the first component in the above inequality, $G$, must have the opposite sign as $P$, or at least one of the two components is zero.

Under the assumption that $P> 0$ (the case of $P\leq0$ leads us to NPE-ASV), like as required for stand-firm bureaucracy (see Step~\ref{step:NPE_gP_SF}) NPE-FSV are sustained if and only if $G<0$, that is, when
\begin{equation}\label{eq:NPE_SV_high_psi}
\psi \geq \tilde \Psi_{Subv} \coloneqq \frac{\beta}{1-\beta} - \frac{1-\lambda}{\lambda(1-\beta)}.
\end{equation}
Intuitively, when $\psi$ is sufficiently high, the probability of ultimately implementing policy $x$ in state $y$ is greater when the politician initially proposes $q_1^P = x$. To see this, note that in this class of equilibria we have
\[
Pr\left(p_1^P=x \mid q_1^P=s_1=y\right) = \beta\lambda,
\]
\[
Pr\left(p_1^P=x \mid q_1^P=x\neq s_1=y\right) = \beta(1-\lambda)+(1-\beta)\left[\psi + (1-\lambda)(1-\psi)\right].
\]
Comparing these two equilibrium probabilities, we obtain
\[
Pr\left(p_1^P=x \mid q_1^P=s_1=y\right)>Pr\left(p_1^P=x \mid q_1^P=x\neq s_1=y\right) \iff \psi <\tilde \Psi_{Subv}.
\]
When $\psi$ exceeds this threshold, good politicians prefer to propose the policy that matches the state, since doing so maximizes the likelihood that the final implemented policy, after accounting for bureaucratic influence, will also be state-matching.

Since $\psi\geq 0$, condition~\eqref{eq:NPE_SV_high_psi} is always met when $\lambda$ is sufficiently low, as
\[
\tilde \Psi_{Subv}\leq 0 \iff \lambda\leq \frac{1}{1+\beta}\in\left(\nicefrac{1}{2},1\right).
\]
Intuitively, when $\lambda$ is very low, the likelihood of implementing policy $x$ through a proposal of $y$ becomes too low. Such an outcome requires a subversive bureaucrat to contest the politician’s proposal $y$ and attempt to overturn it in favor of $x$. With weak bureaucratic influence, this mechanism is ineffective, making a direct proposal of $q_1^P = x$ the more reliable path to achieving $p_1 = x$. By contrast, because $\psi\leq 1$, condition~\eqref{eq:NPE_SV_high_psi} can never be met when $\lambda$ is excessively high, as
\[
\tilde \Psi_{Subv}\geq 1 \iff \lambda>\frac{1}{2\beta}\in(\nicefrac{1}{2},+\infty).
\]

Condition~\eqref{eq:NPE_SV_high_psi}  adds to~\eqref{eq:psi_NPE_SV_beliefs} in Observation~\ref{obs:NPE_SV_beliefs}. Specifically, we have that 
\[
\hat \Psi_{Subv}>\tilde \Psi_{Subv} \iff \lambda<\frac{2}{2+\beta}.
\]
As a result, conditions~\eqref{eq:psi_NPE_SV_beliefs} and~\eqref{eq:NPE_SV_high_psi}, required when $P>0$, can be jointly satisfied if and only if $\lambda<\frac{2}{2+\beta}$. 

Additionally, Observation~\ref{obs:NPE_psi} requires $\psi<\frac{1-\lambda}{\lambda(1-\beta)}$ to hold in NPE. We have that
\[
\tilde \Psi_{Subv} < \frac{1-\lambda}{\lambda(1-\beta)} \iff \lambda<\frac{2}{2+\beta}.
\]
In addition, we have that
\[
\hat \Psi_{Subv} < \frac{1-\lambda}{\lambda(1-\beta)} \iff \lambda<\frac{2}{2+\beta}.
\]


\begin{observation}\label{obs:NPE_conditions}
     The existence of NPE-FSV requires
     $P>0 \; \text{and } \; \hat \Psi_{Subv}>\psi\geq \tilde\Psi_{Subv}.$
\end{observation}
\end{step}


\begin{step}\label{step:NPE_SV_badP_sy}
   Suppose that $\theta^P=b$ and $s_1=y$. Proposing $q_1^P=x$ gives the bad politician the same payoff as in Step~\ref{step:NPE_SF_badP_sy}. By contrast, the payoff from proposing $q^P_1=y$ is
\[
    \beta \left[\lambda \delta V_{bP}^{gB}  +(1-\lambda)r_1^P\right]+(1-\beta) r_1^P.
    \]
Define the function $H$ as
\[
H \coloneqq   1-\frac{\beta\lambda}{1-\lambda\left[ 1-\psi(1-\beta) \right]},
\]
where
\[
H>0 \iff \psi > \tilde\Psi_{Subv},
\]
where $\tilde\Psi_{Subv}$ is defined in \eqref{eq:NPE_SV_high_psi}. Recall that $\tilde\Psi_{Subv}\leq 0$ for $\lambda\leq \frac{1}{1+\beta}$.

The bad politician prefers to propose $q_1^P=x$ if and only if
\begin{equation}\label{eq:NPE_SV_rents}
    H r_1^P \leq H \delta\left( \mu_2^P + E \right)
    - \underbrace{\frac{\delta\beta\rho\lambda(1-2\lambda)}{1-\lambda\left[ 1-\psi(1-\beta) \right]}}_{>0  \iff \lambda<\nicefrac{1}{2}, }\mu_2^P.
\end{equation}
Suppose  that $\psi > \tilde\Psi_{Subv}$ (and thus $H>0$), as it is required by NPE with subversive bureaucracy when $P>0$ and $\lambda>\frac{1}{1+\beta}$ (see Observation~\ref{obs:NPE_conditions}). In such a case, condition~\eqref{eq:NPE_SV_rents} is
\[
r_1^P \leq \delta\left( \mu_2^P + E \right)
    - \frac{\delta\beta\rho\lambda(1-2\lambda)}{1-\lambda\left[ 1+\beta -\psi(1-\beta) \right]}\mu_2^P \triangleq \left(F_1^P\right)^{-1}\left( \gamma(y)  \right).
\]
The probability $\gamma(y)$ is smaller in this case than found for the stand-firm bureaucracy in Step~\ref{step:NPE_SF_badP_sy}. The inequality's right-hand side is always strictly positive when $\psi > \tilde\Psi_{Subv}$ and $\lambda>\frac{1}{2}$, implying $\gamma(y)>0$. Specifically, in this case we have that
\begin{equation}\label{eq:NPE_FP_gammay}
    \gamma(y):=F_1^P\left( \delta\left( \mu_2^P + E \right)
    - \frac{\delta\beta\rho\lambda(1-2\lambda)}{1-\lambda\left[ 1+\beta -\psi(1-\beta) \right]}\mu_2^P   \right).
\end{equation}


\end{step}


\begin{observation}\label{obs:NPE_SV_gammay}
    For NPE with a subversive bureaucracy to exist, we need that $\gamma(y)<1$ (see Observation~\ref{obs:NPESV_Delta}). From Step~\ref{step:NPE_SV_badP_sy}, we obtain that such a condition is met if and only if
     when $P>0$ and $\psi>\tilde\Psi_{Subv}$, then
            \[
        \bar R_1^P > \delta\left( \mu_2^P + E \right)
    - \frac{\delta\beta\rho\lambda(1-2\lambda)}{1-\lambda\left[ 1+\beta -\psi(1-\beta) \right]}\mu_2^P,
    \]

\end{observation}


\bigskip


The collection of Observations~\ref{obs:NPE_SV_beliefs}, \ref{obs:NPE_psi}, \ref{obs:NPESV_Delta},  \ref{obs:NPE_conditions}, and \ref{obs:NPE_SV_gammay} yield the necessary and sufficient conditions for the existence of an NPE-FSV. A few remarks are in order. Under $P>0$,  the conditions on $\psi$ imply that, for the existence of this class of equilibria, it is necessary that
\begin{equation}\label{eq:condition_NPE_psi}
\psi\in\left[ \tilde\Psi_{Subv},\min \left\{ \hat\Psi_{Subv}, \frac{1-\lambda}{\lambda(1-\beta)} \right\} \right).
\end{equation}
The above condition can be satisfied if and only if the following inequalities simultaneously hold,
\[
\tilde\Psi_{Subv}<\min \left\{ \hat\Psi_{Subv}, \; \frac{1-\lambda}{\lambda(1-\beta)} \right\}, \tilde\Psi_{Subv}<1, \; \text{and} \; \hat\Psi_{Subv}>0.
\]
The first inequality holds when $\lambda<\frac{2}{2+\beta}$, while the second holds when $\lambda<\frac{1}{2\beta}$. The threshold $\frac{1-\lambda}{\lambda(1-\beta)}$ is always positive. Before studying the third inequality, recall that, $\hat\Psi_{Subv}>0$ if and only if either $\rho<\hat\rho$ and $\lambda>\tilde\Lambda$, or $\rho\geq\hat\rho$, where $\tilde\Lambda$ is defined by~\eqref{eq:tildelambda}. However,
\[
\tilde\Lambda >\frac{2}{2+\beta},
\]
and therefore $\rho\geq\hat\rho$ is the only relevant condition to ensure $\hat\Psi_{Subv}>0$ in these NPE. Because $\lambda<\min\left\{ \frac{1}{2\beta}, \frac{2}{2+\beta} \right\}$ is necessary, and $\hat\Psi_{Subv}<\frac{1-\lambda}{\lambda(1-\beta)}$ when $\lambda<\frac{2}{2+\beta}$, the condition displayed by Observation~\ref{obs:NPE_psi} is never binding in this case. Furthermore, 
\[
\tilde\Psi_{Subv}\leq 0 \iff \lambda\leq\frac{1}{1+\beta}<\min\left\{ \frac{1}{2\beta}, \frac{2}{2+\beta} \right\},
\]
meaning that the lower bound in~\eqref{eq:condition_NPE_psi} is not tight for an intermediate range of $\lambda$.

The following observation summarizes the conditions for an NPE-FSV to exist. Its conditions correspond to the verbal ones in the statement of the proposition in the main text.
\begin{observation}\label{obs:existence_NPE_SV}
    NPE-FSV exist if and only if

\begin{itemize}[noitemsep]
 \item[i)] $\delta E<  v(y,y)-v(x,y) - \delta\rho( 1-\pi)(1-\lambda)\left[ v(x,x)-v(y,x) \right]$;
 \item[ii)] $\Delta \leq \delta \rho (1-\lambda)\left[ \Pi_B(y,y)-\pi \right];$

        \item[iii)] $\bar R_1^P > \delta\left( \mu_2^P + E \right)
    - \frac{\delta\beta\rho\lambda(1-2\lambda)}{1-\lambda\left[ 1+\beta -\psi(1-\beta) \right]}\mu_2^P;$
    \item[iv)] $\rho \geq \hat\rho;$
    \item[v)] $\lambda<\min\left\{ \frac{1}{2\beta}, \frac{2}{2+\beta} \right\};$
    \item[vi)] $\psi\in\left[ \tilde\Psi_{Subv}, \hat\Psi_{Subv}\right).$
 \end{itemize}
    \end{observation}
\qed

\bigskip
\subsubsection{Proof of Proposition \ref{prop:NPE_ASV_existence}}
The proof of the existence of a NPE-ASV follows steps analogous to those of the NPE-VSF.    Steps \ref{step:NPE_gP_SB} and \ref{step:NPE_SV_badP_sy} differ substantially and have to be revised. All other steps are identical and therefore omitted for brevity. These results complete the proof.

\begin{step}\label{step:NPEASV_gP_SB}
Suppose that $\theta^P=g$ and $s_1=y$. The first part of this step is identical to the one of Step \ref{step:NPE_gP_SB}, so it is skipped for brevity. As we saw in Step \ref{step:NPE_gP_SB}, when $s_1=y$, the good politician's payoff from proposing $q^P_1=y$ is higher than that from proposing $q^P_1=x$ if and only if condition \ref{eq:NPE_SV_condition} holds.

Suppose now that $P\leq 0$, as in PE (see Step~\ref{step:PECB_goodP_x_sy}). If $P=0$, then no additional condition on $\psi$ is required, as any $G$ satisfies~\eqref{eq:NPE_SV_condition}. When $P <0$, the NPE with subversive bureaucracy is sustained if and only if 
\begin{equation}\label{eq:NPE_SV_low_psi}
\psi\leq \tilde \Psi_{Subv}.
\end{equation}  

Intuitively, when $\psi$ is sufficiently low, proposing $x$ is rarely confirmed by bad bureaucrats in state $y$. This results in a low probability of re-election in a context where office rents are attractive. Although the good politician seeks re-election, proposing $x$ is not the most effective way to achieve it. Thanks to a subversive bureaucracy, the politician has a higher chance of re-election by proposing $y$ instead. In this case, politicians rely on bureaucrats to pander on their behalf. By matching the proposal to the state, politicians signal that they are good types—information that subversive bureaucrats use to force pandering by contesting the politician’s proposal. This strategy improves the politician’s re-election prospects, enabling them to enjoy lucrative office rents.\footnote{When condition~\eqref{eq:NPE_SV_condition} fails, the good politician prefers to propose $x$ in state $y$. The analysis conducted here does not apply in these cases because then bureaucrats are sure the politician's type is bad upon observing a proposal $q_1^P=y$. That is, $\Pi_B(y,s_1)=0$, and thus bureaucrats would never be subversive.}
In the case $P<0$, the tightest between~\eqref{eq:psi_NPE_SV_beliefs} and~\eqref{eq:NPE_SV_low_psi} binds, the latter being the tightest when $\lambda<\frac{2}{2+\beta}$. In such cases, we find that $\tilde \Psi_{Subv}\leq 0$ when $\lambda\leq\frac{1}{1+\beta}$, making the condition impossible to be satisfied. By contrast, $\tilde \Psi_{Subv}>1$ if and only if $\lambda>\frac{1}{2\beta}$ and $\beta>\frac{1}{2}$, thus making the restriction automatically satisfied. 

\begin{observation}\label{obs:NPEASV_conditions}
     The existence of NPE-ASV requires
    \[
    \text{either } \; P = 0 \; \text{and } \psi< \hat \Psi_{Subv},
    \]
    \[
     \text{or } \; P< 0 \; \text{and } \psi < \min\left\{ \hat \Psi_{Subv}, \tilde \Psi_{Subv} \right\}.
    \]
   
\end{observation}
\end{step}

\begin{step}\label{step:NPE_ASV_badP_sy}
   Suppose that $\theta^P=b$ and $s_1=y$. The first part of this step is identical to the
one of Step \ref{step:NPE_SV_badP_sy}, so it is skipped for brevity. 

The bad politician prefers to propose $q_1^P=x$ if and only if condition \ref{eq:NPE_SV_rents} holds.
Suppose now that $\psi \leq \tilde\Psi_{Subv}$, as it is possible in NPE with subversive bureaucracy when $P\leq 0$ (see Observation~\ref{obs:NPEASV_conditions}). This scenario requires $\lambda>\frac{1}{1+\beta}$ for $\tilde\Psi_{Subv}>0$. In the knife-edge case where $\psi=\tilde\Psi_{Subv}$, we obtain that $H=0$. As a result, $\gamma(y)=1$ for $\lambda\geq \frac{1}{2}$, and $\gamma(y)=0$ otherwise. However, recall that $\gamma(y)<1$ is necessary for the existence of NPE with a subversive bureaucracy, and thus it must be that either $\psi\neq\tilde\Psi_{Subv}$, or $\lambda<\frac{1}{2}$. 

Consider the case where $\psi<\tilde\Psi_{Subv}$, as required by NPE with a subversive bureaucracy when $P<0$. We find that condition~\eqref{eq:NPE_SV_rents} is
\[
r_1^P \geq \delta\left( \mu_2^P + E \right)
    - \frac{\delta\beta\rho\lambda(1-2\lambda)}{1-\lambda\left[ 1+\beta -\psi(1-\beta) \right]}\mu_2^P \triangleq \left(F_1^P\right)^{-1}\left( 1-\gamma(y)  \right).
\]
In this case, we have that
\begin{equation}\label{eq:NPE_AP_gammay}
\gamma(y):=1-F_1^P\left( \delta\left( \mu_2^P + E \right)
    - \frac{\delta\beta\rho\lambda(1-2\lambda)}{1-\lambda\left[ 1+\beta -\psi(1-\beta) \right]}\mu_2^P   \right).
\end{equation}
To satisfy $\gamma(y)<1$, it must be that 
\[
\delta\left( \mu_2^P + E \right)
    - \frac{\delta\beta\rho\lambda(1-2\lambda)}{1-\lambda\left[ 1+\beta -\psi(1-\beta) \right]}\mu_2^P >0,
\]
which, when $\psi<\tilde\Psi_{Subv}$, gives
\[
\gamma(y)<1 \iff \psi <  \tilde\Psi_{Subv} + \frac{\beta\rho\lambda(1-2\lambda)\mu_2^P}{\lambda(1-\beta)\left(\mu_2^P+E\right)}.
\]
\end{step}

\begin{observation}\label{obs:NPE_ASV_gammay}
    For NPE with a subversive bureaucracy to exist, we need that $\gamma(y)<1$ (see Observation~\ref{obs:NPESV_Delta}). From Step~\ref{step:NPE_ASV_badP_sy}, we obtain that such a condition is met if and only if
     when $P\leq 0$ and $\psi\leq \tilde\Psi_{Subv}$,
    \begin{equation}\label{eq:psi_star_NPE}
    \psi < \tilde\Psi_{Subv}^\star := \tilde\Psi_{Subv} + \frac{\beta\rho\lambda(1-2\lambda)\mu_2^P}{\lambda(1-\beta)\left(\mu_2^P+E\right)}.
    \end{equation}

\end{observation}

The collection of Observations~\ref{obs:NPE_SV_beliefs}, \ref{obs:NPE_psi}, \ref{obs:NPESV_Delta},  \ref{obs:NPEASV_conditions}, and \ref{obs:NPE_ASV_gammay} yield the necessary and sufficient conditions for the existence of an NPE-ASV. As in the case of a NPE-FSV a few remarks are in order. Given  $P\leq 0$, Observations~\ref{obs:NPE_psi} and \ref{obs:NPEASV_conditions}, and the fact that the inequalities $\tilde\Psi_{Subv}<\hat\Psi_{Subv}$, $\hat\Psi_{Subv}<\frac{1-\lambda}{\lambda(1-\beta)}$, $\tilde\Psi_{Subv}<\frac{1-\lambda}{\lambda(1-\beta)}$ all hold true if and only if $\lambda<\frac{2}{2+\beta}$, imply the following existence requirement,
\[
\psi<\min \left\{ \tilde\Psi_{Subv}, \; \frac{1-\lambda}{\lambda(1-\beta)} \right\}.
\]

In this case, the requirement set forth by Observation~\ref{obs:NPE_SV_beliefs} is never binding. To the above condition, we additionally need $\tilde\Psi_{Subv}>0$, which holds true if and only if
\[
\lambda>\frac{1}{1+\beta}\in(\nicefrac{1}{2},1).
\]

Furthermore, Observation~\ref{obs:NPE_ASV_gammay} adds the requirement $\psi<\tilde\Psi_{Subv}^\star$, where
\[
\tilde\Psi_{Subv}^\star<\tilde\Psi_{Subv} \iff \lambda>\frac{1}{2},
\]
which we have seen it must be satisfied in this NPE. Moreover, $\tilde\Psi_{Subv}^\star>0 \iff \lambda>\tilde\Lambda_{Subv}^\star$, where $\frac{1}{2}<\tilde\Lambda_{Subv}^\star <1$ and
    \begin{equation}\label{eq:lambda_star_NPE}
    \tilde\Lambda_{Subv}^\star := \frac{(1+\beta)\left(\mu_2^P + E\right) +\beta \rho \mu_2^P  \left[ 1- \sqrt{1+\frac{\left(\mu_2^P + E\right) \left[(1+\beta)^2 \left(\mu_2^P + E\right) - 2 (3-\beta) \beta   \rho \mu_2^P \right]}{\left(\beta  \rho \mu_2^P  \right)^2 }} \, \right]}{4 \beta \rho  \mu_2^P }.
    \end{equation}
Finally, we have that 
\[
\tilde\Psi_{Subv}^\star<\frac{1-\lambda}{\lambda(1-\beta)} \iff \lambda<\tilde\Lambda_{Subv}',
\]
where $\tilde\Lambda_{Subv}^\star<\tilde\Lambda_{Subv}'<1$, $\tilde\Lambda_{Subv}'>\frac{2}{2+\beta}\in\left(\nicefrac{2}{3},1\right)$, and
\begin{equation}\label{eq:lambda_prime_NPE}
\tilde\Lambda_{Subv}':=\frac{(2+\beta)\left(\mu_2^P + E\right) +\beta \rho \mu_2^P  \left[ 1- \sqrt{1+\frac{\left(\mu_2^P + E\right) \left[(2+\beta)^2 \left(\mu_2^P + E\right) - 2 (6-\beta) \beta   \rho \mu_2^P \right]}{\left(\beta  \rho \mu_2^P  \right)^2 }} \, \right]}{4 \beta \rho  \mu_2^P }.
\end{equation}

As a result, the existence condition on $\psi$ is
\[
\psi<\min \left\{ \tilde\Psi_{Subv}^\star, \; \frac{1-\lambda}{\lambda(1-\beta)} \right\} = \left\{
  \begin{array}{lr}
    \tilde\Psi_{Subv}^\star & \text{ if } \tilde\Lambda_{Subv}^\star<\lambda\leq \tilde\Lambda_{Subv}'\\
    \frac{1-\lambda}{\lambda(1-\beta)} & \text{ if } \lambda > \tilde\Lambda_{Subv}'
  \end{array}
\right..
\]

The requirement $\psi<\tilde\Psi_{Subv}^\star$ ensures that $\gamma(y)<1$, which is necessary for the existence of NPE with a subversive bureaucracy when $P\leq 0$. The requirement $\psi<\frac{1-\lambda}{\lambda(1-\beta)}$ ensures that good politicians propose policy $x$ in state $x$.

The following observation summarizes these results. As in the previous proofs, each condition is the technical counterpart of the verbal conditions in the statement of the proposition in the main text.
\begin{observation}\label{obs:existence_NPE_ASV}
     NPE-ASV  exist if and only if
    \begin{itemize}[noitemsep]
 \item[i)] $\delta E\geq  v(y,y)-v(x,y) - \delta\rho( 1-\pi)(1-\lambda)\left[ v(x,x)-v(y,x) \right]$;
 \item[ii)] $\Delta \leq \delta \rho (1-\lambda)\left[ \Pi_B(y,y)-\pi \right];$

 \item[iii)] $\lambda>\tilde\Lambda_{Subv}^\star;$

        \item[iii)] $\bar R_1^P > \delta\left( \mu_2^P + E \right)
    - \frac{\delta\beta\rho\lambda(1-2\lambda)}{1-\lambda\left[ 1+\beta -\psi(1-\beta) \right]}\mu_2^P;$
    \item[vi)] $\psi<\min\left\{ \tilde\Psi_{Subv}^\star, \frac{1-\lambda}{\lambda(1-\beta)} \right\}.$
 \end{itemize}

    \end{observation}
\qed

\subsection{Proof of Proposition \ref{characterization}}
    By Lemma~\ref{lemma:second}, period-2 equilibrium behavior is identical across all perfect Bayesian equilibria. The proofs of the equilibrium existence propositions proceed by backward induction. Given period-2 behavior, we first characterize bureaucrats' equilibrium strategies as best responses to the politician’s strategy, taking into account anticipated period-2 play. We then consider politicians’ strategies consistent with either PE or NPE. For each case, all possible bureaucratic best responses are examined and incorporated into the respective propositions. Hence, Definitions \ref{def:conj}, \ref{def:PECB_xi_gamma}, \ref{def:non-pandering_SF}, and \ref{def:non-pandering_SV} exhaustively cover all pandering and non-pandering equilibria.

    Now consider the case $\lambda \leq \tilde\Lambda_{Subv}'$, where $\tilde\Lambda_{Subv}'$ is defined by~\eqref{eq:lambda_prime_NPE}. In this range, the necessary condition for the existence of NPE-ASV implies $\psi < \frac{1-\lambda}{\lambda(1-\beta)}$. This inequality is precisely the condition ensuring that a good politician proposes $q_1^P=x$ in state $s_1=x$, as required in NPE. For all other forms of NPE, this condition is automatically satisfied. Only in the case of NPE-ASV it must be explicitly imposed (see Step~\ref{step:NPE_gP_sx} and Observation~\ref{obs:NPE_psi}). However, when $\lambda \leq \tilde\Lambda_{Subv}'$, the condition holds in that case as well (see the part preceding Observation~\ref{obs:existence_NPE_SV}). Hence, in all NPE we have $q_1^P\left(g,x,r_1^P\right)=x$ when $\lambda \leq \tilde\Lambda_{Subv}'$, where $\tilde\Lambda_{Subv}'>\frac{2}{2+\beta}\in\left(\nicefrac{2}{3},1\right)$. It follows that, in such cases, IE can be partitioned into equilibria where $q_1^P\left(g,y,r_1^P\right)=x$ (the PE) and equilibria where $q_1^P\left(g,y,r_1^P\right)=y$ (the NPE), both of which are fully characterized. \qed
    
\subsection{Applications}

\subsubsection{Benchmarks}\label{sec:app_bench}

This section focuses on the benchmark cases where $\lambda=0$ and $\lambda=1$. In doing so, we assume that the good politician retains pandering incentives, while bad politicians have short-term incentives to implement $y$, making the equilibrium informative and one where pandering occurs. Specifically, we require that
\begin{equation*}
\delta E \geq v(y,y)-v(x,y) - \delta\rho( 1-\pi)\left[ v(x,x)-v(y,x) \right],\end{equation*}
and
\begin{equation*}
\bar R_1^P>\delta\left(\mu_2^P+E\right).
\end{equation*}

\noindent{\bf Proof of Propositions~\ref{prop:bench1} and \ref{prop:bench2}.}

Consider the following equilibrium continuation values for the voter.
\[
W_V^{gP} \coloneqq \rho v(x,x) + (1-\rho)v(y,y),
\]
\[
W_V^{bP} \coloneqq \rho v(y,x) + (1-\rho)v(y,y),
\]
\[
W_V^{\pi} \coloneqq \pi W_V^{gP} + (1-\pi)W_V^{bP}.
\]
Denote the equilibrium ex-ante welfare of the voter by $EU^V$. We obtain that the voter's equilibrium welfare in these two benchmarks is
\begin{align*}
    EU^V|_{\lambda=0}= 
    & \rho \bigg\{ \pi \left[ v(x,x) + \delta W_V^{gP} \right] \\
    & + (1-\pi) \left[ \gamma \left( v(x,x) + \delta W_V^{bP} \right) + (1-\gamma)\left( v(y,x)+\delta W_V^\pi \right) \right] \bigg\} \\
    & + (1-\rho) \bigg\{ \pi \left[ v(x,y) + \delta W_V^{gP} \right] \\
    & + (1-\pi) \left[ \gamma \left( v(x,y) + \delta W_V^{bP} \right) + (1-\gamma)\left( v(y,y)+\delta W_V^{\pi} \right) \right]  \bigg\},
\end{align*}
\begin{align*}
    EU^V|_{\lambda=1}=(1+\delta)\left\{ \rho \left[ \beta v(x,x) + (1-\beta)v(y,x) \right] +(1-\rho)v(y,y))\right\}.  
\end{align*}

Define the difference between the two benchmark welfare functions by
\[
    \Delta EU^V_\lambda\coloneqq EU^V|_{\lambda=0}- EU^V|_{\lambda=1}.
\]
In its explicit form, we have
\begin{multline*}
    \Delta EU^V_\lambda=\rho (1-\gamma )\pi  \bigg\{ (1-\pi\delta)  [v(x,x)-v(y,x)]+[v(y,y)-v(x,y)]\bigg\} \\
    -\rho\bigg\{[(\gamma -2) \delta  \pi + (1+  \delta)\beta -\gamma ] [v(x,x)-v(y,x)] -\gamma [v(y,y)-v(x,y)]\bigg\} \\ 
    -[(1-\gamma ) \pi+\gamma ] [v(y,y)-v(x,y)].
\end{multline*}
We now proceed by studying the derivatives of $\Delta EU^V_\lambda$. We find that
\[
\frac{d \Delta EU^V_\lambda}{d \beta}= -\rho (1+\delta) [ v(x,x)- v(y,x)]<0,
\]
meaning that an increase in $\beta$ always favors a dictatorship of the bureaucracy. The comparative statics is not so clear-cut with respect to $\pi$. Indeed, we have that
\begin{align*}
    \frac{d \Delta EU^V_\lambda}{d \pi}=&  \rho  [v(x,x)-v(y,x)] \{ 1-\gamma + \delta  [2-\gamma -2\pi (1-\gamma) ] \} \\ &+(1-\gamma ) (1-\rho ) [v(x,y)-v(y,y)],
\end{align*}
where
\[
\frac{d \Delta EU^V_\lambda}{d \pi} >0 \iff \rho > \rho_\pi\in(0,1).
\]
An increase in $\pi$ favors full political control provided that the likelihood of state $x$, where pandering does not occur, is sufficiently likely. This is intuitive, as good politicians pander, and the inefficiency of pandering is mitigated when the state is likely to be $x$. More specifically, the threshold on $\rho$ is defined by 
\[
\rho_\pi\coloneqq \frac{(1-\gamma) [v(y,y)-v(x,y)]}{ \{ 1-\gamma + \delta  [2-\gamma -2\pi (1-\gamma) ] \} [v(x,x)-v(y,x)]+(1-\gamma ) [v(y,y)-v(x,y)]}.
\]
The next comparative statics concerns $\gamma$. Recall that, when $\lambda\to 0$ and in PECB, we have
\[
\gamma|_{\lambda=0} \triangleq F_1^P\left( \delta \left( \mu_2^P + E \right) \right).
\]
Therefore  $\gamma$ is an increasing function of $\delta$, $\mu_2^P$ and $E$.  We are interested specifically in these two latter parameters. We obtain that
\[
\frac{d \Delta EU^V_\lambda}{d \gamma}= (1-\pi) \left\{\rho  (1-\delta  \pi) [v(x,x)-v(y,x)]-(1-\rho ) [v(y,y)- v(x,y)]\right\}.
\]
As before, an increase in $\gamma$ favors toothless bureaucracy provided that $\rho$ is sufficiently high, and specifically
\[
\frac{d \Delta EU^V_\lambda}{d \gamma}>0 \iff \rho > \rho_\gamma\in(0,1),
\]
where
\[
\rho_\gamma \coloneqq \frac{v(y,y)-v(x,y)}{v(y,y)-v(x,y)+(1-\delta  \pi) [v(x,x)- v(y,x)]}.
\]

Finally, we find that the voter is indifferent between the two benchmark cases at a specific $\beta$ level. That is,
\[
\Delta EU^V_\lambda = 0 \iff \beta = \tilde\beta,
\]
where
\begin{displaymath}
    \tilde\beta \coloneqq \frac{ (1-\delta\pi)[(1-\gamma) \pi+\gamma]  +2\delta  \pi }{(1+\delta) }
    - \frac{(1-\rho)[(1-\gamma) \pi+\gamma ][v(y,y)-v(x,y)]}{(1+\delta) \rho  [v(x,x)-v(y,x)]}.
\end{displaymath}
Together with our previous comparative statics on $\beta$, this implies that the voter prefers a dictatorial to a toothless bureaucracy if and only if $\beta$ is sufficiently high. We have that $\tilde\beta<1$ always, further implying that the voter always prefers dictatorial to toothless bureaucracy for sufficiently high $\beta$. However, such a threshold may be negative. Indeed,
\[
\tilde\beta>0 \iff \rho>\tilde\rho_\beta\in(0,1),
\]
where,
\[
\tilde\rho_\beta\coloneqq \frac{[(1-\gamma) \pi+\gamma ] [v(y,y)-v(x,y)]}{\left\{(1-\delta  \pi)[(1-\gamma ) \pi+\gamma] +2\delta  \pi\right\} [v(x,x)-v(y,x)]+[(1-\gamma ) \pi+\gamma ] [v(y,y)-v(x,y)]}.
\]
As a result, when $\rho$ is excessively low, then the voter prefers dictatorial over toothless regardless of $\beta$. This is because pandering is too inefficient to be tolerated, and voters prefer to give up electoral accountability to get rid of pandering. \qed

\begin{observation}
    The voter prefers a dictatorial over a toothless bureaucracy if and only if $\beta$ is sufficiently high, and always when $\rho$ is excessively low. The welfare difference $\Delta EU^V_\lambda$ is always decreasing in $\beta$, and it is increasing in $\pi$ and $\gamma$ if and only if $\rho$ is sufficiently high.
\end{observation}

\subsubsection{Voter's welfare and political selection}

This section elaborates on the choice of parameters and conditions underpinning the analysis conducted in Section~\ref{sec:cs}. We begin by selecting the policymakers' rents distribution.

\noindent{\bf Assumption.} $F_t^j \sim \mathcal{U}[0,2]$ for every $t=\{1,2\}$ and $j=\{P,B\}$. 

From the above assumption, it follows that $\mu_t^j=1$ for every $t=\{1,2\}$ and $j=\{P,B\}$. Recall that, in both PECB and PEPB, the score $\xi$ is defined by $\xi = F_1^B\left(\delta \pi \rho (1-\lambda)\mu^B_2\right)$. Under our assumptions on the rents' distribution, we obtain
\begin{equation}\label{eq:xi}
    \xi = \frac{\delta\pi\rho(1-\lambda)}{2}\in[0,\nicefrac{1}{2}].
\end{equation}
In PE, it is necessary to have $\xi<\frac{1-\lambda}{\lambda(1-\beta)}$. By substituting for $\xi$ and rearranging, the condition becomes $\delta\pi\rho\lambda(1-\beta)<2$, which is always satisfied for every parameter choice. 

In PECB,  score $\gamma$ is defined by~\eqref{eq:PECB_gamma}. By substituting for $\xi$ and applying our distributional assumption, we obtain\footnote{Score $\gamma$ is not defined for $\lambda=1$ which, as we have seen, renders the politicians' choices irrelevant.}
\begin{equation}\label{eq:gamma}
    \gamma= \min\left\{ 1, \max \left\{ 0, \frac{\delta}{2}\left[ 1 + E - \frac{\beta\rho\lambda(1-\lambda)}{1-\lambda\left[ 1 + \frac{\delta\pi\rho(1-\lambda)(1-\beta)}{2} \right]}\right]\right\}\right\}.
\end{equation}
To ensure that $\gamma<1$, as required for PECB existence, we need a condition on office rents $E$. Such a condition is
\begin{equation}\label{eq:cond_E_1}
E < \left( \frac{2}{\delta} - 1\right) + \frac{\beta\rho\lambda(1-\lambda)}{1-\lambda\left[ 1 + \frac{\delta\pi\rho(1-\lambda)(1-\beta)}{2} \right]}.
\end{equation}

In PEPB,  score $\gamma(y)$ is defined by~\eqref{eq:PEPB_gammay}. By substituting for $\xi$ and applying our distributional assumptions, we obtain
\begin{equation}\label{eq:gamma_y}
    \gamma(y)= \min\left\{ 1, \max \left\{ 0, \frac{\delta}{2}\left[ 1 + E - \frac{\beta\rho\lambda}{1-\lambda(1-\beta)\left[ 1 + \frac{\delta\pi\rho(1-\lambda)}{2} \right]}\right]\right\}\right\}.
\end{equation}

 From Observation~\ref{obs:PE_Deltas}, we have that $\Delta_{PB} > \Delta_{CB}$ if and only if $\gamma>\gamma(y)$, which holds true if and only if $\xi<\frac{1-\lambda}{\lambda}$.
Since, in this configuration, $\xi=\frac{\delta\pi\rho(1-\lambda)}{2}<\frac{1-\lambda}{\lambda}$ always holds, we have that indeed $\gamma>\gamma(y)$ and $\Delta_{PB} > \Delta_{CB}$. Existence of PEPB requires that  $\gamma(y)<1$, which holds since $\gamma>\gamma(y)$ and $\gamma<1$ by condition~\eqref{eq:cond_E_1}.

At the same time, condition $i)$ in Propositions~\ref{PECBexistence} and \ref{PEPBexistence} tells us that office rents must be sufficiently high. Specifically,
\begin{equation}\label{eq:cond_E_2}
 E \geq \frac{1}{\delta} \left[v(y,y)-v(x,y) - \delta\rho( 1-\pi)(1-\lambda)\left( v(x,x)-v(y,x) \right)\right].
\end{equation}
Furthermore, notice that the following condition, required for both PECB and PEPB, is always satisfied,
\[
        \mu_2^B = 1 < \frac{\left(F_1^{B}\right)^{-1}\left( \frac{1-\lambda}{\lambda(1-\beta)}\right)}{\delta\pi\rho(1-\lambda)} =  \frac{2}{\delta\pi\rho\lambda(1-\beta)}.
        \]
To proceed with the analysis, we first need to set our parameters and check that all conditions are satisfied.

\noindent{\bf Assumption.} We set $\delta=v(y,y)=1$, $v(x,y)=v(y,x)=0$, and $v(x,x)=k\geq 1$.

An important consequence of our parametric assumptions is that $\Delta=1/k\leq 1$. Furthermore, the two conditions on the office rents are
\[
E\geq g(\lambda)\coloneqq 1-\rho(1-\pi)(1-\lambda)k,
\]
\[
E< f(\lambda)\coloneqq 1 + \frac{\beta \rho \lambda (1-\lambda)}{1-\lambda \left[1 + \frac{\pi \rho (1-\lambda)(1-\beta)}{2}\right]}.
\]
The functions $g(\lambda)$ and $f(\lambda)$ are strictly increasing in $\lambda$ and such that $g(1)=f(0)=1$.\footnote{Specifically, $\frac{dg(\lambda)}{d\lambda}=\rho(1-\pi)k>0$ and $\frac{df(\lambda)}{d\lambda} = \frac{4 \beta \rho}{\left[2-(1-\beta)\pi\rho\lambda\right]^2}>0$.} By selecting $E=1$, we can perform our numerical comparative statics exercise for every $\lambda\in[0,1)$ because $g(\lambda)<E=1\leq f(\lambda)$. In addition, within the same equilibrium class, the voter's welfare is continuous in $\lambda$ for all $\lambda\in[0,1]$, with no discontinuities at $\lambda=0$ or $\lambda=1$. With $E=1$, the conditions for equilibrium existence outlined by Observations~\ref{obs:PECB_gamma} and \ref{obs:PEPB_bounds} are naturally met for every $\lambda\in(0,1)$.  

\noindent{\bf Assumption.} We set $E=1$ to analyze PECB and PEPB.

To study a PECB, we set $v(x,x)=1$ to ensure that condition $ii)$ in Proposition~\ref{PECBexistence} is always satisfied for every value of $\lambda$. Indeed, by doing so we obtain $\Delta=1>\Delta_{CB}$. By contrast, more care is needed when studying a PEPB. Proposition~\ref{PEPBexistence} tells us that a necessary condition for the existence of a PEPB is 
\begin{equation}\label{eq:condition_PEPB}
\Delta < \Delta_{PB} = \delta\rho(1-\lambda)\left[ \frac{\pi(1-\pi)(1-\gamma(y))}{\pi + (1-\pi)\gamma(y)} \right]
\end{equation}
Since the right-hand side of the inequality is less than one, we need to set $v(x,x)$ such that $\Delta$ is also less than one. We choose $v(x,x)=500$ to set $\Delta=1/500$.

\noindent{\bf Assumption.} We set $v(x,x)=1$ to analyse PECB, and $v(x,x)=500$ to analyse PEPB.

Condition~\eqref{eq:condition_PEPB} is violated for certain values of $\lambda$. As $\lambda\to1$, the right-hand side of the inequality shrinks to zero, violating the necessary condition to be in a PEPB. The same happens as $\gamma(y)\to 1$. From \eqref{eq:gamma_y}, we can see that, under our choice of parameters, $\lim_{\lambda\to 0}\gamma(y) = 1$. As a result, condition~\eqref{eq:condition_PEPB} does not hold for relatively high and relatively low values of $\lambda$. Differently, it holds for all intermediate values of $\lambda$.\footnote{The function $\delta\rho(1-\lambda)\left[ \frac{\pi(1-\pi)(1-\gamma(y))}{\pi + (1-\pi)\gamma(y)} \right]$ is concave in $\lambda$. Therefore, it may give us two thresholds for $\lambda$ in $[0,1]$. An equilibrium is a PEPB only for values of $\lambda$ between those two thresholds, if any.} 
Our comparative statics exercise takes into account that, for extreme values of $\lambda$, the equilibrium we are analyzing is a PECB and not a PEPB.


Next, we examine conditions under which non-pandering equilibria with stand-firm bureaucracy (NPE-SF) exist. The main condition concerns office rents, which must be sufficiently low to deter politicians from pandering. Formally,
\[
\delta E < v(y,y)-v(x,y) - \delta \rho (1-\pi)(1-\lambda)[v(x,x)-v(y,x)].
\]
In our graphical application, we set $\delta=v(x,x)=v(y,y)=1$ and $v(x,y)=v(y,x)=0$. Under our parameter configuration, the condition boils down to
\[
E< \mathcal{E}(\lambda) = 1-\rho(1-\pi)(1-\lambda).
\]
The threshold $\mathcal{E}(\lambda)$ is increasing in $\lambda$ and such that $\mathcal{E}(1)=1$ and $\mathcal{E}(0)=1-\rho(1-\pi)<1$. By rearranging the above condition, we obtain that NPE-SF require a sufficiently high $\lambda$. Specifically,
\[
\lambda > \ell(E,\rho,\pi) \coloneqq  1 - \frac{1-E}{\rho(1-\pi)},
\]
where $\ell(E,\rho,\pi)\in (0,1)$ if and only if $E\in \left(1-\rho(1-\pi),1\right)$. 

Our previous application rules out non-pandering equilibria by setting $E=1$. Finally, by setting $E\in(0,1)$ we obtain that $0<\gamma(x)<\gamma(y)<1$. Specifically, the existence requirement outlined by Observation~\ref{obs:NPE_SF_gammax} is satisfied, as $\gamma(x)<1$, where $\gamma(x)$ in NPE is defined exactly as $\gamma$ in PECB, i.e., by~\eqref{eq:gamma}. Differently, 
in NPE-SF,  score $\gamma(y)$ is defined by~\eqref{eq:NPE_SF_gammay}. Recall that, in NPE-SF, we have that $\psi = \xi$ as defined previously by~\eqref{eq:xi}. By substituting for $\psi$ and applying our distributional assumptions, we obtain
\begin{equation}\label{eq:gamma_y_NPE}
    \gamma(y)= \min\left\{ 1, \max \left\{ 0, \frac{\delta}{2}\left[ 1 + E - \frac{\beta\rho\lambda(1-\lambda)}{1-\lambda\left[ 1 - \frac{\delta\pi\rho(1-\lambda)(1-\beta)}{2} \right]}\right]\right\}\right\}.
\end{equation}

\noindent{\bf Assumption.} We set $E=.85$ to analyze NPE-SF.

To ensure that the selected parameters satisfy the NPE-SF existence conditions, we need to check the requirements outlined by Observation~\ref{obs:NPE_beliefs}. That is, for every $\lambda>\ell(\cdot)$,
\[
\rho>\hat\rho,
\]
\[
\mu_2^B \equiv 1 < \hat\mu := \frac{2\Gamma}{\pi\rho\lambda(1-\beta)}.
\]
Figure~\ref{fig:NPE} illustrates the transition from a PECB to a NPE-SF. The plots use $\pi = 0.7$ and $\rho = 0.85$, and differ only in the choice of $\beta$. In the first plot, with $\beta = 0.9$, we obtain $\hat\rho \approx 0.77$ and $\hat\mu \approx 21.76$ at the tightest $\lambda$. In the second plot, with $\beta = 0.75$, we obtain $\hat\rho \approx 0.79$ and $\hat\mu \approx 6.46$ at the tightest $\lambda$. In both cases, all existence conditions are satisfied for every $\lambda \in (0,1)$.

\bigskip


\subsubsection{Proof of Proposition~\ref{prop:jump}}

Consider the voter's equilibrium welfare functions in PECB and NPE-SF, as defined in Appendix~\ref{app:full_welfare}. We set our parameter configuration as follows:
\begin{itemize}[nosep]
\item $F_j^t\sim \mathcal{U}[0,2] \implies \mu_t^j=1$;
    \item $v(x,x)=v(y,y)=1$ and $v(x,y)=v(y,x)=0$;
    \item $\delta=1$;
    \item $0<E<1$.
\end{itemize}
In the welfare functions, we plug the equilibrium beliefs which have been defined in previous applications using the same parameter configuration. Specifically, these beliefs functions are $\xi$ and $\psi$, with $\xi=\psi$, defined by~\eqref{eq:xi}; $\gamma$ and $\gamma(x)$, with $\gamma=\gamma(x)$, defined by~\eqref{eq:gamma}; and $\gamma(y)$, defined by~\eqref{eq:gamma_y_NPE}. 

We denote the voter's welfare in equilibrium $\omega$ and conditional on the politician being type $\theta^P$ by $EU^V_{\omega}\left(\theta^P\right)$. The voter's welfare is thus 
\[
EU^V_{\omega}=\pi EU^V_{\omega}\left(g\right)+(1-\pi)EU^V_{\omega}\left(b\right).
\]
We are interested in the welfare difference between PECB and NPE-SF. To this end, define
\begin{equation}\label{eq:deltaU}
\Delta EU^V = EU_{PECB}^V - EU_{NPE-SF}^V.    
\end{equation}
Under our parameter configuration, we obtain
\[
\Delta EU^V|_{\left(\theta^P=g\right)}= (1-\lambda ) (1-\rho ) [ (1-\lambda ) (1-\pi) \rho -1],
\]
\[
\Delta EU^V|_{\left(\theta^P=b\right)}= \frac{1}{2} (1-\beta ) (1-\lambda )  (1-\rho) \lambda  \pi\rho  [1+ \pi \rho(1-\lambda)  ]E.
\]
We can thus re-write~\eqref{eq:deltaU} as
\[
\Delta EU^V = \pi \Delta EU^V|_{\left(\theta^P=g\right)} + (1-\pi)\Delta EU^V|_{\left(\theta^P=b\right)}.
\]
We find that
\begin{multline*}
\Delta EU^V = \pi (1-\rho )(1-\lambda) \Bigg\{ (1-\lambda)(1-\pi)\rho -1 + \left( \frac{1-\pi}{2} \right) (1-\beta)\lambda\rho[1+\pi\rho(1-\lambda)]E \Bigg\},
\end{multline*}
and
\[
\Delta EU^V <0 \iff (1-\lambda)(1-\pi)\rho + \left( \frac{1-\pi}{2} \right) (1-\beta)\lambda\rho[1+\pi\rho(1-\lambda)]E<1.
\]

Recall that the threshold in which the equilibrium shifts from PECB to NPE-SF is $\lambda=\ell$, where
\[
\ell  \triangleq  1 - \frac{1-E}{\rho(1-\pi)}.
\]
We choose $E<1$ so that $\ell<1$. We have $\ell>0 \iff E>1-\rho(1-\pi)$, but such a restriction is not required.

We evaluate $\Delta EU^V $ at $\ell$ to obtain the size of the discontinuity jump occurring when shifting from PECB to NPE-SF as a result of a marginal increase in $\lambda$, and we obtain that
\[
\Delta EU^V|_{\lambda=\ell} <0 \iff (1-\beta)(1-\pi E)\left[ \rho(1-\pi) - 1 + E \right]<2(1-\pi).
\]
The left-hand side of the above inequality is decreasing in $\beta$ and increasing in $\rho$. Set $\beta=0$ and $\rho=1$ to obtain
\[
(1-\pi E)\left[  E -\pi \right]<2(1-\pi).
\]
Because the left-hand side's two components satisfy $1-\pi E < 2$ and $E-\pi<1-\pi$, it must be that such an inequality always holds true. Therefore, we obtain that
\[
\Delta EU^V|_{\lambda=\ell} <0,
\]
meaning that such a shift always discontinuously increases the voter's welfare. \qed

\noindent{\bf Selection.} The ex-ante probability that a bad politician is re-elected in a PECB, denoted by $\eta$, is
\begin{align*}
\eta & \coloneqq (1-\pi) \{ \gamma \left[ \rho \left( \beta + (1-\beta)(1-\lambda)\right) + (1-\rho)(1-\lambda)\right] \\
& + (1-\gamma)\left[ \rho(\beta\lambda + (1-\beta)\xi\lambda) + (1-\rho)(1-\beta)\xi\lambda \right] \}.
\end{align*}
The probability $\eta$ is used to plot the left panel of Figure~\ref{fig:selection}.

\subsubsection{Proof of Proposition~\ref{prop:selection}}

Consider the parameter configuration we have used previously to analyze PECB, and leave $E$ positive but otherwise free.  
The ex-ante probability of having a good politician ruling in the second period of a PECB, denoted by $\zeta$, is
\begin{align*}
\zeta & \coloneqq \pi \left\{ \rho \left[ \beta + (1-\beta)\left[\lambda\pi + (1-\lambda)\right] \right] + (1-\rho)\left[\lambda\pi + (1-\lambda)\right] \right\}\\
& + (1-\pi) \{ \gamma \left[ \rho(1-\beta)\lambda\pi + (1-\rho)\lambda\pi \right]\\
& + (1-\gamma) \{ \rho [\beta(1-\lambda)\pi + (1-\beta)[\xi(1-\lambda)\pi + (1-\xi)\pi ]  ] \\
 & \phantom{ + (1-\gamma)\{ }  + (1-\rho)[ \beta\pi + (1-\beta)[\xi (1-\lambda)\pi + (1-\xi)\pi ]  ]  \} \}.
\end{align*}
The probability $\zeta$ can be simplified as follows,
\[
\zeta = \pi \left\{  2-\lambda - \gamma(1-\lambda) - \pi(1-\gamma) \left[ ( 1-\lambda) + (1-\pi)(1-\beta)\lambda\xi   \right]  \right\}.
\]
By plugging in $\gamma$ and $\xi$ (from \eqref{eq:xi} and \eqref{eq:gamma}, respectively) with $\delta=1$, we obtain
\begin{multline*}
\zeta = \frac{\pi}{4}  \{ (1-\lambda) \lambda  (1-\pi) \rho  \left[  (1-\beta) (E-1) \pi-2 \beta \right] \\ 
-2 \left[ E (1-\lambda) (1-\pi) +\pi(1-\lambda)+\lambda-3 \right] \}.
\end{multline*}

We compute the following derivative,
\[
\frac{d \zeta}{d \lambda}=\frac{(1-\pi) \pi}{4}  \left\{ (2 \lambda -1) \rho  \left[  (1-\beta ) (1-E) \pi-2 \beta \right] -2 (1-E) \right\}.
\]
We obtain that
\[
\frac{d \zeta}{d \lambda}> 0 \iff E>E_{\zeta},
\]
where
\[
E_{\zeta}\coloneqq \frac{2+(1-2 \lambda ) \rho  \left[  (1-\beta) \pi-2 \beta \right]}{2+(1-\beta ) (1-2 \lambda ) \pi \rho} >0,
\]
and $E_{\zeta}<1$ if and only if $\lambda<\frac{1}{2}$. Therefore, within a PECB (which requires $\lambda<\ell$), an increase in $\lambda$ improves political accountability as measured by $\zeta$ if and only if office rents are sufficiently high. In the specific configuration we have used before for PECB, where $E=1$, political accountability improves with $\lambda$ if and only if $\lambda<\frac{1}{2}$. \qed


\subsection{Voter's equilibrium  welfare}\label{app:full_welfare}
Denote the voter's payoff in $t=2$ when policymakers types are $\theta^B$ and $\theta_2^P$ by $W_{\theta^B B}^{\theta_2^P P}$. Moreover, define 
\[
W_{\theta^B B}^{\pi}\coloneqq\pi W_{\theta^B B}^{\theta_2^P=g P} + (1-\pi)W_{\theta^B B}^{\theta_2^P=b P}.
\]
The voter's continuation payoffs in every equilibrium are
\[
  W_{gB}^{gP}\coloneqq \rho v(x,x) + (1-\rho)v(y,y),
\]
\[
  W_{gB}^{bP}\coloneqq  \rho\left[ \lambda v(x,x) + (1-\lambda)v(y,x) \right] + (1-\rho)v(y,y),
\]
\[
W_{gB}^{\pi} \coloneqq \rho[\pi+(1-\pi)\lambda] v(x,x) + (1-\rho) v(y,y) + (1-\pi)(1-\lambda)\rho v(y,x),
\]
\[
W_{bB}^{gP}\coloneqq \rho \left[ \lambda v(y,x) + (1-\lambda)v(x,x) \right] + (1-\rho)v(y,y),
\]
\[
W_{bB}^{bP}\coloneqq  \rho v(y,x) + (1-\rho)v(y,y),
\]
\[
W^{\pi}_{bB}\coloneqq \pi\rho (1-\lambda) v(x,x) + (1-\rho)v(y,y) + \rho [(1-\pi) +\pi\lambda] v(y,x).
\]
The voter's expected welfare in an equilibrium $\omega\in\{\text{PECB}, \text{NPE-SF}\}$ is
\[
EU_{\omega}^V = \pi EU^V_{\omega}\left(\theta^P=g\right) + (1-\pi) EU^V_{\omega}\left(\theta^P=b\right).
\]

\subsubsection{Welfare in PECB}
This part outlines the voter's welfare in PECB, conditional on the politician's type. By fixing the politician as being good, we obtain
\begin{align*}
EU^V_{PECB}\left(\theta^P=g\right)&=\beta \bigg\{\rho \left[v(x,x)+\delta W^{gP}_{gB} \right]\\
& + (1-\rho)\left[\lambda \left[v(y,y)+\delta W^{\pi}_{gB} \right] + (1-\lambda) \left[v(x,y)+\delta W^{gP}_{gB} \right] \right] \bigg\}  \\
&+(1-\beta) \bigg\{\rho  \left[ \lambda \left[v(y,x)+\delta W^{\pi}_{bB} \right] + (1-\lambda)\left[v(x,x)+\delta W^{gP}_{bB} \right] \right]\\
& + (1-\rho)\left[ \lambda\left[v(y,y)+\delta W^{\pi}_{bB} \right] + (1-\lambda)\left[v(x,y)+\delta W^{gP}_{bB} \right] \right] \bigg\}.
\end{align*}
By contrast, conditional on the politician being bad, the voter's welfare is
\begin{align*}
EU^V_{PECB}\left(\theta^P=b\right)&= \beta \Bigg\{ \rho \bigg[ \gamma \left[ v(x,x) + \delta W_{gB}^{bP} \right] \\ 
    & + (1-\gamma)\left[ \lambda \left[ v(x,x) + \delta W_{gB}^{bP} \right] + (1-\lambda) \left[ v(y,x) + \delta W_{gB}^\pi \right] \right] \bigg] \\
    & + (1-\rho) \bigg[ \gamma \left[ \lambda\left[ v(y,y) + \delta W_{gB}^\pi \right] + (1-\lambda)\left[ v(x,y) + \delta W_{gB}^{bP} \right] \right] \\
    & + (1-\gamma)\left[ v(y,y) + \delta W_{gB}^\pi \right] \bigg] \Bigg\}\\
    & + (1-\beta)\Bigg\{ \rho \bigg[ \gamma \left[ \lambda \left[ v(y,x) + \delta W_{bB}^\pi \right] + (1-\lambda)\left[ v(x,x) + \delta W_{bB}^{bP} \right]  \right]\\
    & + (1-\gamma) \bigg[ \xi \left[ \lambda\left[ v(x,x) + \delta W_{bB}^{bP} \right] + (1-\lambda)\left[ v(y,x) + \delta W_{bB}^\pi \right] \right]\\
    & + (1-\xi){\left[ v(y,x) + \delta W_{bB}^\pi \right]} \bigg]\bigg]\\
    & + (1-\rho)\bigg[ \gamma \left[ \lambda \left[ v(y,y) + \delta W_{bB}^\pi \right] + (1-\lambda)\left[ v(x,y) + \delta W_{bB}^{bP} \right]  \right]\\
    & + (1-\gamma)\bigg[ \xi \left[ \lambda \left[ v(x,y) + \delta W_{bB}^{bP}  \right] + (1-\lambda)\left[ v(y,y) + \delta W_{bB}^\pi \right]\right]   \\
    & + (1-\xi) \left[ v(y,y) + \delta W_{bB}^\pi \right] \bigg] \bigg] \Bigg\}.
\end{align*}

\subsubsection{Welfare in NPE-SF}
The voter's expected welfare in a non-pandering equilibrium with stand-firm bureaucracy is, conditional on the politician being good,
\begin{align*}
EU^V_{NPE-SF}\left(\theta^P=g\right)&=\beta \bigg\{\rho \left[v(x,x)+\delta W^{gP}_{gB} \right] + (1-\rho)\left[ v(y,y)+\delta W^{\pi}_{gB} \right]\bigg\}  \\
&+(1-\beta) \bigg\{\rho  \left[ \lambda \left[v(y,x)+\delta W^{\pi}_{bB} \right] + (1-\lambda)\left[v(x,x)+\delta W^{gP}_{bB} \right]   \right]\\
& + (1-\rho)\left[v(y,y)+\delta W^{\pi}_{bB}\right] \bigg\}.
\end{align*}
Differently, when the politician is bad we obtain
\begin{align*}
    EU^V_{NPE-SF}\left(\theta^P=b\right)&= \beta \Bigg\{ \rho \bigg[ \gamma(x) \left[ v(x,x) + \delta W_{gB}^{bP} \right] \\ 
    & + (1-\gamma(x))\left[ \lambda \left[ v(x,x) + \delta W_{gB}^{bP} \right] + (1-\lambda) \left[ v(y,x) + \delta W_{gB}^\pi \right] \right] \bigg] \\
    & + (1-\rho) \bigg[ \gamma(y) \left[ \lambda\left[ v(y,y) + \delta W_{gB}^\pi \right] + (1-\lambda)\left[ v(x,y) + \delta W_{gB}^{bP} \right] \right] \\
    & + (1-\gamma(y))\left[ v(y,y) + \delta W_{gB}^\pi \right] \bigg] \Bigg\}\\
    & + (1-\beta)\Bigg\{ \rho \bigg[ \gamma(x) \left[ \lambda \left[ v(y,x) + \delta W_{bB}^\pi \right] + (1-\lambda)\left[ v(x,x) + \delta W_{bB}^{bP} \right]  \right]\\
    & + (1-\gamma(x)) \bigg[ \psi \left[ \lambda \left[ v(x,x) + \delta W_{bB}^{bP} \right] + (1-\lambda)\left[ v(y,x) + \delta W_{bB}^\pi \right] \right]\\
    & + (1-\psi)\left[ v(y,x) + \delta W_{bB}^\pi \right] \bigg]\bigg]\\
    & + (1-\rho)\bigg[ \gamma(y) \bigg[ \psi \left[ v(x,y) + \delta W_{bB}^{bP} \right] \\
    & + (1-\psi)\left[ \lambda \left[ v(y,y) + \delta W_{bB}^\pi \right] + (1-\lambda)\left[ v(x,y) + \delta W_{bB}^{bP} \right]  \right] \bigg]\\
    & + (1-\gamma(y))\bigg[ v(y,y) + \delta W_{bB}^\pi \bigg] \bigg] \Bigg\}.
\end{align*}


\subsection{Plots of PECB}

The following graphs depict the voter’s ex-ante welfare in a PECB as a function of $\lambda$, for relatively low, intermediate, and high values of $\pi$, $\beta$, and $\rho$.

\begin{figure}[h]
    \centering
\includegraphics[width=\linewidth]{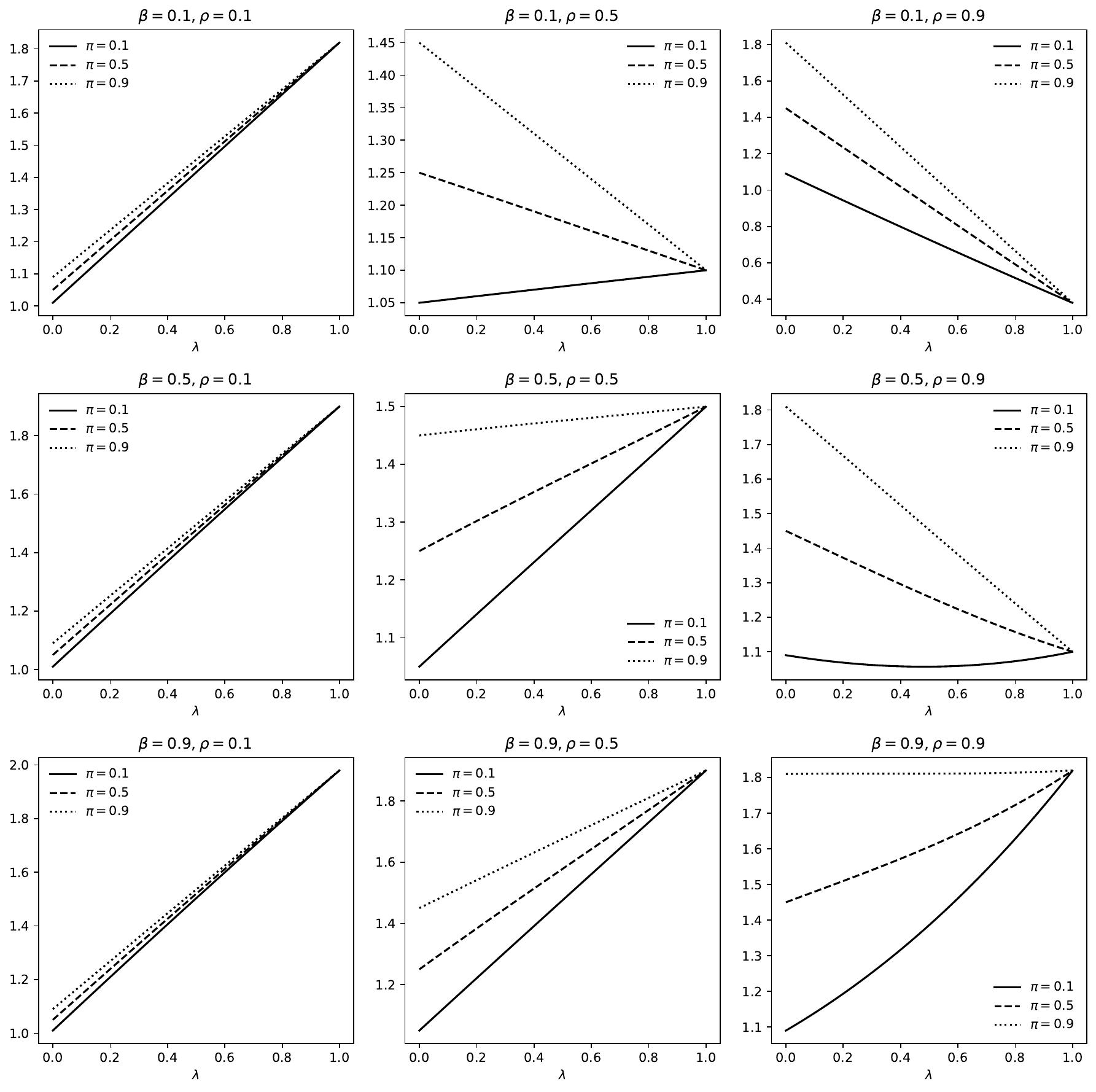}
    \caption{The voter’s expected welfare in a PECB as a function of bureaucratic influence, shown for different parameter combinations.}
    \label{fig:multiple}
\end{figure}

\pagebreak

\clearpage
\addcontentsline{toc}{section}{References}
\bibliographystyle{apacite}
\bibliography{biblio_bureau}
\end{document}